\documentclass[
 reprint,
 superscriptaddress,
nofootinbib,
 amsmath,amssymb,
 aps,
 floatfix
]{revtex4-2}

\usepackage{graphicx}
\usepackage{dcolumn}
\usepackage{bm}
\usepackage{hyperref}
\usepackage{xcolor}
\usepackage{soul}
\usepackage{enumitem}
\usepackage{float}

\newcommand{\PM}[2]{\mbox{($+$#1/$-$#2})}

\newcommand{\MPIfR}{\affiliation{Max-Planck-Institut f\"ur Radioastronomie, Auf dem H\"ugel 69, D-53121 Bonn, Germany}}

\newcommand{\Manchester}{\affiliation{Jodrell Bank Centre for Astrophysics, The
University of Manchester, M13 9PL, United Kingdom}}

\newcommand{\ATNF}{\affiliation{Australia Telescope National Facility, CSIRO Space and Astronomy,  P.O. Box 76, Epping, NSW 1710, Australia}}

\newcommand{\UBC}{\affiliation{ Department of Physics and Astronomy, University of British Columbia, 6224 Agricultural Road, Vancouver, BC
V6T 1Z1, Canada}}

\newcommand{\PI}{\affiliation{Perimeter Institute for Theoretical Physics, 31 Caroline Street North, Waterloo, Ontario, Canada N2L 2Y5}}

\newcommand{\swinburne}{\affiliation{Centre for Astrophysics and Supercomputing, Swinburne University of Technology, P.O. Box 218, Hawthorn, VIC 3122, Australia}}

\newcommand{\ozgrav}{\affiliation{ARC Centre of Excellence for Gravitational Wave Discovery (OzGrav), Australia}}     

\newcommand{\INAF}{\affiliation{INAF - Osservatorio Astronomico di Cagliari, Via della Scienza 5, 09047 Selargius (CA), Italy}}    

\newcommand{\nancay}{\affiliation{Station de Radioastronomie de Nan{\c c}ay, Observatoire de Paris, CNRS/INSU, F-18330 Nan{\c c}ay, France}}

\newcommand{\CNRS}{\affiliation{Laboratoire de Physique et Chimie de l'Environnement et de l'Espace LPC2E CNRS-Universit\'{e} d'Orl\'{e}ans, F-45071 Orl\'{e}ans, France}}

\newcommand{\ASTRON}{\affiliation{ASTRON, Netherlands Institute for Radio Astronomy, Oude Hoogeveensedijk 4, 7991 PD, Dwingeloo, The Netherlands}}

\newcommand{\cagliari}{\affiliation{ Universit\'a di Cagliari, Dipartimento di Fisica, S.P. Monserrato-Sestu Km 0,700 - 09042
Monserrato (CA), Italy}}

\newcommand{\WVU}{\affiliation{Department of Physics and Astronomy, West Virginia University, P.O. Box 6315, Morgantown, WV 26506, USA}}    

\newcommand{\Nashville}{\affiliation{Department of Physics and Astronomy, Vanderbilt University, 2301 Vanderbilt Place, Nashville, TN 37235, USA}}  

\newcommand{\Paris}{\affiliation{LESIA, Observatoire de Paris, Université PSL, CNRS, Sorbonne Universit\'e, Université de Paris, 5 Place Jules Janssen, 92195, Meudon, France}}

\newcommand{\SARAO}{\affiliation{South African Radio Astronomy Observatory, 2 Fir Street, Black River Park, Observatory 7925, South Africa}}

\newcommand{\UEA}{\affiliation{Faculty of Science, University of East Anglia, Norwich NR4 7TJ, UK}}

\newcommand{\UCSD}{\affiliation{Electrical and Computer Engineering, University of California at San Diego, La Jolla, CA 92093, USA}}

\newcommand{\CITA}{\affiliation{Canadian Institute for Theoretical Astrophysics, University of Toronto, 60 St. George Street, Toronto, ON M5S 3H8, Canada}}

\newcommand{\DUNLAP}{\affiliation{David A. Dunlap Department of Astronomy \& Astrophysics, University of Toronto, 50 St. George Street, Toronto, ON, M5S 3H4, Canada}}

\newcommand{\IMAPP}{\affiliation{Department of Astrophysics/IMAPP, Radboud University, P.O. Box 9010, 6500 GL Nijmegen, The Netherlands}}

\newcommand{\IHES}{\affiliation{Institut des Hautes {\'E}tudes Scientifiques, 91440 Bures-sur-Yvette, France}}

\newcommand{\LUTH}{\affiliation{Laboratoire Univers et Th\'eories LUTh, Observatoire de Paris, PSL Research University, CNRS/INSU, Universit{\'e} Paris Diderot, 5 place Jules Janssen, 92190 Meudon, France}}

\newcommand{\arecibo}{\affiliation{Arecibo Observatory, University of Central Florida, HC3 Box 53995, Arecibo, PR 00612, USA}}

\begin{document}

\title{Strong-field Gravity Tests with the Double Pulsar}

\author{M.~Kramer}\email{mkramer@mpifr-bonn.mpg.de}\MPIfR\Manchester
\author{I.~H.~Stairs}\UBC
\author{R.~N.~Manchester}\ATNF
\author{N.~Wex}\MPIfR
\author{A.~T.~Deller}\swinburne\ozgrav
\author{W.~A.~Coles}\UCSD
\author{M.~Ali}\MPIfR\PI
\author{M.~Burgay}\INAF
\author{F.~Camilo}\SARAO
\author{I.~Cognard}\CNRS\nancay
\author{T.~Damour}\IHES
\author{G.~Desvignes}\Paris\MPIfR
\author{R.~D.~Ferdman}\UEA
\author{P.~C.~C.~Freire}\MPIfR
\author{S.~Grondin}\UBC\DUNLAP
\author{L.~Guillemot}\CNRS\nancay
\author{G.~B.~Hobbs}\ATNF
\author{G.~Janssen}\ASTRON\IMAPP
\author{R.~Karuppusamy}\MPIfR
\author{D.~R.~Lorimer}\WVU
\author{A.~G.~Lyne}\Manchester
\author{J.~W.~McKee}\MPIfR\CITA
\author{M.~McLaughlin}\WVU
\author{L.~E.~M\"unch}\MPIfR
\author{B.~B.~P.~Perera}\arecibo
\author{N.~Pol}\WVU\Nashville
\author{A.~Possenti}\INAF\cagliari
\author{J.~Sarkissian}\ATNF
\author{B.~W.~Stappers}\Manchester
\author{G.~Theureau}\CNRS\nancay\LUTH

\date{\today}


\begin{abstract}
Continued timing observations of the Double Pulsar, PSR~J0737$-$3039A/B, which consists of two active radio pulsars (A and B) that orbit each other with a period of 2.45\,hr in a mildly eccentric ($e=0.088$) binary system, have led to large improvements in the measurement of relativistic effects in this system. With a 16-yr data span, the results enable precision tests of theories of gravity for strongly self-gravitating bodies and also reveal new relativistic effects that have been expected but are now observed for the first time. These include effects of light propagation in strong gravitational fields which are currently not testable by any other method. In particular, we observe the effects of retardation and aberrational light-bending that allow determination of the spin direction of the pulsar. In total, we have detected seven post-Keplerian parameters in this system, more than for any other known binary pulsar. For some of these effects, the measurement precision is now so high that for the first time we have to take higher-order contributions into account. These include the contribution of the A pulsar's effective mass loss (due to spin-down) to the observed orbital period decay, a relativistic deformation of the orbit, and the effects of the equation of state of super-dense matter on the observed post-Keplerian parameters via relativistic spin-orbit coupling. We discuss the implications of our findings, including  those for the moment of inertia of neutron stars, and  present the currently most precise test of general relativity's quadrupolar description of gravitational waves, validating the prediction of general relativity at a level of $1.3 \times 10^{-4}$ with 95\% confidence. We demonstrate the utility of the Double Pulsar for tests of alternative theories of gravity by focusing on two specific examples and also discuss some implications of the observations for studies of the interstellar medium and models for the formation of the Double Pulsar system. Finally, we provide context to other types of related experiments and prospects for the future.
\end{abstract}
\pacs{95.30.Sf, 97.60.Gb}

\maketitle


\section{INTRODUCTION: Pulsars as probes of gravitational physics}
\label{sec:intro}

The study of gravitational physics currently benefits from a number of experimental advances which provide unprecedented opportunities for constraining the underlying theory. Many of these methods focus on the study of compact objects, namely neutron stars (NSs) and black holes (BHs), in order to confront the theories to be studied with data obtained under strong-field conditions. A prime example is the availability of ground-based gravitational-wave detectors which provided the first detection of gravitational waves (GWs) with Earth-bound detectors \cite{gwdetect}. Almost 40 years earlier, the first evidence for GWs was provided by observations of the first binary pulsar, PSR~B1913+16 \cite{tfm79,tw82}. Pulsar observations provide unique and complementary experimental constraints, especially thanks to the unrivalled precision enabled by radio astronomical observations and the method of pulsar timing (e.g.~\cite{lk04}), which permit us to trace the orbital evolution of a system over long periods of time.  Pulsar observations furthermore provide valuable information on the structure of NSs, on plasma physics, on  asymmetries in the supernova
explosion of massive stars, and on the ionized content of the interstellar medium.

\subsection{Pulsars}
\label{subsec:intro_bkg}

The discovery of pulsars in 1967 \citep{hbp+68} provided astronomers with natural and extraordinarily stable fly-wheel clocks, opening up the prospect of probing the composition of matter at extremely high densities. Pulsars were soon identified as rotating NSs, exceedingly dense and compact stellar remnants formed in supernova explosions \cite{gol68,lvm68}. Most known pulsars are located within our Galaxy, typically at distances of a few kpc.\footnote{One parsec (pc) is approximately $3.086\times 10^{16}$\,m.} NSs, which typically have a mass of about 1.4 times the mass of the Sun (M$_\odot$) but a radius of only about 12~km, can and in fact are observed to spin very rapidly, up to $\sim 700$ times every second \citep{hrs+06}. 
Precise timing of pulse arrival times at the Earth shows that pulsars can have a rotational stability comparable to that of the best atomic clocks. But not all pulsars have highly stable periods. 
Observations of the rotational instabilities known as ``glitches'' (in
mostly  young pulsars) allow a kind of stellar seismology to probe the NS's solid crust, its liquid interior and the coupling between them (e.g.~\cite{pulsarastronomy, ahcs19}). 

Importantly, many pulsars, especially those with pulse periods in the millisecond range, are both stable rotators and
are in a binary orbit with another star. 
Their great timing stability enables tiny variations in the arrival time due to, e.g., relativistic effects, to be accurately measured, thereby allowing tests of gravitational theories and investigations of many other physical phenomena.

The timing of binary pulsars also delivers measurements of NS masses with unprecedented precision, yielding evidence for NSs with maximum masses of at least 2\,M$_\odot$ \cite{dpr+10,afw+13,cfr+20,fcp+21}. Such large masses are incompatible with a number of theoretical equations-of-state (EoSs)  proposed to describe the properties of super-dense matter; the discovery of even more massive NSs would restrict the families of EoSs even further \cite{Greif_2020}. Other ways of constraining the EoS via pulsar timing include the potential measurement of the moment of inertia (MoI) of a NS.

Soon after the discovery of the first binary pulsar, PSR~B1913+16 in 1975 \citep{ht75a}, it was understood that measurements of 
relativistic effects in binary pulsars would allow investigation of the masses and composition of NSs, with the first attempts at describing relativistic orbital effects as part of a timing model by \citep{sb76} and relativistic spin precession (due to spin-orbit coupling) by \citep{dr74,bo75a}. To first order, measured orbital perturbations are determined by the line-of-sight motion as the pulsar orbits its companion, meaning that motion in the sky plane is not measurable. Consequently, these measurements permit a range of masses and orbital inclination angles.  Multiple relativistic effects must be observed in order to both constrain the masses and begin testing gravitational theories, some of which may depend on the NS composition.  Many
aspects of gravitational theories are best tested when the strongly self-gravitating NSs are 
found as pulsars in systems with short orbital periods, non-zero orbital eccentricities, and orbital planes closely aligned to our line of sight. 

All of these desirable characteristics are present in the still unique Double Pulsar system, PSR J0737$-$3039A/B.  The 2003 discovery \cite{bdp+03} showed it to have a very tight orbit and excellent potential for gravity tests, and shortly thereafter it was shown to be a system
with two orbiting active radio pulsars \cite{lbk+04}. 
The system consists of a ``recycled" 23-ms pulsar (``A'') and the second-born 2.8-s pulsar (``B''). Probably after being ``dead" or at least undetectable for a few million years, the A pulsar was spun up and restored to detectability as its companion star, the progenitor of pulsar B,  evolved and transferred matter and angular momentum to it. Subsequently, the progenitor of B exploded, leading to two NSs in the highly-relativistic, slightly eccentric 2.45-hour orbit that we observe today.  Further details on the  Double Pulsar system are given in Sec.~\ref{sec:systemintro}.  

\subsection{Pulsar Timing}

Pulsar timing analyses begin with measurement of precise pulse times of arrival (ToAs) at the telescope. While many pulsars emit pulsed signals at high energies (X-rays and gamma-rays), it is the radio band which is of most interest here. Two orthogonal polarizations of the incoming electromagnetic wave are recorded by the receiver system of the telescope. These measurements are typically made at radio frequencies of 100's of MHz or several GHz, with signals being Nyquist-sampled at twice the receiver bandwidth.
This sampling is often preceded by a frequency down-conversion and followed by channelisation in frequency using a digital signal processing system that can differ depending on the telescope.
Astronomical radio signals suffer a dispersive delay due to  free electrons in the interstellar medium (ISM), parameterized by the Dispersion Measure (DM), which must be taken into account. Within a given channel bandwidth, e.g.~in our observations 1~MHz, the pulsar signals are ``coherently de-dispersed" to remove this dispersive delay and then folded at the topocentric pulsar period. Again the methods depend on the telescope and the receiver. The folded profiles are averaged over a ``sub-integration" interval, which in our observations is 30~s, resulting in a data cube of pulse amplitude versus pulse phase for each frequency channel.

The next step is to convert the observed pulse profiles into ToAs. This is done by comparing each 
profile, which is time-stamped by the observatory's hydrogen-maser atomic clock, with a carefully prepared template, giving both a ToA and the uncertainty in that ToA \cite{tay92}. The uncertainty due to template matching is independent between ToAs but depends on the strength of the observed pulsar signal, which can vary dramatically due to interstellar scintillation caused by turbulent fluctuations in the ionised ISM density.

However, the uncertainty of the template matching is not the only source of noise. Because of relatively large-scale fluctuations in the ISM electron density, the DM is variable on a time scale of months. 
Pulsars also display intrinsic ``timing noise'' due to variations in spin frequency with a similar time scale and this is then common to all radio
frequency channels.  Both of these noise sources have steep power-law spectra and become negligible on short timescales such as the orbital period of PSR J0737$-$3039A/B.

Finally, the observed ToAs are compared with predictions from a timing model. The model describes
(and in fact, counts) the rotations of the pulsar and accounts for 
physical effects that modify the ToA. The corresponding computations are done
using a timing analysis program, either {\sc Tempo}\footnote{See \url{http://ascl.net/1509.002}} or {\sc Tempo2}\footnote{See Ref. \cite{hem06} and \url{http://ascl.net/1210.015}} in our analysis.

The (usually small) differences between the observed and predicted ToAs are known as ``timing residuals". These residuals are basic to pulsar timing analyses since they reveal effects that are not included in the timing model. The resulting signatures typically have a particular form in the timing residuals, and
some of these may result from previously unknown effects.


\subsection{Tests of theories of gravity} 
\label{subsec:intro_params}

The general theory of relativity, or General Relativity (GR) \cite{ein15}, has passed its experimental tests with flying colours, so far. Despite its successes, it may not be our final answer in describing gravity on a macroscopic scale.
There is a range of parameter space, from the quasi-stationary weak-field regime of the solar system to the strong-field regime of compact objects like NSs and BHs, in all of which one may encounter an experiment, where the theory could be falsified \cite{Wex_2014}. It is therefore important to test different aspects of the predictions of GR and alternative theories with different methods. For instance, observations with gravitational-wave detectors are able to test the highly dynamical strong-field regime and radiative aspects of gravity, but they are not able to test aspects of light-propagation in strong fields. This, on the other hand, and other aspects can be tested with binary pulsars.

The equations of GR, and indeed of alternative theories of gravity, are non-linear and must be approximated for comparison with binary pulsar data.  Damour \& Deruelle \citep{dd85,dd86} provided a leading-order pulsar timing model which includes the effects expected in GR, such as the advance of periastron and Shapiro delay, but which was parameterized in a way that did not assume the validity of GR or any other theory of gravity.  Once measured, these ``Post-Keplerian" (PK) parameters can be used to determine masses (based on an assumed theory) and perform self-consistency tests of theories; this was the approach subsequently taken in timing the Hulse-Taylor pulsar \citep{tw89} and other systems \cite{ksm+06,ks08,kw09,Wex_2014,wk20}. Damour \& Taylor \citep{dt92} expanded the formalism to include parameters based on the pulse profile changes expected in relativistic spin precession, and simulated measurement timescales for several parameters. 

In this framework, any given relativistic theory of gravity provides a description of the PK parameters as functions of the measured Keplerian parameters and the two a priori unknown masses of the binary system. Measuring $n$ PK parameters, where $n>2$, over-determines this system of equations, 
providing $n-2$ independent tests of the studied theory. 
In the work presented here, by measuring more PK parameters than in any other system, and by measuring them also more precisely than usual in any other, we find we need to go well beyond what has been considered in earlier work. As an example, for two of the PK parameters, besides their dependence on the masses and the Keplerian parameters, we need to incorporate their dependence on the MoI of pulsar A in our analyses.

\begin{figure}[htp]
    \centering
    \includegraphics[width=8.6cm]{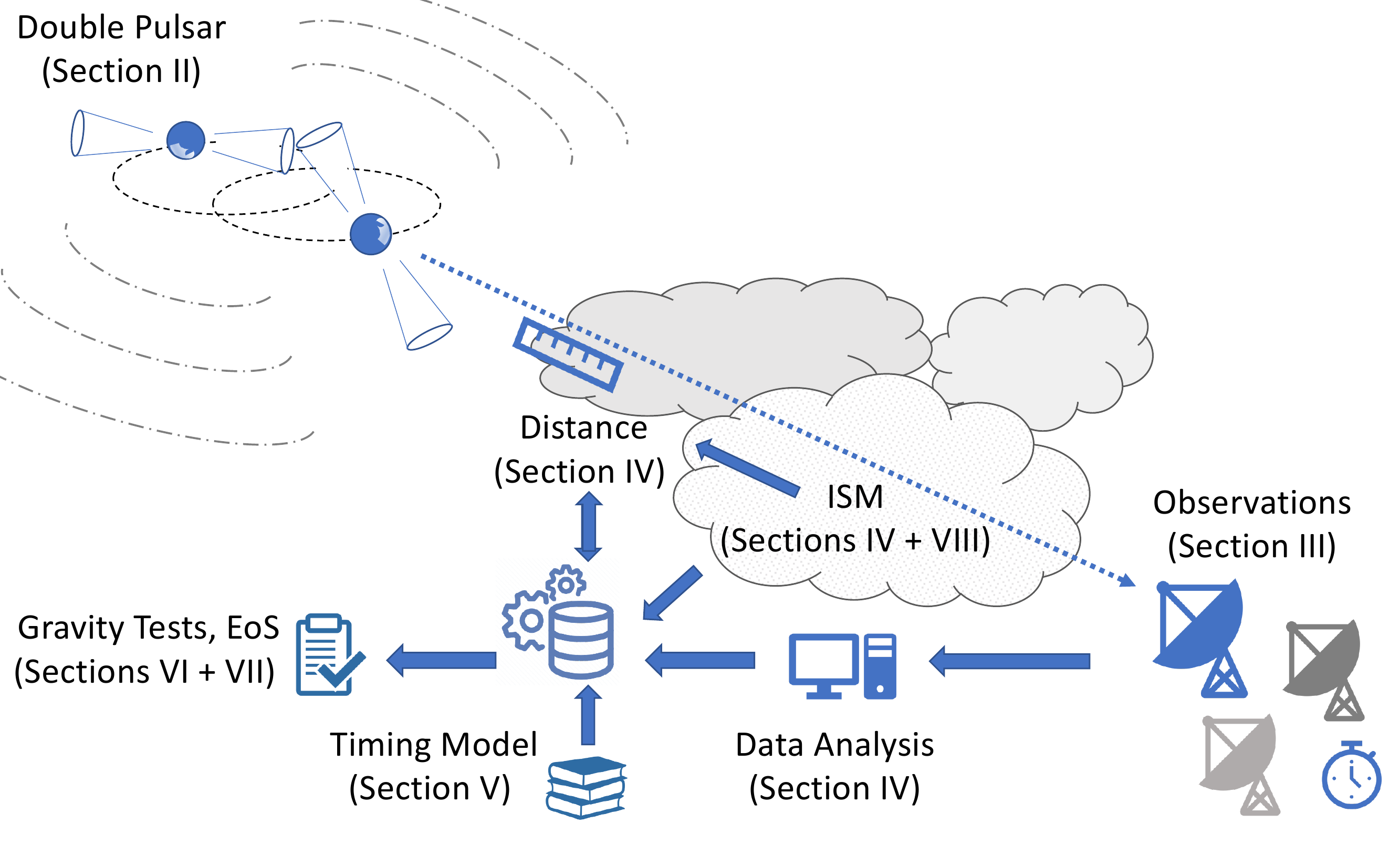}
    \caption{An overview of the experiment described in this work. The named sections provide an orientation to descriptions of the various components.}
    \label{fig:structure}
\end{figure}


\subsection{A Coordinated Gravity Experiment}
\label{subsec_implications}

In many respects we have reached a juncture in the application of binary pulsars to
tests of gravitational physics. As we demonstrate in this paper, from now on we
have to consider a number of effects that could be neglected in the past, but
now require attention and the application of new methods. This is true now for the
Double Pulsar but will eventually also apply for other systems in the future.

The high precision of our measurements that we describe in Section~\ref{sec:obs}
forces us to consider the relativistic mass 
loss of the system due to the pulsar spin-down, while considering the EoS is now 
essential to interpret our observational results. We show explicitly
that relativistic orbital deformation needs to be accounted for in gravity tests based on time-dilation (a combination of second-order Doppler effect and gravitational redshift). Higher-order light propagation effects in strong gravitational fields are also clearly evident in our data.
More specifically, we measure an (higher-order) aberration effect due to the deflection of the radio signal in the gravitational field of the companion, which gives us the rotation sense of A relative to the orbital angular momentum. The advance of periastron, meanwhile, requires two important corrections: a second post-Newtonian term, and a correction due to relativistic spin-orbit coupling that changes the orientation of the orbit over time \cite{ds88,bo75b}. The latter is proportional to the pulsar spin and the (generally negligible) companion spin, and therefore encodes information about the MoI of the pulsar. This offers an opportunity to constrain this NS property, important for determining the EoS, for the first time via pulsar timing.
Consequently, we needed to develop a new timing model to jointly account for all these effects. 

Before we could apply such a model, however, we also needed to determine the distance to the pulsar, as this is an important parameter in many applications of pulsar timing, in particular, in tests of the effects of gravitational-wave emission. Indeed, for the Hulse-Taylor pulsar the limited precision of the known distance (and its acceleration relative to the solar system) has prevented any improvement as a gravity experiment for a considerable time now \cite{dwnc18}. Since the Double Pulsar is relatively close on a Galactic scale, there is a tiny, but measurable, curvature in the signal wavefront.  This results in a small ToA modulation with a period of six months (as we track the source during the Earth's motion around the Sun) which allows us to determine the distance to the Double Pulsar. 

Measurement of this ``timing parallax'' is unfortunately hampered by long time-scale
variations in both the pulsar intrinsic spin and in the intervening ISM, the latter causing variations in the observed DM. The effect of DM variations can be corrected but not the intrinsic spin noise. Hence, we
cannot constrain the pulsar distance using pulsar timing as tightly as we would prefer. These effects are discussed in Sections~\ref{subsec:astrom_timing} \& \ref{subsec:impl_astrometry}.

Fortunately, we have a second independent method of determining the distance to the Double Pulsar. This method uses
high angular resolution imaging of the pulsar with continental-scale radio interferometers, a technique known as ``Very Long Baseline Interferometry" or VLBI. It also depends on the Earth's motion around the Sun, detecting small annual modulations in the pulsar's apparent {\em position} on the sky (``annual geometric parallax"; e.g.\ Ref.~\cite{reid14a}). Since the measurements are made in the plane of the sky, as opposed to pulsar timing which is sensitive to line-of-sight changes, such VLBI measurements are subject to different systematic effects and offer  complementary information. In our analysis,
we adopt the weighted mean of these two independent distance measurements to obtain
the distance with a sufficiently small uncertainty.

With the distance measurement in hand, we develop and present the new
full-relativistic timing model in Sec.~\ref{sec:binary_model} and present
its application to the analysis of our ToAs in Sec.~\ref{sec:bin_params}.
Using the measurement of PK parameters, we obtain precise mass measurements and derive finally a set of independent tests of GR, finding superb internal consistency. 
Our most precise test is one related to GR's prediction for GW emission,  validating this prediction at a level of $1.3\times 10^{-4}$ with 95\% confidence.

The close agreement of the Double Pulsar timing with GR, in turn, means that one can use these observations to place tight constraints on various alternatives to GR. In Sec.~\ref{sec:altgrav} we demonstrate this on the basis of two well-studied gravity theories. The first such theory is ``Damour--Esposito-Far{\`e}se'' (DEF) gravity, a two-parameter mono-scalar-tensor gravity \cite{de93}, containing GR as a limit with the two parameters $\alpha_0 = \beta_0 = 0$. DEF gravity exhibits the typical effects one would expect if the strong equivalence principle (SEP; see e.g.\ \cite{Will:2018}) is violated, for instance the existence of (scalar) dipolar GWs and a location-dependent gravitational constant, both leading to characteristic modifications of the PK parameters \cite{de96}. Moreover, in certain regions of the parameter space, DEF gravity shows genuine non-perturbative strong-field effects, only present in NSs, and therefore not testable in the weak-field regime of the solar system. As shown in detail in Sec.~\ref{sec:altgrav}, with the observations presented here we have limited such deviations from GR, further constraining the $\alpha_0$--$\beta_0$ parameter space of DEF gravity

The second alternative to GR which we confront with our constraints from the Double Pulsar is Bekenstein's tensor-vector-scalar theory (TeVeS) \cite{bek04}, a MONDian relativistic gravity theory that evades the need for Dark Matter in galaxies by a modification of GR. In Sec.~\ref{sec:altgrav} we show that TeVeS is practically incompatible with our observations. Despite the fact that TeVeS has already been falsified by the confirmation that tensor modes of GWs travel with the speed of light \cite{bdkw18}, the Double Pulsar experiment has its own merits. It tests specifically the scalar sector of the theory, and shows that, depending on the details of a MONDian gravity theory, such a theory can be tested by the radiative and strong-field properties of binary pulsars.

As mentioned above, the inhomogeneous turbulent ISM in between the Double Pulsar and Earth leads to slow variations of the DM. Associated refractive variations affect the apparent pulsar position as measured by VLBI with a corresponding impact on the VLBI distance as discussed in Sec.~\ref{subsec:vlbi}.  However, the study of the ISM can also provide independent information on the pulsar distance. These constraints are discussed in detail in Sec.~\ref{sec:implications} where we compare the measured pulsar distance with estimates based on established models of the Galactic ionized gas \citep{cl02,ymw17}. We also investigate the inferred structure of the ISM and bring the discussion full circle, demonstrating consistency with our preferred pulsar distance. As a last consistency check of our results, in Sec.~\ref{sec:DP_formation} we discuss the system's inferred low space velocity and its implications for formation of the binary system.

In Sec.~\ref{sec:prospects} we discuss the future prospects for tests of gravitational physics with the Double Pulsar system and in Section \ref{sec:summary} we conclude with a summary of our results and how these complement other methods for testing gravity. Figure~\ref{fig:structure} provides an overview of both our experiment and the descriptions of its
various components in this paper.


\section{The Double Pulsar}
\label{sec:systemintro} 

In this Section we describe some relevant aspects of the Double Pulsar system in more detail, including the eclipses of the A pulsar emission by the magnetosphere of B and the resultant constraints on the orientation of the spin vectors of the two NSs relative to the orbital angular momentum vector. 

As the original discovery paper \cite{lbk+04} showed, the Double Pulsar system, PSR~J0737$-$3039A/B, is an eclipsing dual-line binary system. Pulsar B is believed to have been formed in a low-kick supernova event, the details of which are still a matter of debate \cite{wkf+06,std+06,tkf+17}. The relatively low eccentricity ($e = 0.088$), the small system transverse velocity ($v_{\rm trans} \sim 10$\,km\,s$^{-1}$) \cite{ksm+06}, and the very small misalignment angle of the spin vector of A relative to the total angular momentum vector ($\delta_{\rm A} < 3.2^\circ$) \cite{mkp+05,fsk+13}  are indeed all signposts of a low-kick birth event of the Double Pulsar.  A retrograde solution ($180^\circ - \delta_{\rm A}$) is possible but considered less likely as it would require a very strong kick with a fine-tuned magnitude and direction.
Furthermore, the prograde rotation of A has been confirmed by using emission properties of B \cite{pmk+18}. In this work, we confirm this independently using the newly seen relativistic effect of aberrational light-bending. 

The eclipses of pulsar A by pulsar B were first detected in the B discovery paper \cite{lbk+04} and shown to have a duration of just 30\,s. The discovery paper already attributed the eclipses to absorption by the magnetosphere of B and this idea was dramatically confirmed when it was shown that the eclipses are modulated at either the rotation rate of B or twice that rate, depending on the orbital phase \cite{mll+04}. This paper also discussed the likelihood that the properties of the eclipse would be affected by relativistic precession \cite{dr74} of the B spin axis about the total angular momentum vector.

In contrast to A, B was born with a significant misalignment of its spin axis relative to the total angular momentum  (which is dominated by the orbital angular momentum). Detailed modelling of the A eclipse modulation by the magnetosphere of B gave a measurement of B's spin misalignment angle of $\delta_{\rm B} \simeq 50^\circ$ \cite{bkk+08}.\footnote{Note, the angle $\theta$ in \cite{bkk+08} is (to a good approximation) $180^\circ - \delta_{\rm B}$, since in the coordinate system of \cite{bkk+08} the orbital angular momentum is closely aligned with the negative $z$-direction.} The rate of the relativistic precession of B's spin vector was also derived from the eclipse modelling: 
The observed rate given by the PK parameter
$\Omega_\mathrm{B}^\mathrm{spin} = 4^\circ.77\pm 0^\circ.66\,\rm{yr}^{-1}$,  is consistent with that predicted by GR (a 70.96-yr precessional period, corresponding to a precessional rate of $5^\circ.073\,\rm{yr}^{-1}$) within the uncertainty of 13\% \citep{bkk+08}. The spin precession causes changes in the observed pulse shapes of B  \cite{bpm+05,ndk+20} and ultimately resulted in B's disappearance from our view in March 2008 \cite{pmk+10}. When the B pulsar returns to our view depends of the actual beam shape as well as the precessional rate \cite{pmk+10,ndk+20}. The beam shape is changing with time, as B's magnetosphere is severely distorted by the wind emerging from A to form a magnetotail on the downstream side \cite{lbk+04,mll+04}. From the length of the eclipse, presumably caused by synchrotron self-absorption of A's radio emission in B's magnetosphere \citep{lt05,bkm+12}, one can estimate that the magnetosphere extends only to about 40\% of B's light-cylinder radius. This is consistent with estimates of the dynamic pressure of A’s wind \cite{mll+04,lyu04}.

The present work builds on the observations and analysis of the Double Pulsar system presented by Kramer {\it et al.} \cite{ksm+06}, together with the measurement of relativistic spin precession by Breton {\it et al.} \cite{bkk+08}. Making use of observations over a 2.5-yr data span, the Kramer {\it et al.} paper presented precise determinations of the relativistic periastron precession of the system, the combined effect of time dilation and gravitational redshift, the Shapiro delay effect due to light propagation in curved spacetime, and the decay of the orbit due to GW emission. These effects are described in more detail in Section~\ref{sec:binary_model}.

Combining these observations led to five strong-field tests of gravity.
These not only made use of six PK parameters previously measured from relativistic effects, but also of the theory-independent\footnote{At least up to 1PN order for boost-invariant gravity theories \cite{dam09}.} 
ratio of the two NS masses uniquely available in this system \citep{ksm+06,bkk+08}. 

Since then, the system has been studied continuously using a number of radio telescopes, with improved data acquisition systems and better sensitivity, resulting in much improved timing precision over time.   Here we report on new results from the timing of A, extending the data span to 16.2\,yr, and describe multiple new phenomena in the system. 


\section{Observations and Signal Processing}
\label{sec:obs}

The vast majority of the results presented in this work are based on extensive pulse timing
experiment made over a 16.2\,yr interval made at six observatories around the world: the Parkes 64-m radio-telescope (``Murriyang") in NSW, Australia, the Green Bank 100-m radio telescope (GBT) in West Virginia, USA, the Westerbork Synthesis Radio Telescope (WSRT) in The Netherlands, the Nan\c{c}ay Radio Telescope (NRT) in France, the Effelsberg 100-m radio telescope in Germany and the Lovell 76-m radio telescope (LT) at Jodrell Bank Observatory in the UK. In addition, we undertook interferometric imaging observations using the Very Long Baseline Array (VLBA) to obtain complementary information on the pulsar's position in the plane of the sky as a function of time.  Details of the data sets from these telescopes and the signal processing methods are described here.

\begin{table*}
\caption{Summary of the datasets. Listed are the corresponding observatory, receiver and instrument used, together with the  center frequency (Ctr freq.), bandwidth (BW), subband bandwidth, the de-dispersion method (DD), the integration time per TOA, the data span, number of TOAs and the root-mean-square residuals (RMS Res).}
\label{tb:datasets}
\begin{tabular}{lcccccccccc} \hline
\noalign{\medskip}
Observatory & Receiver & Instrument & Ctr Freq. & Total BW & Sub-band BW
 & DD$^a$ & $T_s$ & Data span & Nr  &  RMS Res. \\
 & & & (MHz) & (MHz) & (MHz) & & (s) & (MJD) & ToAs & ($\mu$s) \\
 \noalign{\medskip}
 \hline
 \noalign{\medskip}
Parkes & 50cm & AFB  & 680 & 64 & 16 & I & 30 & 53054--55371
   & 2588 &  186.1   \\
Parkes & 50cm & CPSR2 & 732 & 64 & 16 & C & 30 &53954--55100
  & 3480 &  110.6 \\
Parkes & 50cm & DFB3 & 732 & 64 & 16 & I & 30 &55042--56748
   & 32595 &  143.1   \\
Parkes & 50cm & CASPSR$^*$ & 720 & 200 & 12.5 & C & 30 & 55617--58396
   & 2351 &  103.8  \\
Parkes & MB & AFB & 1390 & 256 & 64 & I & 30 &52760--55371 & 111835 & 96.9  \\
Parkes & MB & DFB1 & 1369 & 256 & 64 & I & 30 &54356--54409   & 3179 & 94.5  \\ 
Parkes & MB & DFB2 & 1369 & 256 & 64 & I & 30 &54408--54930   & 9201 & 83.2  \\
Parkes & MB & DFB3 & 1369 & 256 & 64 & I & 30 &54680--56769  & 46853 & 67.9  \\
Parkes & MB & DFB4$^*$ & 1369 & 256 & 64 & I & 30 & 56829--58667 & 37217& 70.1  \\
Parkes & H-OH & DFB4 & 1526 & 512 & 64 & I & 30 &57444--57673 & 11817 & 79.3  \\
Parkes & UWL & DFB4 & 1369 & 256 & 64 & I & 30 &58431--58667 & 12231 & 72.3  \\
Parkes & 10cm & AFB & 3030 &  768 & 768 & I & 30 &52987--54619
                         & 4262  & 94.8  \\
Parkes & 10cm & DFB2 & 3100 & 1024 & 512 & I & 30 &54376--54526 & 690 & 78.6 \\
Parkes & 10cm & DFB3 & 3100 & 1024 & 512 & I & 30 & 54618--56070 & 1120 &  90.8  \\
Parkes & 10cm & DFB4$^*$ & 3100 &  1024 & 512 & I  & 30 &54755--58396  & 19342 & 109.0 \\
Parkes & UWL & DFB4 & 3100 & 1024 & 512 & I & 30 &58264--58619  & 519 & 105.0 \\
GBT & 820\,MHz & GASP$^\dag$  &  820 & 64 & 64 & C &  30 & 53448--55587  & 21531 & 13.6  \\
GBT  & 1400\,MHz & GASP & 1404  & 64 & 64 & C &  30 & 53266--55458 &  7397 & 22.0  \\
GBT  & 820\,MHz & GUPPI &  820 & 200 & 25 & I & 30 &  55002--55431 &  27539 & 18.8 \\
GBT  & 820\,MHz & GUPPI &  820 & 200 & 25 & C & 30 &  55607--58140 & 255301 & 18.2  \\
GBT  & 1400\,MHz & GUPPI &  1500 & 800 & 100 & I & 30 &  55217--56257 &  46521 & 26.7  \\
GBT  & 1400\,MHz & GUPPI &  1500 & 800 & 100 & C & 30 &  56367--58227 &  76173 & 21.3  \\
Nan\c{c}ay & 1.4\,GHz & NUPPI & 1484 & 512 & 64 & C &30 & 55818--58211 & 174173 & 60.8 \\
Nan\c{c}ay & 2.5\,GHz & NUPPI$^\ddag$ & 2520 & 128 & 128 & C   &30 & 56191--57958 & 2594 & 144.9  \\
WSRT & 300cm & PuMa-II$^\ddag$ & 334.6 & 70 & 8.75 & C & 240 &  54519--57119 & 16110 & 89.2 \\
Effelsberg & 20-cm/7-beam & PSRIX$^\ddag$  & 1360 & 150/250 & 15.3/15.9 & C & 160/240 &55722--58328  & 3300 & 40.1 \\
JBO/Lovell   & 1.4\,GHz  & PDFB$^\ddag$ & 1516 & 384 & 12 & I & 240 & 55668--58219  & 33780 & 126.9 \\
All 4-min & -- & -- & -- & -- & -- & --  & 240 & 52760--58667 & 199913  & 13.0 \\
All 30-s & -- & -- & -- & -- & -- & --  & 30 & 52760--58640 & 916648  & 26.0 \\
\noalign{\medskip}
\hline
\noalign{\medskip}
\multicolumn{7}{l}{$^a$ Dedispersion technique: C $=$ coherent, I $=$ incoherent} \\
\multicolumn{7}{l}{$^*$ Band reference instrument for the DM modeling} \\
\multicolumn{7}{l}{$^\dag$ Band reference instrument for the 30-s ToA analysis} \\
\multicolumn{7}{l}{$^\ddag$ Data sets not included in the 30-s ToA analysis}
\end{tabular}
\end{table*}

\begin{figure*}[htp]
    \centering
    \includegraphics[width=17.5cm]{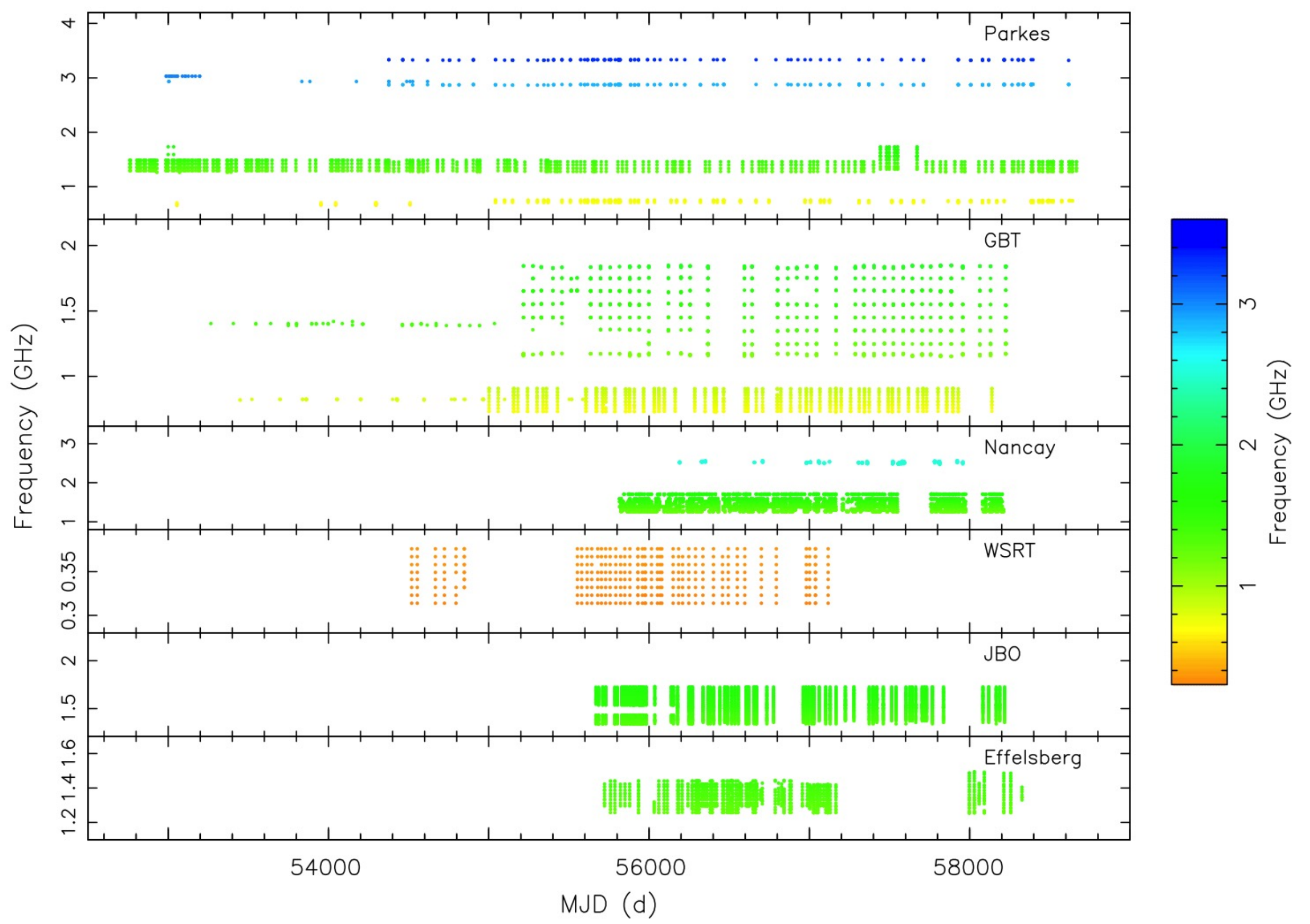}
    \caption{Time and frequency coverage of the 4-min-sampled ToAs from the contributing observatories.}
    \label{fig:obsToAs}
\end{figure*}

Timing observations commenced with the confirmation of the A pulsar search candidate in April, 2003. The Times-of-Arrival (ToAs) included in our timing analysis lie between 2003 May 1, Modified Julian Day (MJD) 52760, and 2019 July 3, MJD 58667, a total span of 16.2 years.  Details of the different data sets used in the analysis are given in Table~\ref{tb:datasets}. This table lists in order, the observatory and/or telescope used, the common designations of the receiver and signal-processing systems, the center frequency and total bandwidth of the data set, the number of frequency channels in the processed data and whether or not the channel data were coherently de-dispersed, the basic sub-integration or ToA sampling time, the data span, the number of ToAs, and the weighted root-mean-square (rms) timing residual of a basic fit to each data set using {\sc Tempo2}. As is discussed in more detail below (Section~\ref{sec:timing_analysis}), two related data sets were used for the timing analyses. For the determination of astrometric parameters and DM variations, 4-min-sampled ToAs were analysed using {\sc Tempo2}; the penultimate line of Table~\ref{tb:datasets} summarizes this data set. Binary and relativistic parameters were determined using a modified version of 
{\sc Tempo}
which uses a new timing model and modifications not available in {\sc Tempo2} (see Section~\ref{sec:binary_model} below for more details) with the 30-s-sampled ToAs. The final line of Table~\ref{tb:datasets} gives the parameters of this combined data set. Figure~\ref{fig:obsToAs} illustrates the time and frequency coverage of the data sets from the different observatories.

In many respects, observations and analyses are identical or similar to those used in Ref.~\cite{ksm+06}. In the following sub-sections, we describe the summaries and details where required for the data-acquisition systems of the contributing observatories.

\subsection{Parkes Telescope}

As summarised in Table~\ref{tb:datasets} the Parkes observations were made in three main bands, centered around 700\,MHz, 1400\,MHz and 3100\,MHz respectively, using a variety of receivers and signal processors. As illustrated in Figure~\ref{fig:obsToAs}, in terms of time coverage the Parkes 1400\,MHz data set is the most complete of those analysed. All data sets had a basic time sampling or sub-integration length of 30\,s. 

The 700\,MHz and most of the 3100\,MHz observations were made using the dual-coaxial 10cm/50cm receiver \citep{gzf+05} while the 1400\,MHz observations primarily used the center beam of the 20cm 13-beam (MB) receiver \citep{swb+96}. For several months in 2016 while the MB receiver was refurbished, the H-OH receiver which covered a 512-MHz band centered at 1526\,MHz was used. Following retirement of the 10cm/50cm receiver in 2018 October, the Ultra-Wideband Low (UWL) receiver \citep{hmd+20} was used to extend the 1400\,MHz and 3100\,MHz data sets. All of these receivers have (or had) orthogonal linearly polarised receptors.   Signal processing systems used with these receivers comprised three groups, analogue filterbanks (AFB), digital filterbanks (DFB) and two coherent de-dispersion systems used for the 700\,MHz band, CPSR \citep{hot07} and CASPSR \citep{vb11}. The AFB systems had channel bandwidths of 0.5\,MHz for the 700\,MHz band and 1\,MHz for the higher-frequency bands and utilised 1-bit digitisers. The DFB systems used 8-bit digitisers and implemented polyphase filterbanks with Field Programmable Gate Array processors (FPGAs). Both the AFB and DFB systems utilised incoherent de-dispersion to sum data across sub-bands. More details of these signal processing systems can be found in \citet{mhb+13}.

\subsection{Green Bank Telescope}

Observations at the Green Bank Telescope (GBT) were carried out at 820\,MHz and 1400/1500\,MHz. Typically, 5-hour observations were conducted with monthly cadence and at alternating frequencies. Twice a year, we usually conducted concentrated observing campaigns in April/May and October/November using observations at 820\,MHz on consecutive days to separate short- from long-term orbital changes. The Green Bank Astronomical Signal Processor (GASP \cite{dem07}) carried out 8-bit Nyquist sampling of the incoming dual-linear-polarization signal, after which it performed coherent dedispersion in software on a Linux-based cluster for each of several 4-MHz channels. Later observations were carried out with the Green Bank Ultimate Pulsar Processing Instrument (GUPPI \cite{drd+08}). Initially, the 8-bit digitized data streams were detected and dedispersed incoherently, while subsequently a coherent de-dispersion mode was available. Coherent dedispersion was implemented later at 1500\,MHz than at 820\,MHz.  GASP and GUPPI were operated in parallel during a transition period in order to carefully calibrate an expected small instrumental time offset between the two systems. For those overlapping epochs, in the final analysis we usually chose the GUPPI data set, unless calibration issues with GUPPI occurred. We note that earlier data sets recorded with the previous SPIGOT and BCPM processors  were not included in the current analysis (cf., Ref.~\cite{ksm+06}). A noise diode was used for flux calibration, with observations of the stable continuum source B1442+10 used to calibrate the diode noise power.  Flux calibration, interference excision and sub-band formation were carried out for GASP using {\sc ASPFITSReader}\footnote{\url{http://ascl.net/1509.003}} \cite{fer08}, while for GUPPI they were carried out using {\sc psrchive} \cite{psrchive}.  The GUPPI interference excision was based on early work carried out by \citet{gra16}.  

\subsection{Nan\c{c}ay Radio Telescope}

Our data set includes the results of high-cadence observations with the Nan\c{c}ay Radio Telescope (NRT), which is a transit telescope of the Kraus-type design with a collecting area equivalent to a 94-m parabolic dish. While observations of the Double Pulsar commenced late 2004, using an earlier backend, we make use of observations conducted since late 2011 with the Nan\c{c}ay Ultimate Pulsar Processing Instrument (NUPPI). NUPPI is a baseband recording system, similar to GUPPI used at the GBT; see, e.g., Ref.~\cite{vcf+20} for more details. Data were taken with the 1.4\,GHz and 2.5\,GHz receivers which have frequency coverage of 1.1--1.8\,GHz and 1.7--3.5\,GHz, respectively. Most of the observations with the 1.4\,GHz receiver were made at a central frequency of 1484\,MHz, while those with the 2.5\,GHz receiver were generally centered at 2520\,MHz.  The effective central frequency for a given observation depended on the excision of radio-frequency interference (RFI) which was determined via visual inspection.  The data were polarization-calibrated using \textsc{psrchive}'s \textsc{SingleAxis} method. Observations of a reference noise diode were made prior to each pulsar observation and the data were calibrated, correcting for differential phase and amplitude between the two polarizations, before being folded into sub-integration profiles.

\subsection{Effelsberg 100-m Radio Telescope}

Observations with the 100-m Radio Telescope at Effelsberg of the Max-Planck-Institut f\"ur Radioastronomie were conducted with two different 20\,cm receiver systems. For most sessions, the central beam of the 7-beam system with a bandwidth of 250\,MHz was used. When this was not available, a single-pixel receiver was used, providing a bandwidth of 150\,MHz. All data streams were coherently dedispersed using the ROACH-based PSRIX signal processor (see Ref.~\cite{lkg+16} for more details). As the Double Pulsar barely rises above the hills surrounding the telescope, full orbital tracks are not possible and the system temperature is significantly increased because of spillover. Consequently, longer integration times of 160s and 240s were used. Standard calibration procedures were applied as described in Ref.~\cite{lkg+16}.

\subsection{Lovell Radio Telescope}

Observations were also conducted with the 76-m Lovell telescope at Jodrell Bank of the University of Manchester. We used a receiver covering a frequency range 1300-1700\,MHz, with a maximum usable bandwidth of 400\,MHz. Data were acquired using a ROACH-based system that is essentially identical to the Effelsberg PSRIX-system (see Ref.~\cite{bjk+16} for details). Gain and polarization calibration were achieved by monitoring a noise-diode signal and manual adjustments to power levels. Similar to the methods used at other observatories, RFI was excised following visual inspection of the data.  

\subsection{Westerbork Synthesis Radio Telescope}

The Westerbork Synthesis Radio Telescope (WSRT) consists of fourteen 25-m dishes arranged in an East-West direction. Observations were made at 334\,MHz, the lowest frequency included in our data set. Dual-polarization signals were acquired with the (nearly real-time) coherently de-dispersing PuMaII instrument \cite{ksv08} using a total bandwidth of 70\,MHz with 8.75-MHZ wide sub-bands. Gain and phase differences between the two polarizations were adjusted during the phased-array calibration of the system  (cf., Ref.~\cite{dcl+16}). 

\subsection{Very Long Baseline Array}

The VLBA consists of ten 25-m dishes spread across the continental USA, Hawaii, and the Virgin Islands. Dual-polarization observations were taken in standard continuum mode with 256 MHz of observing bandwidth distributed over eight 32 MHz wide sub-bands centred at 1.56 GHz.  Further details are given in Sec.~\ref{subsec:vlbi} and Appendix~\ref{appdx:vlbi}.


\section{Pulse Timing and VLBI Data Analyses}
\label{sec:timing_analysis}

In this section we describe how the pulse timing observations discussed in Sec.~\ref{sec:obs} above are processed to form pulse times of arrival (ToAs), how these ToAs are analysed to determine the pulsar astrometric parameters and the long-term DM variations, and the results obtained. We also outline the methods used to analyse the VLBA interferometric data and the astrometric results obtained. A key result from these observations is our best estimate of the distance to the Double Pulsar, an important parameter in the gravitational tests described in Sec.~\ref{sec:binary_model}. Detailed descriptions of the observations, analysis methods and results are deferred to Appendix~\ref{appdx:vlbi}.

For the pulse timing analyses, two sets of ToAs were formed: a) a set with 30-s spacing that were used for the relativistic-parameter analyses described in Sec.~\ref{sec:binary_model} and b) a set with 4-min spacing that were used for the timing-based astrometry and the analysis of DM variations described in this section. 

Analysis of the timing data sets was carried out in two distinct phases: the first phase used {\sc Tempo2} to derive the pulsar astrometric parameters and the DM variations based on the 4-min-sampled ToAs, whereas the second phase used {\sc Tempo} to derive the binary and relativistic parameters based on the 30-s-sampled ToAs. As indicated in Table~\ref{tb:datasets}, data sets from Westerbork, Effelsberg and Jodrell Bank observatories were not included in the {\sc Tempo} analysis since they had insufficient time resolution to correctly trace the Shapiro delay curve. The Nan\c{c}ay 2.5\,GHz data set was also omitted in the second analysis phase as the ToA precision was limited and their number was relatively small. For the astrometric and DM analyses, the binary and relativistic parameters were held fixed and the 4-min sampling made the various Monte Carlo and bootstrap analyses (described in Section~\ref{subsec:astrom_timing} below) more tractable. The 30-s sampling interval chosen for the binary and relativistic analyses is consistent with that used for the Kramer et al. (2006) analysis (Ref.~\cite{ksm+06}) and represents a good compromise between our  ability to resolve orbital effects, signal-to-noise ratio (S/N) considerations (and hence ToA uncertainties) and the potential impact of pulse jitter (see, e.g., Ref.~\cite{sod+14}).

\subsection{Measurement of ToAs}
\label{subsec:toas}

In general, similar procedures for processing observational data and formation of ToAs were adopted for all telescopes and data acquisition systems. After removal of obvious narrow-band and broad-band transient radio-frequency interference (RFI), data were calibrated, in most cases making use of a short observation of an injected noise diode signal that preceded or followed the pulsar observation. The noise diode signal itself was calibrated in flux density units by comparing it to a stable continuum source, normally Hydra A (3C\,218) at Parkes and B1442+10 at Green Bank, and in polarization properties by analysis of rise-to-set tracks of the strong millisecond pulsar PSR\,J0437$-$4715 at Parkes \cite{van04c} and the well-studied calibrator PSR\,B1929+10 at Green Bank. Following calibration, the   data were formed into sub-bands whose width varied according to frequency band as listed in Table~\ref{tb:datasets}, with wider bandwidths used at higher frequencies because of a) the reduced DM smearing, b) to maintain sufficient S/N, and c) so that the timing program ({\sc Tempo} or {\sc Tempo2}) properly accounted for the fact that data at different frequencies received at a given time corresponds to different orbital phases at emission because of the differential dispersion delay (see e.g.\ Refs.~\cite{ksm+06,hem06}). The sub-bandwidths chosen were sufficiently small to ensure that this effect was not significant. 

For the Parkes data sets, ToAs were obtained using the {\sc Psrchive} routine {\sc pat} with the Fourier phase gradient algorithm \cite{tay92}. Separate pulse profile templates derived were used for the three main receiver bands (50\,cm, 20\,cm and 10\,cm). These were analytic templates aligned in phase and
first used in Ref.~\cite{ksm+06}, derived by fitting Gaussian components to pulse profiles that were obtained from averaging a large number ($>$100,000) of individual pulses. This helps to avoid ``self-mirroring'',
i.e.~using a template based on the same pulse profiles that are to be timed, which leads to
potential biases in ToA uncertainties due to correlated noise contributions \citep{hbo05,ksm+06}. These same templates were used to derive ToAs for data from the Green Bank GASP instrument and from the WSRT, Effelsberg and JBO/Lovell telescopes. The wider-bandwidth GUPPI and NUPPI data generally have higher S/N than those from GASP and use of the analytic templates resulted in systematics in the residuals that worsened with increasing S/N. For GUPPI and NUPPI we therefore constructed separate templates based on multiple observations aligned and smoothed using {\sc ASPFITSReader} and {\sc psrchive} utilities such as {\sc psrsmooth} \cite{dfg+13}. 

As mentioned above, 4-min-averaged ToAs were used for the astrometric and DM analyses. The method used to obtain the 4-min ToAs differed for the different observatories. For Parkes, the original 30-s-sampled data sets were directly summed to form 4-min-sampled data sets using standard {\sc Psrchive} routines and ToAs determined as for the 30-s data. For data from Green Bank and Nan\c{c}ay, the 30-s ToAs were averaged over 4-min intervals using the {\sc Tempo2} {\sc averageData} routine. This routine determines the weighted mean residual and its uncertainty for ToAs in the 4-min sample and applies the necessary pulse phase offset to a central ToA. Data from WSRT, Effelsberg and JBO/Lovell were used directly. 

Because of the short sampling time and the effects of pulse jitter, pulse profiles often varied greatly in shape, e.g., having the interpulse stronger than the main pulse. Consequently ToAs were sometimes very discrepant in phase. These discrepant ToAs were simply removed from the data set to produce a ``clean'' set for each telescope and instrument. Next, we determined scaling factors ``EFAC'' and ``EQUAD'' for each instrument (see, e.g., Ref.~\cite{hem06}). These factors  modify the initially derived uncertainty of each ToA, $\sigma_i$, according to $\sigma_i' = {\rm EFAC} \times \sqrt{ \sigma_i^2 + {\rm EQUAD}^2} $ such that the reduced $\chi^2$-value obtained after fits of the initial timing model is close to unity. This ensures an appropriate weight for a given data set in the final combined least-squares fit to our timing model. For well-behaved data sets free of systematic errors or radio interference signals, one expects the ToA uncertainties to reflect the true measurement uncertainties, hence EFAC~$\sim 1$ and EQUAD~$\sim 0$. For the 30-s data set, we determined EQUAD values for each observing epoch and excluded those where the resulting EQUADs showed large deviation from zero. Then, a single EFAC value was assigned to the each of the data sets listed in Table~\ref{tb:datasets}. For the 4-min data sets, EFACs and EQUADs were determined for each instrument and each band for each observatory using the {\sc efacEquad} function of {\sc Tempo2}. Because of the preliminary ``cleaning'', the EFACs were generally close to unity and in most cases less than 1.2. EQUADs were in most cases 10\,$\mu$s or less, although for some systems they were up to 40\,$\mu$s. 

Because of profile frequency evolution and the different templates used for different RF bands, there are systematic offsets between ToAs for  different bands. For Parkes data, for each RF band, we determined ToA offsets for each instrument relative to a reference instrument using the
{\sc Tempo} and {\sc Tempo2}
``JUMPS'' facility. The reference instrument for each band is indicated in Table~\ref{tb:datasets}. In most cases, there was significant overlap between pairs of data sets, which greatly improved the precision of the JUMP determination.

For Green Bank GUPPI data, Nan\c{c}ay 1.4\,GHz data and WSRT data, the fractional bandwidths are much larger, and it was necessary to determine ToA offsets between sub-bands. In the 4-min ToA analysis, in each case a sub-band close to the center of the band was chosen as reference. Since these jumps are best determined from individual data sets, they were held fixed in the subsequent analyses. In the 30-s ToA analysis, we determined phase
offsets between the analytic templates and the separately derived GUPPI and Nan\c{c}ay templates. We accounted for these offsets by applying phase shifts to the corresponding ToAs prior to applying the timing model, thereby effectively aligning all templates in phase.

\subsection{Timing astrometry and DM variations}
\label{subsec:astrom_timing}

The 4-min-sampled ToAs from all observatories were analysed using {\sc Tempo2} to determine the astrometric parameters for the pulsar and the variations in DM across the total data span. All analyses used the DE436 solar-system ephemeris\footnote{\url{https://naif.jpl.nasa.gov/pub/naif/JUNO/kernels/spk/de436s.bsp.lbl}} to transfer ToAs to the SSB frame.  Studies of the effects of using different solar-system ephemerides for this transfer (see e.g.~\cite{vts+20,cgl+18}) show differences much below the ToA precision of our data sets. 

Initial parameters for the {\sc Tempo2} fit were obtained from the {\sc Tempo} analysis of the 30-s ToAs. Pulsar astrometric parameters (right ascension, declination, proper motions and annual parallax), the pulsar rotational frequency ($\nu$) and its first time derivative ($\dot\nu$), and the secular orbital terms ($\dot\omega$ and $\dot P_b$) were fitted. All other binary and relativistic parameters are better determined by the 30-s ToA and, hence, were held fixed and updated in following iterations.

DM and common-mode (i.e. wavelength-independent) variations were fitted for, typically at 100-day intervals in the central part of the data set and at somewhat longer intervals at the ends where the frequency coverage is poorer, using the methods described by Keith et al. \cite{kcs+13}. In addition, jumps relative to the Parkes 20cm data set were fitted for all observatory bands listed in Table~\ref{tb:datasets}. DM variations are relative to the reference DM, held at 48.917208\,pc\,cm$^{-3}$. The reference solar-wind electron density was set to zero for all fits. Because of the high ecliptic latitude ($-51.2^\circ$) the annual DM variations due to the solar wind are quite smooth, so they are mostly absorbed by the fitted DM offsets. Transient solar events, such as coronal mass ejections, are not absorbed in the fit but are smaller and short-lived. We note that the worst-case solar-wind DM contribution is negligible compared with the errors in estimating DM($t$) (see Appendix~\ref{appdx:dmvar}). Transient events also have a negligible effect on measured ToAs.
Although ToA offsets between observatory bands resulting from DM variations are absorbed by the jumps, variations within each data set are preserved and these determine the derived DM variations. Higher-order rotational-frequency derivatives up to the fourth ($\ddddot\nu$) were held fixed at values determined in the {\sc Tempo} analysis of the 30-s data set because of covariances with the common-mode variations. This whole process was iterated as the system parameters were refined. 

Astrometric parameters and DM/common-mode variations were initially determined by iterated fits to the entire data set for the default 100-day sampling of the DM and common-mode variations. Because of the known covariance between astrometric parameters and the sampling of the DM/common-mode variations, we then undertook a Monte Carlo analysis, randomly varying the spacing and phase of DM/common-mode sampling within defined limits, normally 64 -- 256 days, for the spacing and for phase within $\pm 0.5$ of the current spacing. After 4096 iterations, the distribution for each parameter was fitted by a Gaussian function to give estimates of the parameter and its 1-$\sigma$ uncertainty. 

To test for systematic effects related to data from a particular observatory, the Monte Carlo analysis was repeated on data sets that a) excluded the GBT data, b) excluded the Parkes data, and c) contained just the GBT and Parkes data. These tests had 2048 iterations and they were analysed in the same way as the full data set. 

As an additional check on the reliability of the parameter uncertainties, we performed a bootstrap analysis (with replacement) of the full data set using the 100-day DM/common-mode sampling. In a similar manner to the Monte Carlo analysis, a Gaussian function was fitted to the parameter distributions after 2048 iterations. 

At time scales longer than a few months, steep-spectrum or ``red" noise in the intrinsic pulsar spin rate which has a power spectrum of the form $f^{\alpha}$ (where $\alpha\ll 0$) becomes important. In the initial analysis, much of this red noise was removed by fitting for the first four pulse frequency derivatives. However, proper analysis of timing data sets in the presence of red noise requires the estimation of a noise covariance matrix and the use of generalized least squares (GLSQ) approach when fitting the timing model \cite{chc+11}. The DM/common mode analysis does not require GLSQ because the common mode absorbs all the red noise and the residuals are ``white", i.e., flat-spectrum. In this case, a weighted least squares analysis is sufficient. For the GLSQ analysis, the covariance matrix has $N_{\rm ToA}^2$ elements, and it is not feasible to perform a full Monte Carlo analysis. We chose to estimate the effects of the GLSQ analysis compared to the approximate method described above by forming smaller averaged data sets with 16-min and 32-min sampling using the {\sc Tempo2} {\sc averageData} routine. Even for these smaller data sets, it was necessary to modify the indexing of the covariance matrix and eliminate unnecessary matrix inversions in {\sc Tempo2}. The best-fit value of $\alpha$ was $-3.0$ but changes of $\pm0.3$ made very little difference in the parameters or their uncertainties. The results showed that the initial analysis  under-estimated the parallax uncertainty by about 75\%, but did not significantly bias the value. For the other astrometric parameters, the increase in uncertainty was less than 50\% (zero for the position terms), but there were some biases, in all cases less than 50\% of the combined uncertainty.

Results for the astrometric parameters from the Monte Carlo analysis of the full 4-min-sampled data set modified to reflect the biases and uncertainty changes indicated by the GLSQ analyses of the shorter data sets are listed in the second column of Table~\ref{tb:astrom_results}\footnote{We note that we have also explored alternative methods, e.g., using the ``DMX'' method implemented in {\sc Tempo}, in order to determine the astrometry and its uncertainties. Using simulated timing data with parallaxes of known sizes injected into our datasets, we determined that the method implemented here appears to be the most reliable in recovering the correct astrometry with appropriate uncertainties.}.
The final three columns of Table~\ref{tb:astrom_results} list Monte Carlo results from the partial data sets. All results are within or close to twice the combined uncertainty of the final result, with the parameters from the GBT-omitted data set differing the most. This shows that the GBT data set with its high S/N has the greatest influence on the final results. It is notable though that the measured parallax is not significantly biased by the data set from any one observatory.

Results from the bootstrap analysis are given in the third column of Table~\ref{tb:astrom_results}.  Uncertainties from the bootstrap analysis are typically about half those from the Monte Carlo (MC) analysis. The mean values are also offset, although generally within the combined uncertainties. Both of these differences result from the covariance of the astrometric parameters and the sampling of the DM/common-mode variations.

\begin{figure}[htp]
    \centering
    \includegraphics[width=8.5cm]{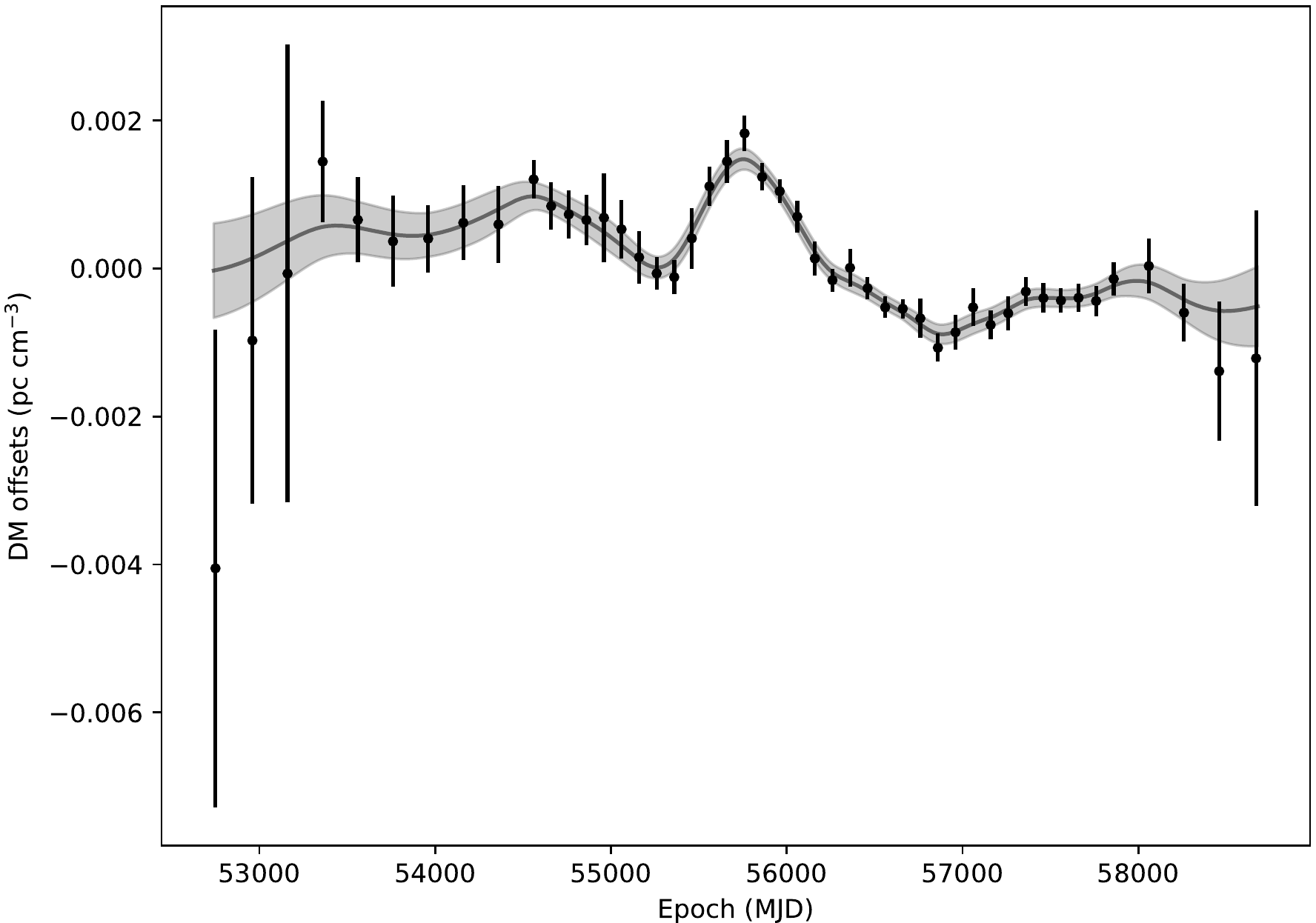}
    \caption{Variations in DM relative to the reference value, 48.917208\,pc\,cm$^{-3}$, derived by averaging over values from the Monte Carlo analysis at 100-day intervals (points with error bars). The solid line is the result of using a Gaussian Learning Process to describe the data with resulting uncertainties indicated by the grey band.
    }
    \label{fig:dmcurve}
\end{figure}

Figure~\ref{fig:dmcurve} shows the DM variations derived by averaging the Monte Carlo DM-offset results for the 4-min-sampled data set in 100-day bins across the data span. As noted above, the DM results are unaffected by the GLSQ analysis.   Apart from a few points with large uncertainties near the ends of the data span where there was less radio-frequency coverage, the DM shows fluctuations on timescales of a few hundred days with amplitude of up to 0.0015\,pc\,cm$^{-3}$ and a downward trend of approximately $-1.41\times10^{-4}$\,pc\,cm$^{-3}$\,yr$^{-1}$ over the data span.

For the final {\sc Tempo} analysis of the 30-s ToAs, the astrometric parameters and DM variations from the final {\sc Tempo2} analysis of the 4-min data were applied using a method described in Section~\ref{sec:timpar}. The final analysis takes into account the DM variations, which we can describe by fitting a smoothly varying curve that reflects the uncertainties of the DM measurements. This was achieved using a Gaussian Learning Process as implemented in the Python library {\sc George} \cite{rw06}. We have explored a number of stationary kernels with very similar results. We finally adopted the {\sc matern32kernel} leading to the modeling of the DM offsets as shown in Figure~\ref{fig:dmcurve}. Each ToA was modified with a DM offset dithered according to the interpolated DM uncertainty at that epoch.

The observed DM variations imply the presence of transverse phase gradients which will shift the apparent position of the pulsar on a time scale of months. These  shifts can be estimated from the measured intensity scintillation \cite{rcn+14}. They are of the same order as the parallax uncertainty and will affect both timing and VLBI astrometry, which
we present below in Section~\ref{subsec:vlbi}. We discuss these effects in Section~\ref{subsec:impl_astrometry}. We believe that the interstellar electron density fluctuations responsible for both the DM variations and the scintillation occur in or close to the Gum Nebula, as discussed in Section~\ref{subsec:impl_ism}. To avoid the poorly sampled data at the ends of the full data span, this latter discussion is based on a restricted data set between MJDs of 53500 and 58300.

\begin{table*}
\caption{Astrometric results from the analysis of the 4-min timing datasets (see the text for details of the different datasets and analysis procedures, i.e. bootstrap (boot.) and Monte-Carlo (MC)). Numbers in parentheses are 1-$\sigma$ uncertainties relative to the last quoted digit of the parameter value. For the right ascension we only show the seconds of $07^{\rm h}37^{\rm m}ss.s^{\rm s}$, and for the declination only the arcseconds of $-30^\circ 39'ss.s''$.}
\label{tb:astrom_results}

\begin{tabular}{lr@{\hspace{2.5em}}rrrr}
\hline
\hline
\noalign{\smallskip}
Data set & {\bf All data} & All data Boot. & All ex. GBT MC & All ex. Pks MC & Pks+GBT MC  \\
\noalign{\smallskip}
\hline
\noalign{\smallskip}
Right ascension, $\alpha$ (J2000)  (s)  & 51.248115(10) & 51.248111(6) & 51.248125(10)  &51.248105(10)  & 51.248105(9) \\ 
Declination, $\delta$ (J2000)  ($''$) & 40.70485(17) & 40.70496(11) & 40.70544(21) &  40.70529(23)  & 40.70525(18) \\
Proper motion R.A., $\mu_\alpha$  (mas\,yr$^{-1}$)   & $-$2.567(30)  & $-$2.570(16) & $-$2.618(15)  &$-$2.581(16)  & $-$2.582(16) \\
Proper motion Dec., $\mu_\delta$  (mas\,yr$^{-1}$)  & 2.082(38)  &  2.081(19) & 2.178(35) & 2.102(35)  & 2.105(25) \\
Parallax, $\pi_{\rm t}$  (mas)   & 2.15(48) & 2.40(11) & 2.41(10) & 2.01(22) & 2.22(24) \\
Position epoch (MJD) & 55045 & 55045 & 55045 & 55045 & 55045 \\                              
Data span  (MJD) & 52760--58667 & 52760--58667 & 52760--58667 & 53266--58328  & 52760--58667\\
$N_{\rm ToA}$   & 199913 & 199913 &129787 &  151883 & 118156 \\ 
\noalign{\smallskip}
\hline
\noalign{\smallskip}
\multicolumn{5}{l}{$^*$ See text for details of the different data sets and analysis procedures.}
\end{tabular}
\end{table*}

\subsection{VLBI astrometry}
\label{subsec:vlbi} 

Astrometric terms in the pulsar timing model can be compared to independent estimates for these parameters obtained via radio imaging.  
PSR~J0737$-$3039A/B was previously targeted by an astrometric program using the Australian Long Baseline Array \cite{dbt09} which measured an annual geometric parallax of $0.87\pm 0.14$ mas. This measurement is in tension with the timing parallax presented in Table~\ref{tb:astrom_results}.  In an attempt to resolve this, we undertook an extended astrometric program with the VLBA.  The details of these observations, data reduction, and robustness testing using in-beam reference sources are given in Appendix~\ref{appdx:vlbi}, and the resultant astrometric parameters are summarised in Table~\ref{tab:shortvlbiresults}. 

For this analysis, the proper motion measured by pulsar timing (and listed in Table~\ref{tb:astrom_results}) was applied as a prior to the VLBI fit.  While some previous studies have highlighted substantial disagreement between timing and VLBI proper motions (e.g.\ Ref.~\cite{dvk+16}), the uncertainty in the timing proper motion for PSR~J0737$-$3039A/B is sufficiently small that even an error exceeding those seen previously for other millisecond pulsars would not materially affect the VLBI parallax uncertainty.  Details of the effects of proper motion fitting are included in Appendix~\ref{appdx:vlbi}.  The timing and VLBI reference positions are consistent when propagated to the same reference epoch, although the precision of the VLBI reference position for PSR~J0737$-$3039A/B is relatively low, for the reasons discussed in Appendix~\ref{appdx:vlbi}.

\begin{table}
	\centering
	\caption{VLBA astrometric results. Listed are right ascension (R.A.), declination (Dec.), proper motion in both coordinates, parallax and position epoch.}
	\label{tab:shortvlbiresults}
	
	\begin{tabular}{lr}
		\hline
		\hline
		\noalign{\smallskip}
R.A., $\alpha$ (J2000) & $07^{\rm h}37^{\rm m}51.\!\!^{\rm s}247(1)$ \\
Dec, $\delta$ (J2000)  & $-30^{\circ}39'40.\!\!''68(1)$ \\
Parallax, $\pi_{\rm v}$  (mas)      & $1.30^{+0.13}_{-0.11}$ \\
Proper motion in R.A., $\mu_\alpha$ (mas yr$^{-1}$)$^*$ & $-2.567\pm0.030$  \\
Proper motion in Decl., $\mu_\delta$ (mas yr$^{-1}$)$^*$ & $\phantom{-}2.082\pm0.038$ \\
Position epoch (MJD)                & 58000 \\
		\noalign{\smallskip}
		\hline
		\noalign{\smallskip}
	\end{tabular}
	\begin{minipage}[b]{0.95\linewidth}
		$^*$ A prior based on the timing proper motion was applied as described in Appendix~\ref{appdx:vlbi}.
	\end{minipage}
\end{table}

The quantity of most interest, the parallax, lies between the previous VLBI result and the current timing value, but is more precise than either.  We discuss the difference between the distance implied by a straightforward inversion of the VLBA parallax ($770 \pm 70$\,pc) and the timing parallax $465^{+134}_{-85}$\,pc in Section~\ref{subsec:impl_astrometry}.  For the reasons discussed in Appendix~\ref{appdx:vlbi}, we adopt the weighted mean of our new VLBA parallax and the timing parallax as our best estimate of the system parallax, and use the resultant distance ($735 \pm 60$\,pc) and transverse velocity ($11.5 \pm 1.0$\,km\,s$^{-1}$) for the remainder of the paper.  We stress that the precision of the distance measurement does not (yet) affect the gravitational tests presented below.

\section{Binary and relativistic-parameter analysis methods }
\label{sec:binary_model}

In this section we describe the methods used to derive the relativistic terms that form the basis of the tests of gravitational theories, the key results of this paper, from the observational data and results described in Sec.~\ref{sec:obs} and Sec.~\ref{sec:timing_analysis}. The astrometric, pulsar spin, binary and PK relativistic parameters, derived using a Monte Carlo analysis to allow for the uncertainties in the derived DM variations (Sec.~\ref{sec:timing_analysis}), are presented in this section. The results of the gravitational tests are described in the following Sec.~\ref{sec:bin_params} for tests of GR and in Sec.~\ref{sec:altgrav} for tests of alternative gravity theories.

A timing model describing the rotational properties of pulsar A, its position and apparent movement on the sky, the propagation delays in the ISM and the orbital motion and relativistic effects that cause deviations from an arrival time expected for a simple Keplerian orbit is applied to ToA measurements for pulsar A. Before the timing model can be applied, the measured topocentric arrival times need to be transferred to the SSB to account for the varying position of our telescopes during the course of the year and for relativistic effects in the solar system \cite{lk04}. Before we present the results of such a procedure, we explain the required relativistic binary timing model. We do this in some detail as we introduce a number of new considerations, which will eventually also become relevant for other systems in the future. 

The barycentric arrival time $t_{\rm b}$ of a pulsar signal is related to the proper time $T$ of the pulsar (modulo a constant factor) by a phenomenological timing model that accounts for all relevant contributions related to the orbital motion of the pulsar and the signal propagation in and near the binary system \cite{dd86,dt92}:
\begin{equation}
    t_{\rm b} - t_0 = \frac{1}{D} \left[T + \Delta_{\rm R}(T) + 
        \Delta_{\rm E}(T) + \Delta_{\rm S}(T) + \Delta_{\rm A}(T) \right] \,.
    \label{eq:timingmodel}                                   
\end{equation}
The time $t_0$ denotes a (chosen) reference epoch. The different $\Delta_a$ ($a=\mathrm{R,E,S,A}$) on the right hand side denote respectively, the {\em R{\o}mer delay} resulting (solely) from the orbital motion of the pulsar, the {\em Einstein delay} caused by time dilation along the pulsar's worldline, the {\em signal propagation delay} due to relativistic contributions to the propagation of the radio signal, and the {\em aberration delay} due to the fact that the radio signals originate from a rotating moving source \cite{bt76,dd86,sb76}. These contributions depend on the proper time $T$ and a set of Keplerian and PK parameters, which we will explain in detail below. $D$ is the Doppler factor due to the velocity ${\bf v}_{\rm b}$ of the pulsar system with respect to the SSB \cite{dd86}:
\begin{equation}
    D = \frac{1 - \hat{\bf K}_0 \cdot {\bf v}_{\rm b}/c}
             {\sqrt{1 - {\bf v}_{\rm b}^2/c^2} } \,,
    \label{eq:D}
\end{equation}
which to leading order is dominated by the unknown radial velocity $v_{\rm R} = \hat{\bf K}_0 \cdot {\bf v}_{\rm b}$ ($>0$ if the Earth-pulsar distance increases). $\hat{\bf K}_0$ is the LoS unit vector pointing from the SSB to the binary system.
Only temporal changes of $D$ are of interest here, and therefore $D \equiv 1$ can be chosen at a given epoch (see discussion in \cite{dd86}). The proper time of the pulsar is related to the rotational phase $\phi$ of the pulsar by
\begin{equation}
    \frac{\phi(T)}{2\pi} = N_0 + \nu (T - t_0) + \frac{1}{2}\dot\nu  (T - t_0)^2
           + \frac{1}{6}\ddot\nu (T - t_0)^3 + \dots
    \label{eq:phiT}
\end{equation}
where $N_0$ is the pulse number at the reference epoch $t_0$. In the following, we will discuss the individual contributions $\Delta_a$ and all significant effects that need to be accounted for in those contributions when analyzing the timing data presented here. In terms of theory-specific interpretation of the different $\Delta_a$ and their PK parameters, we primarily focus on GR. As it turns out, for several contributions we need to account for next-to-leading-order (NLO) terms in order to achieve a correct interpretation of our timing data. In the following sections we therefore discuss the individual contributions to the SSB arrival times in detail.

\subsection{R{\o}mer delay}
\label{subsec:Roemer}

In terms of the orbital motion, pulsar timing analysis of binary pulsars is based on a particularly compact solution of the first post-Newtonian (PN) two-body problem (Damour \& Deruelle solution \cite{dd85}). As it turns out, this quasi-Keplerian parametrization of the orbital motion is valid for a wide class of alternative theories, in particular the class of (mass-less) scalar-tensor theories \cite{dd86,dt92}. The {\em R{\o}mer delay}, $\Delta_{\rm R}$ in Eq.~(\ref{eq:timingmodel}), related to the orbital motion of the pulsar, is given by \cite{dd86}
\begin{equation}
    \Delta_{\rm R} = x \left[\sin\omega(\cos u - e_r) + 
                     \sqrt{1 - e_\theta^2}\cos\omega\sin u\right] \,,
    \label{eq:Roemer}
\end{equation}
where $x = a_{\rm p} \sin i/c$ is the projected semi-major axis of the pulsar orbit. The angle $\omega$ is the longitude of periastron (measured from the ascending node\footnote{Depending on the literature, this is called the {\em argument of periastron}. We follow the naming that is more common in pulsar astronomy, and for instance also used in \cite{pw14} (see their Figure~3.2).}) and $u$ is the ``relativistic'' {\em eccentric anomaly}, related to the proper time of the pulsar via Kepler's equation
\begin{equation}
    u - e_T \sin u = 2\pi \left[\left(\frac{T - T_0}{P_{\rm b}}\right) - 
    \frac{\dot{P}_{\rm b}}{2}\left(\frac{T - T_0}{P_{\rm b}}\right)^2\right] \,.
    \label{eq:Kepler}
\end{equation}
The $\dot{P}_{\rm b}$ term accounts for any secular (intrinsic or apparent) change in the orbital period $P_{\rm b}$. It therefore accounts for the decay in the orbital period due to GW damping, which enters the GR equations of motion at the 2.5PN level, i.e.\ order $v^5/c^5$. The parameter $T_0$ denotes a time of periastron passage. As will become clear below, for the Double Pulsar it is important to distinguish between the three different eccentricities $e_r$, $e_\theta$, and $e_T$ that enter the two equations (\ref{eq:Roemer}) and (\ref{eq:Kepler}), and are different at their first PN correction \cite{dd86}
\begin{equation}
\label{eq:ecc}
    e_r      \equiv e_T(1 + \delta_r) \,, \quad 
    e_\theta \equiv e_T(1 + \delta_\theta) \,.
\end{equation}
While the parameter $\delta_r$ is not observable in the Double Pulsar (it is  absorbed by higher order frequency derivatives), the ``relativistic deformation of the orbit'' $\delta_\theta$ has to be accounted for in the analysis (Section~\ref{subsubsec:test_dth}). The ``time eccentricity'' $e_T$ corresponds to the fitting parameter called {\em ``eccentricity of orbit''} in the {\sc Tempo} and {\sc Tempo2} implementation of the Damour \& Deruelle (DD) timing model \cite{mtp+15,hem06,ehm06}, and therefore can be considered as the observed eccentricity in pulsar timing, a Keplerian parameter.

The longitude of periastron $\omega$ that enters the R{\o}mer delay (\ref{eq:Roemer}) changes over time, due to the relativistic precession of the apsidal line of the Double Pulsar. In the quasi-Keplerian parametrization of the orbital motion, the longitude of periastron is simply related to the  ``relativistic'' {\em true anomaly} $\theta$ \cite{dd85}, i.e.
\begin{eqnarray}
    \omega &=& \omega_0 + k\,\theta \,,
    \label{eq:omega} \\
    \theta &=& 2 \arctan\left[\sqrt{\frac{1 + e_\theta}{1 - e_\theta}} 
             \tan \left(\frac{u}{2}\right)\right] \,.
    \label{eq:trueanomaly}
\end{eqnarray}
By adding or subtracting multiples of $2\pi$, the angle $\theta$ is adjusted such that it matches the number of full phase cycles of $u$, as determined by Eq.~(\ref{eq:Kepler}). The structure of the R{\o}mer delay presented above is expected to be the same for many viable relativistic theories of gravity. What changes is how the PK parameters ($k$, $\dot{P}_{\rm b}$, $\delta_r$, $\delta_\theta$) depend on the Keplerian parameters ($P_{\rm b}$, $x$, $e_T$) and the inertial masses $m_{\rm A}$ and $m_{\rm B}$ of A and B respectively. Detailed expressions for GR can be found in \cite{dd86,lk04}, where one also finds expressions for the PK parameters which we will introduce below, in combination with the Einstein and Shapiro delays in the signal propagation.

Instead of the PK parameter $k$, in pulsar timing one generally quotes the value for the (secular) {\em periastron advance} parameter defined as $\dot\omega \equiv n_\mathrm{b}k$, where $n_\mathrm{b} \equiv 2\pi/P_{\rm b}$ is the measured orbital frequency (cf.\ Eq.~(5.8) in \cite{ds88}). Given the high numerical precision of $n_\mathrm{b}$ normally obtained in pulsar timing observations, $\dot\omega$ and $k$ can be used synonymously. For that reason, whenever comparison with former publications is required, we will 
use $\dot\omega$ for convenience.

\subsubsection{Second post-Newtonian corrections}
\label{subsubsec:2PN}

The high precision achieved for several PK parameters raises the question of whether higher-order corrections need to be taken into account. In terms of the R{\o}mer delay, 2PN corrections to the quasi-Keplerian parametrization have been calculated in \cite{ds88,sw93,sw93a,wex95}. The result is a generalized quasi-Keplerian description of the orbital motion, in which Kepler's equation and the equation for the true anomaly are augmented by periodic terms of order $v^4/c^4$. These 2PN periodic terms, however, are well below the timing precision so far achieved in the Double Pulsar system, and can be ignored. This also means that we can replace $e_\theta$ with $e_T$ in Eq.~(\ref{eq:trueanomaly}) when calculating the longitude of periastron $\omega$ via Eq.~(\ref{eq:omega}).

Apart from these additional periodic terms, parameters that are already present in the 1PN solution get extended by 2PN corrections. Of particular importance for the analysis presented here is the secular advance of periastron. The (intrinsic) advance of periastron to 2PN approximation was first calculated within GR in \cite{ds87,ds88}. When expressed as function of the masses and the (directly observed) Keplerian parameters, it is given by
\begin{equation}
  k = k^\mathrm{1PN}  + k^\mathrm{2PN} + \mathcal{O}(\beta_\mathrm{O}^6)\,,
\label{eq:k2PN}
\end{equation}
where
\begin{equation}
    k^\mathrm{1PN} = \frac{3 \beta_\mathrm{O}^2}{1 - e_T^2} \,,\quad
    k^\mathrm{2PN} = \frac{3 f_{\rm O} \beta_\mathrm{O}^4}{1 - e_T^2} \,,
\end{equation}
and
\begin{eqnarray}
  \beta_\mathrm{O} &=& \frac{(G M n_{\rm b})^{1/3}}{c} \,,\label{eq:bO}\\
  f_{\rm O} &=& \frac{1}{1-e_T^2} \left(\tfrac{39}{4}X_{\rm A}^2 + \tfrac{27}{4}X_{\rm B}^2 + 15X_{\rm A}X_{\rm B}\right) \nonumber\\ && \qquad\;
  - \left(\tfrac{13}{4}X_{\rm A}^2 + \tfrac{1}{4}X_{\rm B}^2 + \tfrac{13}{3}X_{\rm A}X_{\rm B} \right) \,,
\end{eqnarray}
with $M \equiv m_{\rm A} + m_{\rm B}$, $X_{\rm A} \equiv m_{\rm A}/M$, and $X_{\rm B} \equiv m_{\rm B}/M$ (cf.\ Eq.~(5.18) in \cite{ds88}). Note, $X_{\rm A} + X_{\rm B} = 1$. With the numbers from the timing parameter table~\ref{tab:params} one finds $\dot{\omega}^\mathrm{2PN} \equiv n_\mathrm{b}k^\mathrm{2PN}\simeq 4.39 \times 10^{-4}\,{\rm deg\,yr^{-1}}$, which is about 35 times the measurement error of $\dot\omega$, quoted in table~\ref{tab:params}. More importantly, other PK parameters, in particular $s$ and $\dot{P}_\mathrm{b}$, have now reached a precision that, for the first time in a binary pulsar, we have to account for $k^\mathrm{2PN}$ in order to perform, for instance, consistent gravity tests.

The 3PN corrections to $k$ have been calculated in \cite{djs00,mgs04}. They are of the order of $\beta_\mathrm{O}^2 \approx 4 \times 10^{-6}$ times the 2PN terms, and therefore absolutely negligible.

\subsubsection{Spin-orbit contribution and equation of state}
\label{subsubsec:LT}

So far we have ignored the influence of spin, i.e.\ the proper rotations of A and B, on the orbital dynamics. Several effects related to spin can lead to significant modifications of the equations of motion of a binary system, in particular the relativistic spin-orbit and spin-spin coupling, and the rotationally induced mass quadrupole moments \cite{bo75b}. For the analysis in this paper, only the coupling between the spin of A and the orbital motion is of any relevance (the spin of B is about 135 times smaller). It is numerically of 2PN order \cite{kg05,ior09,kw09}, which is typical for binary pulsars in relativistic double NS systems \cite{wex95}. Furthermore,  since the spin of A is practically parallel to the orbital angular momentum (see Section~\ref{sec:systemintro}) there is only a contribution to the precession of the periastron that is of relevance here. We refer the reader to \cite{kw09,hkw+20} for more details.

In order to incorporate spin-orbit coupling in our analysis, Eq.~(\ref{eq:k2PN}) needs to be extended by the Lense-Thirring (LT) term, i.e.
\begin{equation}
    k = k^\mathrm{1PN}  + k^\mathrm{2PN} + k^\mathrm{LT,A} \,,
    \label{eq:k_theory}
\end{equation}
where within GR the LT contribution is given by \cite{bo75b,ds88}\footnote{For the precision needed, we can safely use $g_{S_{\rm A}}^\parallel$ instead of $g_{S_{\rm A}}$ in \cite{ds88}, since $\delta_\mathrm{A} < 3.2^\circ$, while $g_{S_{\rm A}} \simeq g_{S_{\rm A}}^\parallel \cos\delta_\mathrm{A}$ for the Double Pulsar (cf.\ \cite{kw09}).}
\begin{equation}
  k^\mathrm{LT,A} = 
    -\frac{3\beta_\mathrm{O}^3\,\beta_{S_{\rm A}}}{1 - e_T^2} \,
    g_{S_{\rm A}}^\parallel \,,
    \label{eq:kLT}
\end{equation}
with 
\begin{eqnarray}
  \beta_{S_{\rm A}} &=& 2\pi\nu \frac{c I_{\rm A} }{Gm_{\rm A}^2} \,,\\
  g_{S_{\rm A}}^\parallel &=& 
    \frac{X_{\rm A}(1 + \frac{1}{3}X_{\rm A})}{\sqrt{1 - e_T^2}} \,.
\end{eqnarray}
Apart from the MoI $I_{\rm A}$, all quantities in the above equations are known with high precision. $I_{\rm A}$ depends on the EoS for NS matter, which is still afflicted by considerable uncertainty. Consequently, there is a range in the prediction for $k^{\rm LT,A}$. From Eq.~(\ref{eq:kLT}) one finds the numerical expression
\begin{equation}
    \dot{\omega}^\mathrm{LT,A} \equiv n_\mathrm{b} k^\mathrm{LT,A}  
    \simeq -3.77 \times 10^{-4} \times I_{\rm A}^{(45)} \, {\rm deg\,yr^{-1}} \,,
    \label{eq:omdotLTnum}
\end{equation}
where $I_{\rm A}^{(45)} \equiv I_{\rm A}/(10^{45}\,{\rm g\,cm^2})$. Using the multi-messenger constraints on the radius that can be inferred from \cite{dcp+20} 
(probability distribution function F) \footnote{Although the mass of A (1.34\,M$_\odot$) is slightly below  1.4\,M$_\odot$, within the accuracy needed here, the radius constraints in \cite{dcp+20} for a 1.4\,M$_\odot$ NS can also be applied to pulsar A.}, in combination with the radius-MoI relation for A given in \cite{Lattimer:2019}, we find a range of $I_{\rm A}^{(45)} \approx 1.15$--$1.48$ (95\% confidence)\footnote{There are other distributions that have been derived for the MoI of pulsar A, see e.g. \cite{lhs19,lk18,shcy20}. However, for the results of this paper this does not make any difference.}. Alternatively, Eq.~(\ref{eq:omdotLTnum}) can be used to infer limits for the MoI of A purely from the timing observations of the Double Pulsar, if combined with two other suitable PK parameters \cite{ls05,kw09,hkw+20}. A corresponding analysis will be given in Section~\ref{subsubsec:test_EoS_LT}.

\subsubsection{Proper motion contributions}

The proper motion of a binary pulsar leads to a change in its orientation with respect to the observer on Earth. Such a change leads to an apparent change in the longitude of periastron $\omega$ and the orbital inclination $i$ \cite{ajrt96,kop96}. The change in $\omega$ leads to a proper motion related offset $\dot{\omega}^{\rm pm}$ between the intrinsic and the observed advance of periastron. Using the proper motion and orbital inclination from Table~\ref{tab:params}, in combination with the longitude of the ascending node obtained from scintillation measurements \cite{rcn+14}, one finds $\dot{\omega}^{\rm pm} \approx -4 \times 10^{-7}\,{\rm deg\,yr^{-1}}$ (see also \cite{hkw+20}). This is about a factor of 30 smaller than the current measurement error for $\dot{\omega}$ (see Table~\ref{tab:params}) and can therefore be ignored. As a consequence, there is no need to distinguish between the observed and the intrinsic $\dot\omega$.

The change in the orbital inclination enters the timing model through a temporal change in the projected semimajor axis of the pulsar orbit, showing up as a $\dot{x}$ in the timing solution, if significant. However, this contribution is even smaller than the contribution to the advance of periastron, since it is greatly suppressed by the fact that $i$ is close to $90^\circ$ ($\dot{x}^{\rm pm} \propto \cot i \approx 0.01$; see Table~\ref{tab:params}).

\subsubsection{Next-to-leading-order contributions in the mass function}

The inclination of the binary orbit is linked to the projected semimajor axis $x$ in Eq.~(\ref{eq:Roemer}) via the binary mass function. In Newtonian gravity one finds (see e.g.\ \cite{dt92,lk04})
\begin{equation}
    \sin i = \frac{n_{\rm b} x}{\beta_\mathrm{O} X_{\rm B}} \,,
    \label{eq:mf}
\end{equation}
where $n_{\rm b}$ and $x$ are both (observable) Keplerian parameters, generally known to very high precision for a binary pulsar. As we will discuss later (Section~\ref{subsec:Shapiro}), the measurement of the Shapiro delay in the Double Pulsar gives access to $\sin i$, and therefore Eq.~(\ref{eq:mf}) leads to an additional constraint for the two masses $m_{\rm A}$ and $m_{\rm B}$.

In the 1PN approximation Kepler's third law, which enters the derivation of Eq.~(\ref{eq:mf}), gets modified by an additional term (see Eq.~(3.7) in \cite{dd85} and Eq.~(3.7) in \cite{dt92}). Consequently, Eq.~(\ref{eq:mf}) gets modified as well at the 1PN level. Using the 1PN expression for Kepler's third law one finds
\begin{equation}
    \sin i = \frac{n_{\rm b} x}{\beta_\mathrm{O} X_{\rm B}} \left[1 + \left(3 - \tfrac{1}{3}X_{\rm A}X_{\rm B}\right) \beta_\mathrm{O}^2\right] \,.
    \label{eq:mf1PN}
\end{equation}
We have used the fact that for the Damour \& Deruelle solution the Newtonian relation between the semimajor axis of the pulsar orbit and the semimajor axis of the relative orbit also holds at the 1PN level, i.e.\ $a_{\rm A} = (m_{\rm B}/M)a_R + {\cal O}(v^4/c^4)$ (see \cite{dd85,dd86} and the discussion in \cite{dam09}). With $P_{\rm b}$, $x$ and the masses from Table~\ref{tab:params} one finds for the 1PN correction in Eq.~(\ref{eq:mf1PN}) approximately $1.3 \times 10^{-5}$, which is only about a factor of 1.3 larger than the error for $\sin i$ in Table~\ref{tab:params}. For that reason, we will use the full 1PN mass function (\ref{eq:mf1PN}). This is the first time that 1PN corrections to the mass function become relevant for any binary pulsar.

\subsubsection{Secular changes in orbital period}

The observed change in the orbital period is a combination of effects intrinsic to the system and apparent changes related to a temporal change in the Doppler factor $D$ in Eq.~(\ref{eq:timingmodel}) \cite{dt91}. For the Double Pulsar, by far the dominant contribution to $\dot{P}_{\rm b}$ is the orbital period decay due to the emission of GWs. In GR, GW damping enters at the 2.5PN level in the equations of motion (see \cite{Blanchet_LRR} and references therein). The explicit expression for the leading order changes due to GW emissions for an eccentric orbit have been worked out in \cite{pet64} (see \cite{lk04} for a more pulsar-astronomy-adapted expression). The Hulse-Taylor pulsar was the first binary system where the leading order GW damping has been tested \cite{tfm79,tw89}.

The NLO correction to the change in the orbital period corresponds to the 3.5PN terms in the equations of motion, and hence to the 1PN corrections in the radiation reaction force \citep{Iyer_1995,Pati_2002, Konigsdorffer_2003}. It has been calculated in \cite{bs89}. For the Double Pulsar this contribution amounts to about $-1.75 \times 10^{-17}$ \cite{hkw+20}. This is about a factor of 4.5 smaller than the error in $\dot{P}_{\rm b}$ (see Table~\ref{tab:params}). Although this higher-order correction is in principle still negligible, we will include it in our analysis. This is of particular interest for the comparison with the LIGO/Virgo results in Section~\ref{subsubsec:test_gw}. In the near future, however, that contribution will become of importance (see \cite{hkw+20}).

Yet another intrinsic effect that changes the orbital period in the Double Pulsar is the mass loss related to the spin-down of the pulsars \cite{dt91}. This mass loss is a result of Einstein's energy-mass equivalence in the sense that here one is seeing the loss of mass associated with the loss of rotational (kinetic) energy of the pulsar. In \cite{hkw+20} these contributions have been calculated based on Eq.~(4.1) in \cite{dt91}. While for B this is negligible, for A one has \cite{hkw+20}
\begin{equation}
     \dot{P}_{\rm b}^{\dot{m}_{\rm A}} = 2.3 \times 10^{-17} \times I_{\rm A}^{(45)} \,.
     \label{eq:Pbdotmassloss}
\end{equation}
For two reasons it is important to include this contribution in the analysis below. First, given the range for $ I_{\rm A}^{(45)}$ (see end of Section~\ref{subsubsec:LT}), $\dot{P}_{\rm b}^{\dot{m}_{\rm A}}$ can be as large as $3.1 \times 10^{-17}$, which is a fair fraction of the measurement error of $\dot{P}_{\rm b}$ (see Table~\ref{tab:params}). Second, and more importantly, when estimating a MoI constraint solely based on the Double Pulsar observations, $\dot{P}_{\rm b}^{\dot{m}_{\rm A}}$ and its dependence on $I_{\rm A}$ has a significant impact on the result (see discussion in \cite{hkw+20}).

Finally, there are the external contributions to the observed $\dot{P}_{\rm b}$, related to a temporal change of $D$, where $\dot{P}_{\rm b}/P_{\rm b} = \dot{D}/D$ \cite{dt91}. There is an apparent radial acceleration due to the transverse motion of the Double Pulsar with respect to the SSB, which leads to the so called Shklovskii effect \cite{shk70}
\begin{equation}
     \dot{P}_{\rm b}^{\rm Shk} = \frac{(\mu_\alpha^2+\mu_\delta^2) \, d}{c} \, P_{\rm b} \,,
\end{equation}
and a physical radial acceleration due to the Galactic gravitational potential, leading to a $\dot{P}_{\rm b}^{\rm Gal}$. Both effects depend on the distance $d$, which is the main source of uncertainty in those corrections (see Section~\ref{subsec:impl_astrometry} below). For the Galactic potential, we use the model provided in \cite{McMillan:2017}. We have also tested the analytic correction of \cite{dt91} (adapted for $b < 0$) with the latest Galactic parameters from \cite{GRAVITY:2019}. The result is practically the same. Numerical details for these contributions will be given in Section~\ref{subsubsec:test_gw}.

\subsection{Einstein delay}
\label{subsec:Einstein}

The Einstein delay $\Delta_{\rm E}$ in Eq.~(\ref{eq:timingmodel}) links the proper time of the pulsar $T$ to the coordinate time of the binary system $t$. To leading order it can be viewed as a combination of the second-order Doppler effect and the gravitational redshift caused by the companion. If $T$ is renormalized such that the orbital period is the same as measured in $t$ one finds to leading order the following expression \cite{bt76,dd86}
\begin{equation}
    \Delta_{\rm E} = \gamma_{\rm E} \sin u \,.
\end{equation}
The {\em amplitude of the Einstein delay} $\gamma_{\rm E}$ is a PK parameter that in GR can be calculated according to
\begin{equation}
    \gamma_{\rm E} = e\frac{P_{\rm b}}{2\pi}  
                     X_{\rm B}(1 +  X_{\rm B}) \beta_\mathrm{O}^2 \,.
\end{equation}

In alternative theories of gravity, $\gamma_{\rm E}$ is generally more complicated, since it also includes a periodic variation of the pulsar rotation due to a variation of the pulsar's MoI along the orbit \cite{ear75,wz89,de96}. The variation of the MoI is caused by a variation of the local gravitational constant along the (eccentric) orbit, a result of a violation of the SEP \cite{Will:2018}.

NLO order contributions have been calculated for GR in \cite{wr95,wex95}. Similarly to the periodic 2PN terms in the orbital motion (see Section~\ref{subsubsec:2PN}), these contributions are currently absolutely negligible. They are of order 10\,ns, which is an order of magnitude below the  precision for $\gamma_{\rm E}$ (see Table~\ref{tab:params}). They are not expected to be of relevance before the full SKA\footnote{see \url{www.skatelescope.org}} is in operation.

\subsection{Signal propagation delay}
\label{subsec:Shapiro}

For binary pulsars which are sufficiently edge-on, i.e. $i$ close to 90$^\circ$, the curved spacetime of the companion star has a significant effect on the propagation of the pulsar's radio signal. To leading order, we have the well-known Shapiro delay \cite{sha64}, which for binary pulsars is expressed in the following form \cite{bt76,dd86}:
\begin{eqnarray}
    \Delta_{\rm S}^{\rm (LO)} &=& -2 r \ln \Lambda_u \,, 
    \label{eq:ShaLO} \\
    \Lambda_u &=&  1 - e_T \cos u - s \big[\sin\omega(\cos u - e_T) \nonumber\\
              &&   \qquad\qquad\quad + (1 - e_T^2)^{1/2} \cos\omega\sin u \big] \,,
    \label{eq:Lambda_u}
\end{eqnarray}
with the two PK parameters $r$ and $s$, called the {\em range} and the {\em shape} of the Shapiro delay respectively. The Shapiro shape can quite generally be identified with the sine of the orbital inclination, i.e. $s = \sin i$ \cite{will93,dt92}. The range of the Shapiro delay is linked to the mass of the companion. In GR one finds $r = Gm_{\rm B}/c^3$ \cite{bt76,dd86}.

The leading order expression for the Shapiro delay in a binary pulsar has been obtained in \cite{bt76}, by integrating along a straight line (in an isotropic coordinate system) and by assuming a static mass distribution while the signal propagates away from the pulsar (static limit). Relaxing the first assumption and accounting for the fact that the radio signal propagates along a curved path due to the deflection in the gravitational field of pulsar B, leads to a modification of Eq.~(\ref{eq:ShaLO}) \cite{sch90,lr05}. This {\em lensing correction} to the Shapiro delay, given the current timing precision in the Double Pulsar, is however not yet observable due to a strong covariance with $s$. We have tested this against the real data as well as in mock data simulations\footnote{{Simulations are based on the integration of null geodesics in a Schwarzschild spacetime, co-moving with pulsar B. Since additive constant terms can be ignored, the use of Eddington-Finkelstein coordinates turned out to be of great advantage in the numerical integration for the propagation time \cite{Ali:2011}.}\label{foot:sim}}, and found a shift in $s$ which is yet insignificant (less than 0.5$\sigma$). One has to keep in mind that these effects are strongest around conjunction, in a part of the orbit where the timing precision is significantly reduced due to the (partly intermittent) eclipses $\sim \pm 1^\circ$ around superior conjunction, caused by the rotating plasma-filled magnetosphere of B \cite{krb+04,mll+04,bkk+08}.\footnote{The (intermittent) eclipses of A by the rotating magnetosphere of B lead to a significant reduction in the number of pulses available for the integrated pulse profile. This in turn, reduces its S/N, leading to significantly fewer viable ToAs, which at the same time have considerably larger errors.} In spite of that, we have extended the Shapiro delay implementation in {\sc Tempo} by an adapted version of the elegant approximation in Eq.~(73) of \cite{kz10}, i.e.\ $\Lambda_u \rightarrow \Lambda_u + \delta\Lambda_u^{\rm len}$ with $\delta\Lambda_u^{\rm len} = 2rc/a_R$, which accounts to good approximation (only few \% error) for the delay related to a curved signal path, and therefore removes nearly completely the already insignificant bias to $s$. It is interesting to note that the lensing correction leads to a reduction in the (calculated) propagation time, which is a result of Fermat's principle (see e.g.\ \cite{per14}).

A higher-order signal propagation effect that became relevant for the Double Pulsar timing in the meantime, and cannot be absorbed into other timing parameters, is the {\em retardation effect}, i.e.\ specifically the velocity-dependent contributions in the Shapiro time delay (1.5PN corrections) \cite{ks99,rl06:prd}. It results from the fact that the ``lens'', i.e.\ pulsar B, moves while the signal of A propagates across the binary system towards Earth. The result derived in \cite{ks99} is absolutely sufficient for modelling the ToAs around superior conjunction\footnote{Lensing corrections to the retardation effect as discussed in \cite{rl06:prd} are totally negligible. The corresponding post-fit residuals are below 0.02\,$\mu$s, restricted to a narrow range around conjunction where, because of the eclipse, the timing precision is anyway greatly compromised.} 
and one has
\begin{equation}
    \Delta_{\rm S} = -2 r \ln (\Lambda_u + \delta\Lambda_u^{\rm len} + \delta\Lambda_u^{\rm ret}) \,, 
    \label{eq:ShapRet}
\end{equation}
where retardation correction $\delta\Lambda_u^{\rm ret}$ is of order $x/P_{\rm b}$ and can directly be taken from Eq.~(130) in \cite{ks99}. As in \cite{ks99}, we have omitted the velocity-dependent term in front of the logarithmic function, since its contribution is negligible (less than about 0.015\,$\mu$s).

For systems with $i$ close to 90$^\circ$, like the Double Pulsar, one often obtains rather asymmetric error bars for the Shapiro shape parameter $s$. For that reason, in \cite{ksm+06a} the {\em logarithmic Shapiro shape parameter} was introduced, 
\begin{equation}
    z_s \equiv -\ln(1 - s) \,.
\end{equation}
It not only leads to a more Gaussian distribution for the {\em fitted} parameter, i.e. $z_s$ instead of $s$, it also ensures that the a priori free parameter $s$ always fulfills $s < 1$, which is important when interpreted as the sine of the inclination angle.

\subsection{Aberration}
\label{subsec:Aberration}

The term $\Delta_{\rm A}$ in Eq.~(\ref{eq:timingmodel}) collectively refers to delays that are related to the fact that the pulsar is acting as a moving ``light-house'' with beamed radio emission. These effects (with one exception) are proportional to the rotational period of the pulsar, $P \equiv 1/\nu$, and would be absent if the variability of the radio signal were caused by an (intrinsic) oscillation/variability instead of the rotation of the NS. For pulsar timing, aberration means that the proper angle of emission of an observed pulse changes along the orbit \cite{sb76,dd86}. To leading order the aberration delay can be written as
\begin{equation}
    \Delta_{\rm A}^{\rm (LO)} =
        {\cal A} \, [\sin\psi + e_T \sin\omega] + 
        {\cal B} \, [\cos\psi + e_T \cos\omega] \,.
   \label{eq:aberration}
\end{equation}
where $\psi \equiv \omega + \theta$ is the longitude of the pulsar with respect to the plane of the sky. The aberration coefficients, ${\cal A}$ and ${\cal B}$, are proportional to $P$ and depend on the orientation of the pulsar with respect to the orbit and the observer. Since, as discussed in Section~\ref{sec:systemintro}, the spin of A is (practically) parallel to the orbital angular momentum, one has ${\cal B} = 0$ and 
\begin{equation}
    {\cal A} = \frac{P \, x}{P_{\rm b}(1 - e_T^2)^{1/2}\sin^2i}
            \simeq 3.65\,\mu{\rm s} \,.
    \label{eq:aberrationA}
\end{equation}
In practice, ${\cal A}$ is a non-observable parameter. If ${\cal A}$ is not provided in the timing model, it gets absorbed by a shift in various timing parameters (see \cite{dd86,dt92} for details). For that reason, we will a priori add the aberration coefficient ${\cal A}$ to our model, as a fixed parameter with the value given in the equation above.

Classical aberration, as expressed in Eq.~(\ref{eq:aberration}), assumes a flat spacetime in which the signal propagates towards Earth. However, this is only a first approximation, which is no longer sufficient for an adequate description of the Double Pulsar timing observations, in particular for ToAs near the superior conjunction of pulsar A. It is necessary to account for the deflection of the radio signal in the gravitational field of pulsar B. This gravitational signal deflection leads to a small change in the proper angle of emission, which in turn leads to a lensing correction to the classical (``longitudinal'') aberration of Eq.~(\ref{eq:aberration}). The potential importance of such a correction in close to edge-on binary systems has first been discussed in \cite{dk95}. For the specific situation of A (spin aligned with orbital angular momentum), one finds
\begin{equation}
    \Delta_{\rm A} = {\cal A}\,[\sin\psi + e_T \sin\omega]
                   + {\cal D}\,\frac{\cos\psi}{\Lambda_u} \,,
    \label{eq:aberrationBend}
\end{equation}
where
\begin{equation}
    {\cal D} = \frac{P}{\pi}\,\frac{r}{x + x_\mathrm{B}} \,. 
    \label{eq:aberrD}
\end{equation}
The quantity $x_{\rm B}$ denotes the projected semi-major axis of pulsar B, which is an observed Keplerian parameter in the Double Pulsar \cite{ksm+06}. The NLO contribution in $\Delta_{\rm A}$ contains the same two PK parameters as the Shapiro delay in Section~\ref{subsec:Shapiro}, i.e. the range $r$ ($\mathcal{D} \propto r$) and the shape $s$ (in $\Lambda_u$, see Eq.~(\ref{eq:Lambda_u})). Like in the Shapiro delay (see Section~\ref{subsec:Shapiro}), also here we have accounted for retardation in our TEMPO implementation, i.e.\ used the position of B when the signal reaches the minimum distance to B. This leads to small corrections in the last term of Eq.~(\ref{eq:aberrationBend}), i.e.\ a corresponding shift in $\psi$ defined by the new position of B (and the position of A at emission) and a correction of $\Lambda_u$ like in Eq.~(\ref{eq:ShapRet}). Although these corrections are not yet significant, we have implemented them along with the retardation in the Shapiro delay, for completeness.

In \cite{rl06:apj,rl06:prd} it has been pointed out, that the approximation for this lensing correction to the aberration delay, as given in \cite{dk95}, loses its validity if the impact parameter for the radio signal becomes comparable to the Einstein radius $R_{\rm E} \approx \sqrt{4Gm_{\rm B}a_R/c^2} \approx 2500$\,km
\footnote{Note, the expressions in \cite{rl06:apj,rl06:prd} are based on the lensing equation, which makes the assumption of small angles (cf.\ e.g.\ \cite{esf92}), and therefore they are strictly speaking only valid near conjunction, which also limits their usability in a {\sc Tempo} implementation.}. 
Our tests, based on simulated and the real timing data, shows however, that Eq.~(\ref{eq:aberrationBend}) is absolutely sufficient for the current Double Pulsar timing analysis. In fact, the main difference between the (retardation-corrected) approximation (\ref{eq:aberrationBend}) and the full calculations  (cf.\ footnote \ref{foot:sim}) is below 50\,ns (post-fit) and confined to a region very close to superior conjunction where, as explained above, the timing precision is anyway significantly compromised. One has to keep in mind that for outside the eclipse region one finds that the impact parameter $\gtrsim 7\,R_{\rm E}$, which for the deflection angle means $\lesssim 0.026$\,deg. 

Finally, there is the lensing correction to the ``latitudinal'' aberration delay, related to a potential change in the observed pulse profile, as the LoS cuts a different part of pulsar A's magnetosphere due to the signal deflection \cite{rl06:prd,rl06:apj}. This effect does not directly depend on the rotational period of pulsar A, but is linked to A's rotation by having the LoS crossing the emission region of the pulsar. At this stage it is not clear if there is at all a significant profile evolution across the range of the LoS variation. This is a question currently under investigation, utilizing the improved data quality provided by the MeerKAT telescope \cite{ksv+21}. Due to a strong covariance with $s$ and the eclipses near conjunction caused by the magnetosphere of pulsar B, this effect is not (yet) expected to be of any relevance 
\footnote{Periodic variations in the pulsar profile due to a latitudinal shift in the LoS caused by aberration have been observed in PSR~B1534+12 \cite{sta04}}.

\subsection{Fitting for NLO contributions in the Shapiro and aberration delay}
\label{subsec:NLOfit}

As it turns out, the NLO contributions in the Shapiro and aberration delays cannot be tested separately in the Double Pulsar. We have mentioned already in Section~\ref{subsec:Shapiro} that the lensing correction to the propagation delay is covariant with the Shapiro shape $s$. Furthermore, the NLO contribution to $\Delta_{\rm S}$ is quite similar to the NLO contribution to $\Delta_{\rm A}$, which renders a separation of $\delta\Lambda_u^{\rm ret}$ and ${\cal D}$, at least under current timing precision, impossible. For that reason, in order to test for the significance of NLO signal propagation contributions in our data, we collectively re-scale these contributions with a common factor $q_{\rm NLO}$, that can be fitted for in our new timing model, i.e. for the retardation effect in the Shapiro delay
\begin{eqnarray}
    \delta\Lambda_u^{\rm ret} &\rightarrow& 
    \delta\Lambda_u^{\rm ret} \times q_{\rm NLO} \,,\label{eq:Rret}
\end{eqnarray}
and for effects related to signal deflection
\begin{eqnarray}
    \delta\Lambda_u^{\rm len} &\rightarrow& 
    \delta\Lambda_u^{\rm len} \times q_{\rm NLO} \,,\label{eq:Rlen}\\
    {\cal D} &\rightarrow& 
    {\cal D} \times q_{\rm NLO} \,.\label{eq:Rben}
\end{eqnarray}
In GR one has $q_{\rm NLO} = 1$. The NLO corrections above depend on Keplerian parameters and the PK parameters $r$ and $s$ (note, $a_R = c(x+x_\mathrm{B})/s$). Therefore $q_{\rm NLO}$ simply measures the relevance of these NLO corrections, and is not considered as a new PK parameter.


\section{Binary and relativistic-parameter results}
\label{sec:bin_params}

In this section, we present key results from this work relating to tests of Einstein's General Relativity. These results are based on the methods described in Sec.~\ref{sec:binary_model} above and include a number of previously undetected relativistic effects. All observed effects are shown to be fully consistent with GR, a remarkable confirmation of Einstein's extraordinary intuition and vision. In the first Timing Parameters subsection, we give a full list of the parameters derived from the analysis of the 30-s sampled data set. The next subsection presents results depending on our best determined values for the masses of the A and B pulsars, including constraints on the MoI of pulsar A ($I_{\rm A}$) and hence on the neutron-star EoS, and the newly detected relativistic effects. Using these masses, we can constrain rate of change of the orbital period due to GW damping $\dot{P}^{\rm GW}_{\rm b}$ and probe light propagation effects in the gravitational field of a NS, key tests of GR, to unprecedented levels. We compare limits of deviation from GR in GW emission obtained with the Double Pulsar with limits from the LIGO/Virgo observations of BH-BH and NS-NS coalescenses, showing that, at leading PN order (quadrupole formula), the Double Pulsar is more sensitive by three orders of magnitude, whereas at high PN orders, LIGO/Virgo gives more stringent limits, demonstrating the complementarity of these experiments.

\subsection{Timing parameters}
\label{sec:timpar}

For the fit of the timing model to the ToAs one does not use Eq.~(\ref{eq:timingmodel}) in its form $t_{\rm b} = T + \Delta(T)$ directly. What is needed is the inverted timing model, i.e.\ something of the form $T = t_{\rm b} - \tilde\Delta(t_{\rm b} )$. In \cite{dd86} an analytic inversion of the timing model was worked out, giving an analytic expression for $\tilde\Delta(t_{\rm b})$. However, the approximation used in \cite{dd86}, in particular with the Shapiro delay, is no longer sufficient for the Double Pulsar, mainly due to the higher order signal propagation and aberration delays. For that reason, we have modified the standard routine of the DDS model\footnote{The DDS timing model is a modification of the DD model \cite{dd86}, where the Shapiro shape parameter $s$ is replaced by the logarithmic Shapiro shape parameter $z_s$ via $s \equiv 1 - \exp(-z_s)$ (see \cite{ksm+06a} for details). It is part of the {\sc Tempo} standard distribution.} to perform a numerical inversion of model Eq.~(\ref{eq:timingmodel})  (for details see the public {\sc Tempo} distribution after November 27$^{\rm th}$, 2015). 

\begin{figure}[htp]
    \centering
    \includegraphics[width=8.5cm]{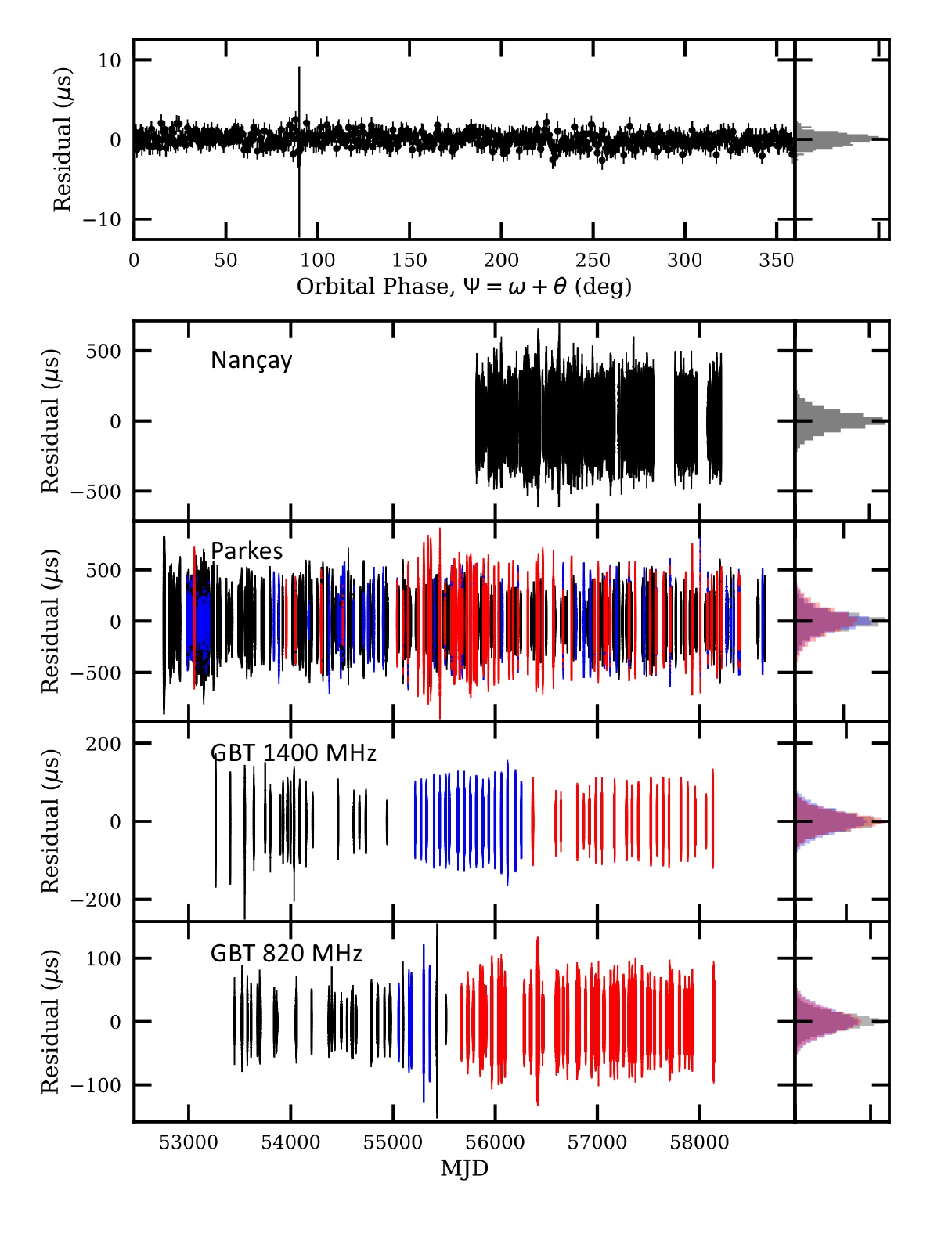}
    \caption{Residuals obtained after applying the timing model to the 30-s data set. The panels show from bottom to top: GBT 820 MHz (black: GASP, blue: GUPPI incoherent, red: GUPPI coherent), GBT 1400 MHz (same colours as for 820 MHz), Parkes (black: 50cm, blue: 20cm, red: 10cm) and Nan\c{c}ay. The top most panel shows residuals as function of orbital phase, averaged to 15-s-long phase bins. The large uncertainties at $\Psi=90$ deg are the result of the presence of the eclipse. For each panel, we add a histogram of the distribution of residuals on the right hand side.}
    \label{fig:residuals}
\end{figure}

The results of applying the timing model are shown in Table~\ref{tab:params} and Figure~\ref{fig:residuals}. In addition to the binary parameters introduced and discussed in detail above, we fit for spin frequency and higher order derivatives ($\nu$, $\dot\nu$, $\ddot\nu$, $\dddot\nu$ and $\ddddot\nu$), FD parameters \cite{zsd+2015,abb+15},  and time offsets of the ToA data sets relative to the reference data sets as indicated in Table~\ref{tb:datasets}.

FD parameters 
mitigate the effect of possible unmodelled frequency evolution of the pulse profile by allowing for a time delay between different sub-bands according to $\Delta t_{\rm FD} = \sum_{i=1}^{i=N} c_i (\log f)^i$ \cite{abb+15}. Here, $c_i$ are the FD parameters and $f$ the sub-band frequency in GHz, when $\Delta t_{\rm FD}$ is expressed in seconds. Following Ref.~\cite{abb+15}, we employed an F-test to determine that three FD parameters are needed.  Similarly, we employed an F-test to determine that four frequency derivatives are required to model the spin-evolution of pulsar A.

Non-zero FD parameters suggest that future analyses of wide-band data of the Double Pulsar will benefit from deploying smoothly-varying, frequency-dependent templates \cite{ldc+14,pen19}. But, in general, profile evolution and dispersion measure are highly covariant, with a fit for DM being able to (partly) absorb the effects of profile evolution and vice versa. Indeed, allowing the DM reference value to float in our final fit reduces the number of significant FD parameters to one. We note that the described choice does not have an impact on any of the binary timing parameters. We choose to fix the reference DM to the value given in Table~\ref{tab:params} and determine the DM variations as described in Section~\ref{subsec:astrom_timing}.

The parameters shown in Table~\ref{tab:params} are the result of a large number of MC runs. Before each run, we select a random realisation of the astrometry as determined in Section~\ref{subsec:astrom_timing}, i.e.~position, proper motion and parallax that are consistent with the determined distributions. While these parameters are kept fixed for the given MC run, we use a modified version of {\sc Tempo} to fit the described timing model to the remainder of the 20 timing parameters (see Table~\ref{tab:params}) and 22 constant time offsets (JUMPS) between the different data sets (see Section~\ref{subsec:toas}). In each run, the DM value of a given ToA epoch is varied according to the DM curve and the estimated (time-varying) uncertainties as determined in Section~\ref{subsec:astrom_timing}, thereby accounting for the particular DM value and its uncertainty as determined by the Gaussian Learning Process. The entries in Table~\ref{tab:params} are the mean of the resulting distributions of each parameter after a few thousand MC runs. The uncertainties are the standard deviation of the distributions or the maximum {\sc Tempo} error encountered in all MC runs, whichever is larger. 


\begin{table}[H]
\centering
\caption{
Timing parameters for PSR~J0737$-$3039A in TDB units (see text). Except for astrometry and DM, the parameters  were derived using {\sc Tempo} with the 30-s ToA data set. Numbers in parentheses are 1$\sigma$ uncertainties referred to the last quoted digit. The overall reduced $\chi^2$ is 0.97.}
\begin{tabular}{l@{\hspace{1.5em}}l} \hline
\hline
\noalign{\smallskip}
{\bf Parameter} & {\bf Value} \\ 
\noalign{\smallskip}
\hline
\noalign{\smallskip}
Right ascension, $\alpha$ (J2000)  &\hspace{-4mm} $07^{\rm h}37^{\rm m}51.\!\!^{\rm s}248115(10)^\dagger$ \\
Declination, $\delta$ (J2000)      &\hspace{-4mm} $-30^\circ 39'40.\!\!''70485(17)^\dagger$ \\
Proper motion R.A., $\mu_\alpha$ (mas\,yr$^{-1}$) &\hspace{-4mm} $-$2.567(30)$^\dagger$ \\
Proper motion Dec., $\mu_\delta$ (mas\,yr$^{-1}$) &\hspace{-4mm} 2.082(38)$^\dagger$ \\
Parallax, $\pi_{\rm c}$ (mas)      &\hspace{-4mm} 1.36($+0.12$,$-0.10$)$^\dagger$ \\
Position epoch (MJD)               &\hspace{-4mm} 55045.0000 \\[0.7mm]
Rotational frequency, $\nu$ (Hz)   &\hspace{-4mm} 44.05406864196281(17)$^\ddagger$\hspace{-6mm} \\
First freq.\ derivative, $\dot{\nu}$ (Hz s$^{-1}$) &\hspace{-4mm} $-3.4158071(11) \!\times\! 10^{-15}$$^\ddagger$\hspace{-6mm} \\ 
Second freq.\ derivative, $\ddot{\nu}$ (Hz\,{\rm s}$^{-2}$)  &\hspace{-4mm} $-2.286(29) \!\times\! 10^{-27}$ $^\ddagger$ \\
Third freq.\ derivative, $\dddot{\nu}$ (Hz\,{\rm s}$^{-3}$)  &\hspace{-4mm} $1.28(26) \!\times\! 10^{-36}$ $^\ddagger$\\
Fourth freq.\ derivative, $\ddddot{\nu}$ (Hz\,{\rm s}$^{-4}$) &\hspace{-4mm} $4.580(86) \!\times\! 10^{-43}$ $^\ddagger$ \\
Timing epoch, $t_0$ (MJD)  &\hspace{-4mm}  55700.0\\[1mm]
Profile evolution, FD parameter $c_1$  &\hspace{-4mm}  $ 0.0000180(75)$ \\
Profile evolution, FD parameter $c_2$  &\hspace{-4mm}  $-0.0001034(10)$ \\   
Profile evolution, FD parameter $c_3$  &\hspace{-4mm}  $ 0.0000474(26)$ \\ [1mm]
Dispersion measure, DM (pc\,cm$^{-3}$) &\hspace{-4mm}  48.917208 \\[1mm]
Orbital period, $P_{\rm b}$ (day)         &\hspace{-4mm} 0.1022515592973(10) \\
Projected semimajor axis, $x$ (s)         &\hspace{-4mm} 1.415028603(92) \\ 
Eccentricity (Kepler equation), $e_T$     &\hspace{-4mm} 0.087777023(61) \\
Epoch of periastron, $T_0$ (MJD)          &\hspace{-4mm} 55700.233017540(13) \\
Longitude of periastron, $\omega_0$ (deg) &\hspace{-4mm} 204.753686(47) \\[1mm]
Periastron advance, $\dot\omega$ (deg\,yr$^{-1}$) $^\#$  &\hspace{-4mm} 16.899323(13)\\ 
Change of orbital period, $\dot{P}_{\rm b}$        &\hspace{-4mm} $-1.247920(78) \!\times\! 10^{-12}$\hspace{-6mm} \\
Einstein delay amplitude, $\gamma_{\rm E}$ (ms)    &\hspace{-4mm} 0.384045(94) \\
Logarithmic Shapiro shape, $z_s$                   &\hspace{-4mm} 9.65(15) \\
Range of Shapiro delay, $r$ ($\mu$s)          &\hspace{-4mm} 6.162(21) \\
NLO factor for signal prop., $q_{\rm NLO}$         &\hspace{-4mm} 1.15(13) \\
Relativistic deformation of orbit, $\delta_\theta$ &\hspace{-4mm} $13(13)\!\times\!10^{-6}$\\
Change of proj.\ semimajor axis, $\dot{x}$         &\hspace{-4mm} $8(7)\!\times\!10^{-16}$ \\
Change of eccentricity, $\dot{e}_T$ (s$^{-1}$)     &\hspace{-4mm} $3(6)\!\times\!10^{-16}$ \\[1mm]
{\em Derived parameters} & \\[1mm]
$\sin i = 1 - \exp(-z_s)$                          &\hspace{-4mm} 0.999936(+9/$-$10) \\
Orbital inclination, $i$ (deg)                     &\hspace{-4mm} 89.35(5) or 90.65(5)  \\
Total mass, $M$  (M$_\odot$)$^{\ast}$              &\hspace{-4mm} {2.587052\PM{9}{7}} \\
Mass of pulsar A, $m_{\rm A}$ (M$_\odot$)$^{\ast}$ &\hspace{-4mm} {1.338185\PM{12}{14}} \\
Mass of pulsar B, $m_{\rm B}$ (M$_\odot$)$^{\ast}$ &\hspace{-4mm} {1.248868\PM{13}{11}} \\
Galactic longitude, $l$ (deg)                      &\hspace{-4mm} 245.2357  \\
Galactic latitude,  $b$ (deg)                      &\hspace{-4mm} $-$4.5049 \\
Proper motion in $l$, $\mu_l$ (mas\,yr$^{-1}$)     &\hspace{-4mm} {$-$3.066(35)} \\
Proper motion in $b$, $\mu_b$ (mas\,yr$^{-1}$)     &\hspace{-4mm} {$-$1.233(31)} \\
Distance from $\pi_{\rm c}$, $d$ (pc)              &\hspace{-4mm}  {$735(60)$} \\
Transverse velocity, $v_{\rm T}$ (km\,s$^{-1}$)    &\hspace{-4mm} {$11.5(10)$} \\
\noalign{\smallskip}
\hline
\noalign{\smallskip}
\multicolumn{2}{l}{$^\dagger$ \footnotesize
See Sec.~\ref{subsec:astrom_timing} \& \ref{subsec:vlbi} for the derivation of these values.}\\
\multicolumn{2}{l}{$^\ddagger$ \footnotesize See footnote \ref{foot:F0F1}.}\\
\multicolumn{2}{l}{$^\#$ \footnotesize $\dot\omega \equiv 2\pi k/P_\mathrm{b}$. $k$ is the PK timing parameter in Eq.~(\ref{eq:omega}).} \\
\multicolumn{2}{l}{$^\ast$ \footnotesize See footnote \ref{foot:Gm}.} \\
\end{tabular}
\label{tab:params}
\end{table}


In order to allow direct comparisons with previous publications (especially~Ref.~\cite{ksm+06}), parameters in Table~\ref{tab:params} were measured within the timescale known as ``Barycentric Dynamical Time'' (TDB) as implemented in {\sc Tempo}. TDB runs at a slower rate than the ``Barycentric Coordinate Time'' (TCB), which is recommended by  IAU 2006 Resolution B3 \cite{IAU06B3}. This choice does not have any consequences for the gravity tests or discussions presented below, as all (dimensionful) parameters determined from ToAs measured using TDB are multiplied by a constant factor, which either drops out or is (still) too small to be relevant for the discussion of masses or PK parameters. In order to transfer from TDB to TCB, parameters with units of time shown in Table~\ref{tab:params} need to be divided by $\kappa = (1 - 1.550519768\times 10^{-8})$ and the values adjusted accordingly \cite{hem06,IAU06B3}\footnote{Even though the factor $(1-\kappa)$ is small, such correction is required before using the listed timing results for predictions of the folding parameters with {\sc Tempo2}, which has implemented TCB. According to IAU Resolution 2006 B3, the conversion takes place as: ${\rm TDB} = {\rm TCB} - (1-\kappa) \times ({\rm JD}_{\rm TCB} - 2443144.5003725) \times 86400 - 6.55 \times 10^{-5})$, where JD$_{\rm TCB}$ is the relevant Julian Date. See \url{https://www.iau.org/static/resolutions/IAU2006_Resol3.pdf}. Note that this described conversion needs to be applied as given to the quoted epoch of periastron passage ($T_0$) value before it is used with {\sc Tempo2}.}. 
As for the astrometric timing, the transfer of the ToAs from the topocentric to the barycentric reference frame was made using the DE436 solar-system ephemeris.

While our timing results are perfectly consistent with those based on 2.5 years of timing data presented earlier \cite{ksm+06},
the increased length and density of our data set leads to unprecedented precision in the measured parameters. For instance, the orbital period is measured with a precision of 86 nanoseconds. Most importantly, the Keplerian and PK-parameters have reached a precision that leads to a very significant improvement in our ability to conduct precision tests of strong-field gravity, including radiative and light-propagation aspects, as shown in the following sections. The need to fit up to the fourth spin frequency derivative (cf.~Section~\ref{sec:binary_model}) reflects on one hand the exceptional duration and density of our data set but also indicates a certain degree of timing noise, but at a level that is consistent with other pulsars of this age \cite{hlk10}\footnote{The rotational spin frequency parameters were estimated with a standard {\sc Tempo} analysis. As discussed in Ref.~\cite{chc+11}, this procedure underestimates the true uncertainty in the presence of un-modelled red timing noise. Comparison of the standard analysis with the results of a full generalised least-squares analysis shows that the true uncertainty in the spin frequency parameters is about an order of magnitude larger than the values quoted in Table~\ref{tab:params}. We emphasize that the un-modelled red noise has no effect on the orbital parameters as expected because of their much shorter timescale.\label{foot:F0F1}}.
  
The observed improvements in PK parameters are in line with the expectation based on our earlier measurements \cite{kw09}, although a detection of the relativistic deformation of the orbit, described by PK parameter $\delta_\theta$, has occurred somewhat earlier than predicted. A consequence is a slightly smaller improvement in the precision of PK parameter $\gamma_{\rm E}$, due to a correlation between the two parameters (see Section~\ref{subsubsec:test_dth}). The precision in the measurement of the periastron advance, PK parameter $\dot\omega$ (or $k$), has improved beyond the level of 2PN-contributions, making it necessary for the first time to include considerations of the Lense-Thirring effect and EoS of super-dense matter in our analysis (see Section~\ref{subsubsec:test_EoS_LT}). Similarly, we have to take higher-order effects into account when studying light propagation in the Double Pulsar system (see Section~\ref{subsubsec:test_Shapiro}). Here we point out that the derived orbital inclination angle $i = 89^\circ.35\pm 0^\circ.05$ (or $180^\circ - i$) is consistent with the value derived in our earlier timing analysis, i.e.~$i_{\rm KSM06} = 88.69\PM{50}{76}$  \cite{ksm+06}, but now also in perfect agreement with the value obtained from modelling of the eclipse pattern ($i_{\rm eclipse} = 89^\circ.3\pm 0^\circ.1$ \cite{bkk+08,breton09}). In contrast, both values are in tension with measurements based on the changes in the intensity scintillation pattern as a function of orbital phase ($i_{\rm scint} = 88^\circ.1\pm 0^\circ.5$ \cite{rcn+14}). The latter method has the potential to resolve the $i \rightarrow 180^\circ - i$ ambiguity caused by the Shapiro delay's dependency on $\sin i$. Ongoing MeerKAT measurements promise to improve on the scintillation results to resolve this difference \cite{ksv+21}.

\subsection{Masses and Tests of General Relativity}
\label{subsec:GRtests}

\subsubsection{Mass measurement}
\label{subsubsec:masses}

\begin{figure}[H]
    \centering
    \includegraphics[width=8.5cm]{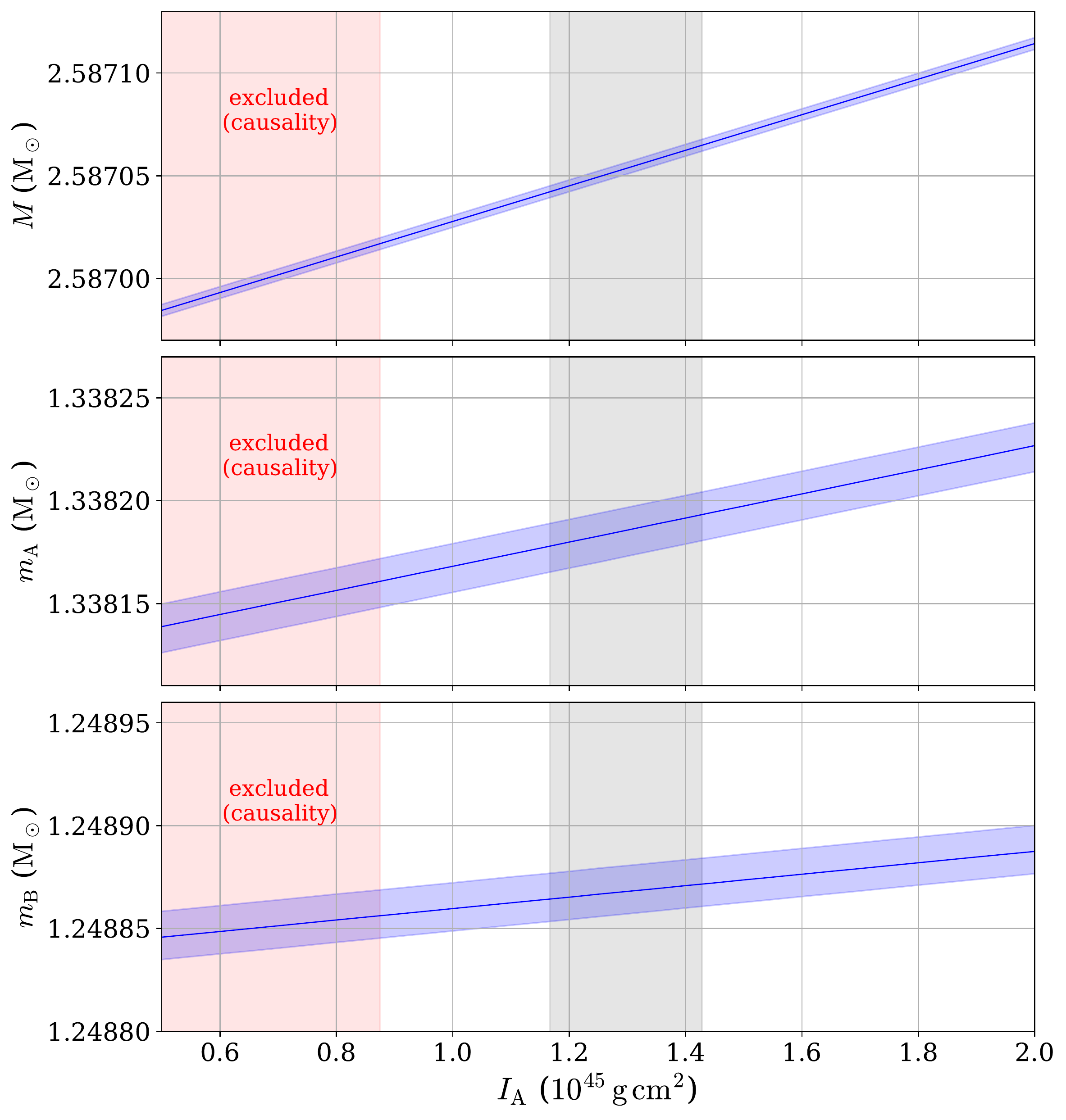}
    \caption{Dependence on the masses of the Double Pulsar system on the MoI of pulsar A, $I_{\rm A}$. Mass calculations are based on GR and the two PK parameters $k(m_{\rm A},m_{\rm B},I_{\rm A})$ [Eq.~(\ref{eq:k_theory})] and $s(m_{\rm A},m_{\rm B})$ [Eq.~(\ref{eq:mf1PN})]. The figure shows the total mass $M = m_{\rm A} + m_{\rm B}$ (upper panel) and the masses of A (middle panel) and B (lower panel) as a function of $I_{\rm A}$. The blue bands indicate the 68.3\% confidence range in the masses for each given $I_{\rm A}$. The grey bands indicate the likely MoI range, as suggested by the radius limits of \cite{dcp+20} (90\% confidence), when using the radius-MoI relation of \cite{Lattimer:2019}. The red areas are excluded by the causality condition for the EoS \cite{Lattimer:2019AIPC}.}
    \label{fig:massdet}
\end{figure}

The standard procedure to obtain the masses of a (``clean'') relativistic binary pulsar system is the measurement of two PK parameters. Under the assumption of GR, and using the well-measured Keplerian parameters, one can calculate the two a priori unknown masses \cite{lk04}. Alternatively, one can fit the DDGR model \cite{tw89}, which is based on GR and has the companion and total masses as free parameters. For the Double Pulsar the situation is more complicated, since the most precisely known PK parameter, $\dot\omega \equiv n_\mathrm{b}k$, has a contribution from the Lense-Thirring effect due to the rotation of pulsar A, which is more than 30 times larger than the measurement error (see Eq.~(\ref{eq:omdotLTnum}) and $\dot\omega$ in Table~\ref{tab:params}). However, a calculation of the Lense-Thirring contribution requires the knowledge of the MoI of pulsar A, $I_{\rm A}$, which comes with a significant uncertainty due to our imperfect knowledge of the EoS for NS matter (cf.\ \cite{lp01,ls05}). Figure~\ref{fig:massdet} demonstrates how the masses, determined from $k$ and the Shapiro shape $s$ (the second most precise PK parameter) depend on $I_{\rm A}$. To constrain $I_{\rm A}$ one can use constraints on the EoS obtained from the double NS merger GW170817 \cite{GW170817}. As a first approximation, we use the limits on the radius of a NS obtained in \cite{dcp+20} and convert them into a probability distribution for $I_{\rm A}$ with the help of the radius-MoI relation for A given in \cite{Lattimer:2019}. Doing so, one obtains from Eq.~(\ref{eq:omdotLTnum}) 
\begin{equation}
   \dot\omega^{\rm LT,A} = -4.83\PM{29}{35}\times 10^{-4}\,{\rm deg\,yr^{-1}} \;,
\end{equation}
and in combination with the PK parameter $s$ one then gets for the (Doppler-shifted) masses\footnote{Mass determination from PK parameters yields the product $Gm_i$ ($i={\rm A,B}$). Consequently, whenever we express the masses in terms of solar masses M$_\odot$, the numerical value is calculated via $Gm_i / (\mathcal{GM})_{\odot}^\mathrm{N}$, where $(\mathcal{GM})_{\odot}^\mathrm{N} \equiv 1.3271244 \times 10^{26} \,{\rm cm^3\,s^{-2}}$ is the {\em nominal solar mass parameter} as defined by the IAU 2015 Resolution B3 \citep{mamajek2015iau}. \label{foot:Gm}}
\begin{eqnarray}
    m_{\rm A} &=& 1.338185\PM{12}{14} \, {\rm M}_\odot, \,\label{eq:mA}\\ 
    m_{\rm B} &=& 1.248868\PM{13}{11} \, {\rm M}_\odot, \,\label{eq:mB}\\
    M         &=& 2.587052\PM{9}{7}   \, {\rm M}_\odot.\label{eq:M}
\end{eqnarray}
In these calculations, for both of the PK parameters $k$ and $s$ we have accounted for NLO contributions as given in Section~\ref{sec:binary_model}. While for $k$ this 2PN correction is highly significant ($\sim 35\,\sigma$), for $s$ the 1PN term amounts to a correction of only $\sim$ 1-$\sigma$. It is interesting to note, while to leading order a measurement of $k$ can directly be converted into a measurement of the total mass $M$, this is no longer the case once 2PN and Lense-Thirring corrections need to be accounted for (cf.\ Eqs.~(\ref{eq:k2PN}) and (\ref{eq:kLT})). 

Concerning the precision obtained in Eqs.~(\ref{eq:mA}) to (\ref{eq:M}), one has to keep in mind that strictly speaking they need to be re-scaled with the unknown Doppler factor $D$ of Eq.~(\ref{eq:timingmodel}), in order get the actual (intrinsic) inertial masses \cite{dd86,dt92}\footnote{{This Doppler-shift re-scaling , $m^{\rm measured} = m^{\rm intrinsic}/D$, between the pulsar-timing measured masses and the intrinsic ones \cite{dd86} is due to the same effect which causes the masses directly deduced from gravitational-wave-template fitting in LIGO/Virgo data analysis to differ from the intrinsic, source-frame masses by a redshift factor: $m^{\rm obs} = (1+z) m^{\rm source}$ \cite{ks87}.}}.
To calculate $D$ one would need the radial velocity $v_{\rm R}$ of the Double Pulsar system with respect to the SSB, which is unknown. Given the small transverse velocity $v_{\rm T}$ (see Table~\ref{tab:params} and also Section~\ref{sec:DP_formation}) one would naturally expect $v_{\rm R} \sim 11$\,km\,s$^{-1}$ (cf.\ Figure~\ref{fig:Gal_motion}), which converts into a mass uncertainty of $\sim 3 \times 10^{-5}\,{\rm M}_\odot$. It is important to note, that in gravity tests the unknown factor $D$ drops out and is therefore irrelevant \cite{dd86,dt92}.

The PK parameter $\gamma_{\rm E}$ (within the approximation required) is not affected by the uncertainty in the EoS. However, a mass determination based on $s$ and $\gamma_{\rm E}$ yields considerably larger errors in the masses: {$m_{\rm A} = 1.3393(19)\,{\rm M}_\odot$} and {$m_{\rm B} = 1.2494(9)\,{\rm M}_\odot$}. The same is true for a mass determination based on $s$ and $R$ (latter taken from \citep{ksm+06}, where $R \equiv m_{\rm A}/m_{\rm B} = 1.0714(11)$ is given): $m_{\rm A} = 1.3379(29)\,{\rm M}_\odot$ and $m_{\rm B} = 1.2487(14)\,{\rm M}_\odot$.  Note that a mass measurement using $\dot{P}_{\rm b}$ is to some extent also EoS dependent, due to the spin-down mass loss of A (see Eq.~(\ref{eq:Pbdotmassloss})), although currently this is still at a negligible level. Combining $s$ and $\dot{P}_{\rm b}$ (from Section~\ref{subsubsec:test_gw}) gives {$m_{\rm A} = 1.338176(68)\,{\rm M}_\odot$} and {$m_{\rm B} = 1.248842(34)\,{\rm M}_\odot$}, which is obviously less precise than the masses in Eqns.~(\ref{eq:mA}) and (\ref{eq:mB}). In summary, in spite of the uncertainty in the EoS, $s$ and $k$ give the by far best estimate for the Double Pulsar masses.

\subsubsection{Gravitational wave emission}
\label{subsubsec:test_gw}

With the (GR) masses (\ref{eq:mA}) and (\ref{eq:mB}), the binary system  is fully determined, and any further PK parameter can be used for a test of GR. Of particular importance is the test of GW damping. In their back-reaction on the system, GWs extract orbital energy and angular momentum from the binary motion, leading to a secular change of the Keplerian parameters $P_{\rm b}$, $x$ and $e$. Temporal changes in the latter two are still well below the measurement precision, which in terms of $x$ means that the shrinkage of the orbit due to GW damping is not yet seen directly through a change in the size of the orbit. The (observed) change of the orbital period $\dot{P}_{\rm b}$, on the other hand, is highly significant, measured with a fractional precision of $6\times 10^{-5}$ (see Table~\ref{tab:params}). This is nearly an order of magnitude better than for the Hulse-Taylor pulsar \cite{wh16}, and as in that system it is dominated by the GW damping. 

For a GW test one needs the intrinsic change of the orbital period, $\dot{P}_{\rm b}^{\rm int}$.  Therefore external contributions, $\dot{P}_{\rm b}^{\rm ext}$, resulting from differential accelerations between the SSB and the Double Pulsar system, need to be subtracted from the observed orbital period change \cite{dt91}. The Shklovskii effect \cite{shk70} leads to an apparent acceleration due to the transverse motion of the Double Pulsar system, and is given by
\begin{equation}
    \dot{P}_{\rm b}^{\rm Shk} 
    = \frac{d(\mu_\alpha^2 + \mu_\delta^2)}{c} \, P_{\rm b}
    = 1.72(15) \times 10^{-16} \,. 
    \label{eq:PbdotShk}
\end{equation}
The distance $d$ has been calculated from the weighted mean probability distribution of Figure~\ref{fig:pxpdfs}. The differential acceleration in the Galactic gravitational potential leads to a correction of similar magnitude. If one takes the potential of \cite{McMillan:2017} to calculate the Galactic gravitational accelerations of the SSB and the Double Pulsar, $\vec{g}_{\rm SSB}$ and $\vec{g}_{\rm DP}$ respectively, then one finds
\begin{eqnarray}
     \dot{P}_{\rm b}^{\rm Gal}
     &=& \frac{\vec{K}_0\cdot (\vec{g}_{\rm DP} - \vec{g}_{\rm SSB})}{c} \, P_{\rm b}
     \nonumber\\
     &=& -3.40\PM{25}{24} \times 10^{-16} \,,
     \label{eq:PbdotGal}
\end{eqnarray}
where $\vec{K}_0$ is a unit vector directed from the SSB to the Double Pulsar. Instead of the potential in \cite{McMillan:2017}, one can also use a more analytic approach for the Galactic correction, like in \cite{dt91,nt95,lwj+09}, which in combination with the latest Galactic parameters based on the results in \cite{GRAVITY:2019,rb20,GRAVITY:2021} gives very similar numbers. Other external contributions to $\dot{P}_{\rm b}$, as for instance discussed in the context of the Hulse-Taylor pulsar in \cite{dt91}, are still negligible (see also the discussions in \cite{kwkl18,hkw+20}). Combining (\ref{eq:PbdotShk}) and (\ref{eq:PbdotGal}), while accounting for correlations, gives
\begin{equation}
      \dot{P}_{\rm b}^{\rm ext} 
      = \dot{P}_{\rm b}^{\rm Shk} + \dot{P}_{\rm b}^{\rm Gal} 
      = -1.68\PM{11}{10} \times 10^{-16} \,,
      \label{eq:PbdotExt}
\end{equation}
and consequently
\begin{equation}
      \dot{P}_{\rm b}^{\rm int} 
      = \dot{P}_{\rm b} - \dot{P}_{\rm b}^{\rm ext} 
      = -1.247752(79) \times 10^{-12} \,.
      \label{eq:PbdotInt}
\end{equation}
At this point it is important to note, that the error in the calculated $\dot{P}_{\rm b}^{\rm int}$ is still clearly dominated by the error in the observed orbital period change $\dot{P}_{\rm b}$ (cf.\ Table~\ref{tab:params}). In other words, the uncertainty in our knowledge of the external contributions to the observed change in the orbital period (Eq.~(\ref{eq:PbdotExt})) is still significantly smaller than the measurement error of the timing parameter $\dot{P}_{\rm b}$.

The second largest contribution to $ \dot{P}_{\rm b}^{\rm int}$, besides GW damping, is related to the mass loss of A as given in Eq.~(\ref{eq:Pbdotmassloss}). Using the limits of \cite{dcp+20} and the radius-MoI relation in \cite{Lattimer:2019} gives
\begin{equation}
     \dot{P}_{\rm b}^{\dot{m}_{\rm A}} = 2.9(2) \times 10^{-17} \,, 
     \label{eq:PbdotMAdot}
\end{equation}
which is still a factor of three smaller than the error of $\dot{P}_{\rm b}^{\rm int}$. Nevertheless, we will include the mass loss contributions in our calculations, in particular since this is relevant in Section~\ref{subsubsec:test_EoS_LT} below, where instead of using a priori constraints on $I_{\rm A}$ we use $\dot{P}_{\rm b}^{\rm int}$ to obtain a probability distribution for the MoI of A. The measured change in the orbital period due to GW emission amounts to
\begin{equation}
    \dot{P}_{\rm b}^{\rm GW} 
    = \dot{P}_{\rm b}^{\rm int} - \dot{P}_{\rm b}^{\dot{m}_{\rm A}}
    = -1.247782(79) \times 10^{-12} \,,
    \label{eq:PbdotGW}
\end{equation}
meaning that GW damping in the Double Pulsar has been measured with a fractional precision of $6\times 10^{-5}$. This improves the last published value for the Double Pulsar \cite{ksm+06} by more than a factor of 200. This is also about a factor of 25 more precise than in the Hulse-Taylor pulsar \cite{wh16}, which is presently limited by the uncertainties in the correction for the external effects. 

\begin{figure}[htp]
    \centering
    \includegraphics[width=8.6cm]{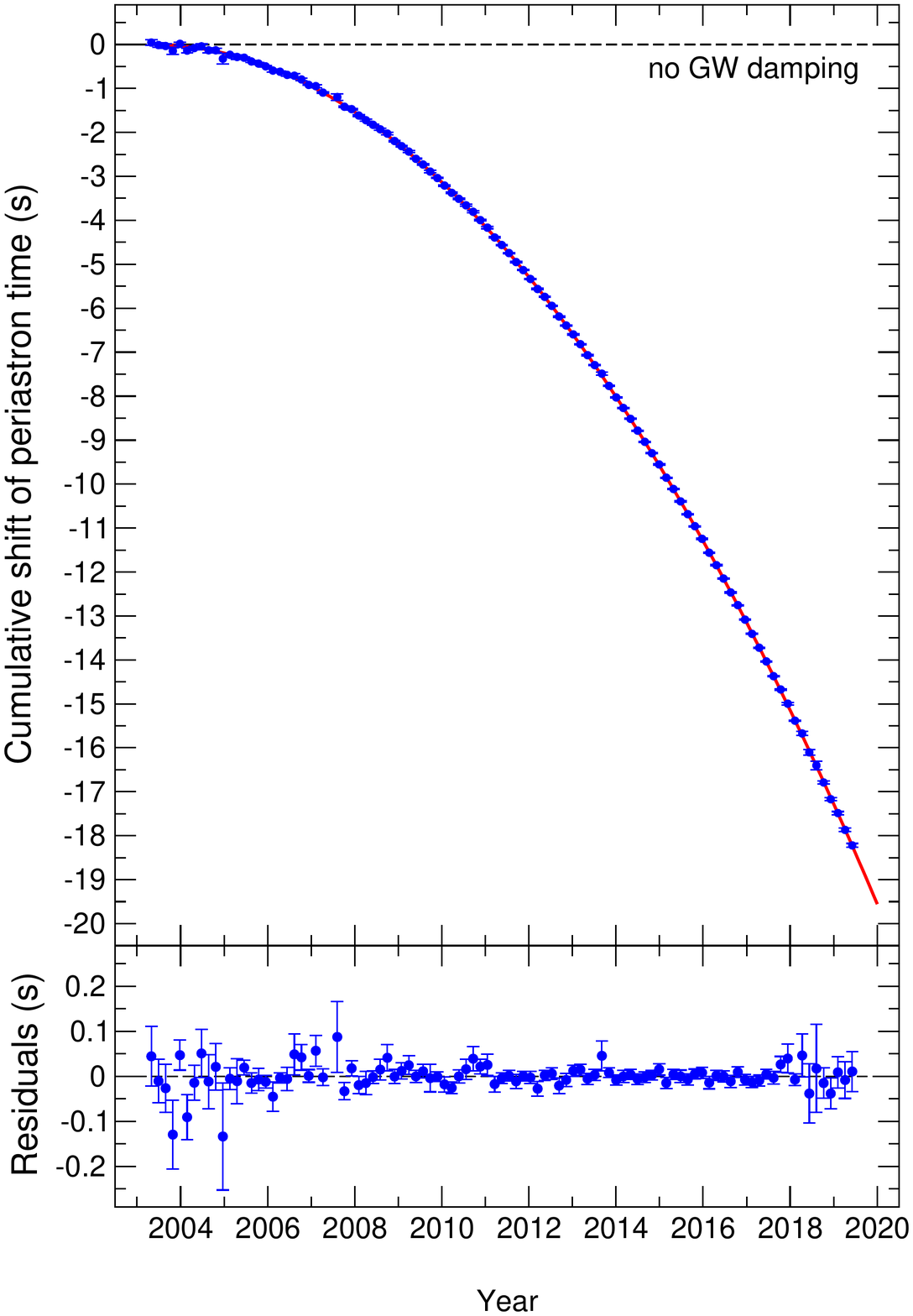}
    \caption{Cumulative shift of the times of periastron passage relative to a non-dissipative model. Each data point covers a time span of 60 days. To each of these sub-sets, we fit
    a Keplerian orbit optimising only for the orbital period and the time and angle of periastron passage. We include the advance of periastron, light-propagation effects and the Einstein delay in the orbital model but keep the values fixed to those in Table~\ref{tab:params}. We plot the difference between the measured time of periastron and a periastron time near its discovery (i.e.~MJD 52759.89, or 2003.33).
    The red curve in the top panel is the GR prediction based on the masses of Section~\ref{subsubsec:masses}. The bottom panel shows the deviation from this prediction, characterised by a normalized $\chi^2=0.76$.}
    \label{fig:parabola}
\end{figure}

The high precision in Eq.~(\ref{eq:PbdotGW}) leads to important tests of the radiative properties of gravity theories. In Section~\ref{sec:altgrav} we will use this for constraining two alternatives to GR. In this section we only provide a test for GR. In GR, to leading order $\dot{P}_{\rm b}^{\rm GW}$ is given by the quadrupole formula \cite{pet64}, which corresponds to the 2.5PN contributions in the equations of motion. Using the masses in Eqs.~(\ref{eq:mA}) and (\ref{eq:mB}) one obtains
\begin{equation}
   \dot{P}_{\rm b}^{\rm GW,GR(2.5PN)} = -1.247810\PM{6}{7} \times 10^{-12} \,.
   \label{eq:Pbdot2.5PN}
\end{equation}
The next higher correction (3.5PN in the equations of motion) has been calculated in \cite{bs89}, and amounts to
\begin{equation}
   \dot{P}_{\rm b}^{\rm GW,GR(3.5PN)} = -1.75 \times 10^{-17} \,,
   \label{eq:Pbdot3.5PN}
\end{equation}
which is larger than the error in Eq.~(\ref{eq:Pbdot2.5PN}), and therefore we include it in our calculation for the predicted value:
\begin{eqnarray}
    \dot{P}_{\rm b}^{\rm GW,GR} 
    &=& \dot{P}_{\rm b}^{\rm GW,GR(2.5PN)} + \dot{P}_{\rm b}^{\rm GW,GR(3.5PN)} 
    \nonumber\\
    &=& -1.247827\PM{6}{7} \times 10^{-12} \,. \label{eq:PbdotGR}
\end{eqnarray}
This value is in perfect agreement with the observed value given in Eq.~(\ref{eq:PbdotGW}), leading to a GR test of
\begin{equation}
    \dot{P}_{\rm b}^{\rm GW}  / \dot{P}_{\rm b}^{\rm GW,GR} 
    = 0.999963(63) \,. 
    \label{eq:GR_GW_test}
\end{equation}
This is by far the most precise test of GW emission, about a factor of 25 better than in the Hulse-Taylor pulsar \cite{wh16}. The agreement with GR is further demonstrated in Figure~\ref{fig:parabola}, which shows the cumulative shift in periastron time due to the acceleration in the orbital phase evolution as a result of $\dot{P}_{\rm b}$ (cf.\ Eq.~(\ref{eq:Kepler}). Since this is the most precise of the different tests one obtains from the Double Pulsar, one can say that GR is in agreement with the Double Pulsar at a level of $1.3\times 10^{-4}$ with 95\% confidence. We emphasize that the result (\ref{eq:GR_GW_test}) has been calculated in a Monte-Carlo run, where we have simultaneously randomized the observed parameters and $I_{\rm A}$ (probability distribution like for Eq.~(\ref{eq:PbdotMAdot})) in order to calculate the masses (from $k$ and $s$) and all the $\dot{P}_{\rm b}$ contributions. 

We have also tested the dependence of result (\ref{eq:GR_GW_test}) on our assumptions about the NS EoS. In a separate Monte-Carlo run, instead of the limits in Ref.~\cite{dcp+20} we have assumed a uniform distribution for the MoI, in the (rather extreme) range of $I_{\rm A}^{(45)} = 0.875$ (causality limit \cite{Lattimer:2019AIPC}) and  $I_{\rm A}^{(45)} = 1.981$ ($R_{\rm A} = 15$\,km) \cite{lp01,Lattimer:2019}. The upper limit with its large radius is in clear tension with, e.g., LIGO/Virgo observations \cite{GW170817}. Nevertheless, even for such a distribution for $I_{\rm A}$ we obtain {$\dot{P}_{\rm b}^{\rm GW} / \dot{P}_{\rm b}^{\rm GW,GR} = 0.999958(64)$}, which shows that  (\ref{eq:GR_GW_test}) depends only very weakly on the EoS uncertainty. 

\begin{figure}[htp]
    \centering
    \includegraphics[width=8.5cm]{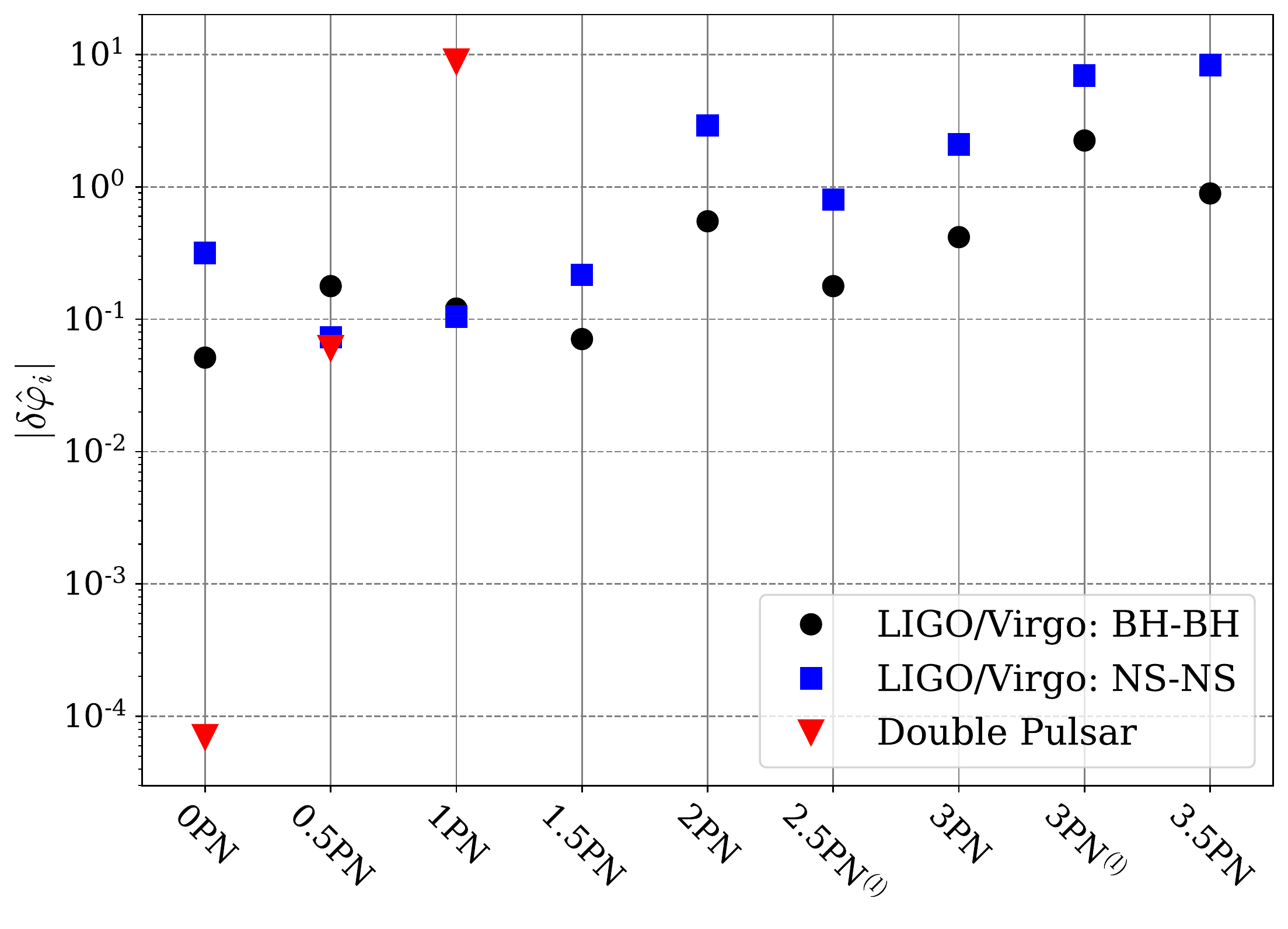}
    \caption{Update of Figure~6 in \cite{GW150914_GR_test} (including data from \cite{LIGO_Virgo_GR_2019,GWTC-2_GR_test}) which shows the 90\% upper bounds on the absolute magnitude of the GR violation parameters $\delta\hat\varphi_i$, from 0PN through 3.5PN (``relative'' order) in the inspiral phase (see e.g. \cite{khh+16} for the definition of the PN phase coefficients and \cite{GW150914_GR_test} for further details on the method). As discussed in \cite{GW150914_GR_test}, the 0.5PN parameter is zero in GR, and therefore understood not as a relative but as an absolute shift. Black circles show the combined limits from the double BH mergers, blue squares are the limits from the double NS merger GW170817, and red triangles give the limits derived from the Double Pulsar GW test in this paper. The PN order on the $x$-axis is in the GR radiation reaction, where the leading contribution (0PN) corresponds to the dissipative 2.5PN term in the equations of motion. Note, such a comparison of tests with different compact objects (BHs vs.\ NSs) as well as different gravity regimes (mildly relativistic vs.\ highly relativistic strong field) does come with a caveat, which is explained in more detail in the text.}
    \label{fig:DP_vs_LIGO}
\end{figure}

In the following we will use the accelerated phase evolution in the orbital motion as shown in Figure~\ref{fig:parabola} to update the PN parameter test introduced by the LIGO/Virgo Collaboration in \cite{GW150914_GR_test}. In particular, it is the tests from the early-inspiral stage, where the phase evolution of the merger of compact objects is analytically described by the PN approximation \cite{Blanchet_LRR}, that can be used in such a comparison with the GW damping measured in the Double Pulsar. Most importantly, the measurement precision of $\dot{P}_{\rm b}$ has greatly improved compared to Ref.~\cite{ksm+06}, which is the data that had been used in \cite{GW150914_GR_test}. Figure~\ref{fig:DP_vs_LIGO} gives an updated version of Figure~6 in  Ref.~\cite{GW150914_GR_test} (including the new results from Refs.~\cite{LIGO_Virgo_GR_2019,GWTC-2_GR_test}\footnote{We have taken the combined results of these references, obtained from the {\sc IMRPhenomPv2} waveform family.}). The bounds are plotted for the different PN levels, allowing for possible GR violations at different PN levels (i.e.\ different powers of frequency), one at a time. Note, Figure~\ref{fig:DP_vs_LIGO} uses the ``relative'' PN order in the radiation reaction (i.e.\ PN order beyond the Einstein quadrupole formula), where the leading order, i.e.\ 0PN, occurs at the 2.5PN order in the binary equations of motion (see e.g.\ \cite{Blanchet_LRR} for a detailed discussion). Due to the many orbits since 2003 ($\sim$60,000), which can be tracked with high precision in a phase-coherent timing solution, the Double Pulsar leads to considerably tighter constraints at low PN orders, whereas it becomes very quickly less constraining for higher PN orders, due to its comparatively small velocity ($v \sim 0.002\,c$).  

While Figure~\ref{fig:DP_vs_LIGO} certainly serves as a comparison on how much a given PN parameter of the inspiral phase evolution can (each at a time) deviate from its GR value in the different experiments, that figure has to be taken with a grain of salt when it comes to interpreting these bounds as limits on deviations from GR predicted by alternative theories of gravity. First, such a comparison mixes tests from two different types of compact objects, i.e.\ NSs and BHs, which might behave quite differently depending on how GR is broken. Hence constraints from experiments with material bodies might not apply to BH dynamics and vice versa. Particularly obvious cases are alternative theories where BH binaries behave like in GR (e.g. \cite{mw13}), or alternative theories where NSs do not carry any scalar charge, while BHs do \cite{ysy16}. Second, the Double Pulsar tests a different gravity regime (mildly relativistic strong field) compared to the GW merger events (highly relativistic strong field). For instance, the Double Pulsar test would generally be insensitive to modifications of GR that only lead to short-range effects, e.g.\ \cite{bppl13,abwz12}\footnote{Short-range modifications of GR that significantly affect the MoI, could in principle show up in the Lense-Thirring test of Eq.~(\ref{eq:LTtest}), see discussion in \cite{hkw+20}.}.
Nevertheless, at least to some extent such a comparison illustrates the complementarity of binary pulsar experiments and merger observations by GW detectors, as long as one keeps in mind the qualitative differences of the various experiments, which are closely linked to the details of a given theory of gravity.

\subsubsection{Lense-Thirring Effect and Equation of State}
\label{subsubsec:test_EoS_LT}

\begin{figure}[htp]
    \centering
    \includegraphics[width=8.5cm]{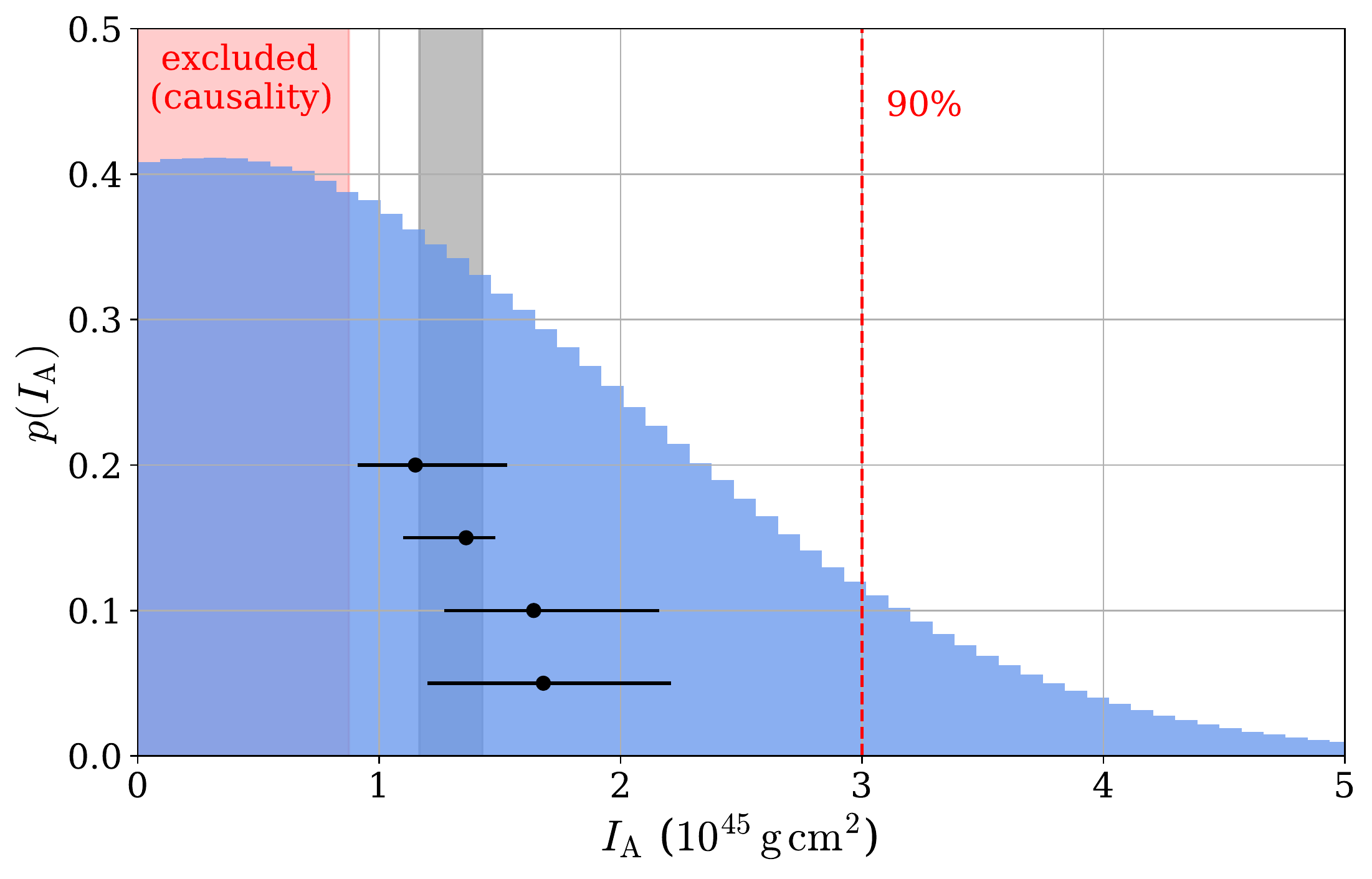}
    \caption{Probability distribution for the MoI $I_{\rm A}$ of A derived from the $k-s-\dot{P}_{\rm b}$ test. The vertical dashed red line indicates the upper bound with 90\% confidence. $I_{\rm A} > 0$ has been assumed as a prior. The light grey band shows the 90\% credible interval one obtains with the limits from \cite{dcp+20} using the radius-MoI relation of \cite{Lattimer:2019}. As a comparison, the horizontal black lines indicate different 90\% ranges derived from (top to bottom): tidal-deformability constraints from GW170817 \cite{lk18}, Bayesian modeling of a range of EOSs \cite{lhs19}, and two different constraints from NICER observations \cite{shcy20}. The red area is excluded by the causality condition for the EoS \cite{Lattimer:2019AIPC}.}
    \label{fig:pdf_MoI}
\end{figure}

In Section~\ref{subsubsec:masses} we have used constraints on the MoI of pulsar A, $I_{\rm A}$, derived from the multi-messenger analysis in Ref.~\cite{dcp+20}, in order to obtain the best mass estimates for the Double Pulsar, as given in Eqs. (\ref{eq:mA})--(\ref{eq:M}). In this section, at first, we ignore any existing constraints on the EoS of NSs, and simultaneously determine $m_{\rm A}$, $m_{\rm B}$ and $I_{\rm A}$, following the procedure outlined in \cite{hkw+20}. As in Section~\ref{subsubsec:masses}, we assume GR to be the correct theory of gravity, and use the three best PK parameters to simultaneously calculate the individual masses of the Double Pulsar and the MoI of A. From the calculations in Section~\ref{subsubsec:masses} it is already obvious that the combination of the PK parameters $k$, $s$, and $\dot{P}_{\rm b}$ is expected to give by far the best results. In a way, we use $s$ and $\dot{P}_{\rm b}$ to determine the masses $m_{\rm A}$ and $m_{\rm B}$, and then use $m_{\rm A}$ to extract $I_{\rm A}$ from the observed advance of periastron $k^\mathrm{obs}$ (see $\dot\omega \equiv n_\mathrm{b}k$ in Table~\ref{tab:params}), a procedure already proposed for the Double Pulsar in \cite{kw09}. In practice the calculations are slightly more complicated, as $\dot{P}_{\rm b}$ also has a contribution proportional to $I_{\rm A}$ (see Eq.~(\ref{eq:Pbdotmassloss})). Although that contribution is still smaller than the error in $\dot{P}_{\rm b}$, we will nevertheless account for it, and follow the procedure of \cite{hkw+20}, i.e.\ calculate $m_{\rm A}$, $m_{\rm B}$ and $I_{\rm A}$ by simultaneously solving the three equations $k^\mathrm{obs} = k(m_{\rm A},m_{\rm B}, I_{\rm A})$, $s^\mathrm{obs} = s(m_{\rm A},m_{\rm B})$, and $\dot{P}_{\rm b}^\mathrm{int} = \dot{P}_{\rm b}(m_{\rm A},m_{\rm B}, I_{\rm A})$. By this we obtain probability distributions for the Double Pulsar masses and the MoI of pulsar A. For the MoI we find {$I_{\rm A} < 3.0 \times 10^{45}\,{\rm g\,cm^2}$} with 90\% confidence. Figure~\ref{fig:pdf_MoI} compares our result with those derived from the GW170817 LIGO/Virgo merger and from NICER X-ray timing. Using a universal relation, like the one in \cite{Lattimer:2019}, one can convert the probability distribution of $I_{\rm A}$ into a probability distribution for A's radius. With 90\% confidence, this gives an upper limit for A's radius of $22$\,km, a value outside any physically valid EoS, and clearly exceeding the range used in \cite{Lattimer:2019}.

While these numbers cannot compete with those obtained from LIGO/Virgo and NICER observations (see e.g.\ \cite{lk18,lhs19,shcy20,dcp+20}), they show that the Double Pulsar constraints are narrowing in on realistic values for NS radii. This is expected to improve considerably over the next years (see \citet{hkw+20}). For the masses we get {$m_{\rm A} = 1.338183\PM{78}{52}\,{\rm M}_\odot$} and {$m_{\rm B} = 1.248869\PM{38}{27}\,{\rm M}_\odot$}. These masses are clearly more uncertain than in Eqs.~(\ref{eq:mA}) and (\ref{eq:mB}), but they do not require any assumption about the EoS for matter at supranuclear densities.

Instead of using the Lense-Thirring effect to constrain the MoI, one can conversely use existing constraints on the NS EoS to test the contribution of spin-orbit coupling to the precession of the pulsar orbit (cf.\ Sec.~7 in \cite{hkw+20}). For this, we introduce a scaling factor $\lambda_\mathrm{LT}$ for the Lense-Thirring part in Eq.~(\ref{eq:kLT}), i.e.
\begin{equation}
    k = k^\mathrm{1PN}  + k^\mathrm{2PN} + \lambda_\mathrm{LT} \, k^\mathrm{LT,A} \,,
    \label{eq:k_lam_LT}
\end{equation}
For GR, one has $\lambda_\mathrm{LT} = 1$. Similar to the procedure above, we can now calculate $m_{\rm A}$, $m_{\rm B}$ and $\lambda_\mathrm{LT}$ by simultaneously solving the three equations $k^\mathrm{obs} = k(m_{\rm A}, m_{\rm B}, \lambda_\mathrm{LT}| I_{\rm A})$, $s^\mathrm{obs} = s(m_{\rm A},m_{\rm B})$, and $\dot{P}_{\rm b}^\mathrm{int} = \dot{P}_{\rm b}(m_{\rm A},m_{\rm B}| I_{\rm A})$, where this time in our Monte Carlo runs $I_{\rm A}$ is randomly chosen from the a probability distribution based on the EoS constraints in \cite{dcp+20} (like in Section~\ref{subsubsec:masses}, when determining the masses from $k$ and $s$). In a sense, we determine the masses of A and B from the PK parameters $s$ and $\dot{P}_\mathrm{b}$ to extract the Lense-Thirring contribution form the observed periastron advance $k$, by assuming a range of values for $I_\mathrm{A}$. We obtain
\begin{equation}
    \lambda_\mathrm{LT} = 0.7 \pm 0.9 \quad \mbox{(68\% C.L.)} \,.
    \label{eq:LTtest}
\end{equation}
This is consistent with GR, but admittedly not very constraining, at least when compared to the weak-field (test-particle type) frame-dragging experiments in the solar system (Lense-Thirring precession of satellite orbits \cite{cpp+19} and Pugh-Schiff precession of gyroscopes \cite{GPB}). Moreover, the above result is not generic, in the sense of the PPN tests in the solar system. In order to extract the Lense-Thirring contribution to $k$, we have assumed that all three, the advance of periastron without Lense-Thirring contribution (i.e.\ Eq.~(\ref{eq:k2PN})), the GW damping, and the Shapiro shape can be calculated from GR. This is generally not expected to be the case in alternative gravity theories. As a result, one would have to implement the equivalent analysis for any other gravity theory, in order to obtain a fully consistent Lense-Thirring test. Given the low precision of the LT test, one may ask if, in view of the other tests with the Double Pulsar, mostly based on very precisely measured PK parameters, this then yields any additional constraints for the gravity theories under consideration. For the alternative theories discussed in Sec.~\ref{sec:altgrav}, at least, spin-orbit effects observed in the Double Pulsar (including $\Omega_\mathrm{B}^\mathrm{spin}$) do not contribute to the constraints. See \cite{hkw+20} for a more detailed discussion, in particular on this test in the context of near field modifications of GR.

Finally, from the analysis outlined above, one can also extract a test of the total advance of periastron $k$, in an $s$--$\dot{P}_\mathrm{b}$--$k$ test. From this one obtains $k^\mathrm{obs}/k^\mathrm{GR} = 1.000015(26)$.

\subsubsection{Relativistic deformation of the orbit}
\label{subsubsec:test_dth}

\begin{figure}[htp]
    \centering
    \includegraphics[width=8.5cm]{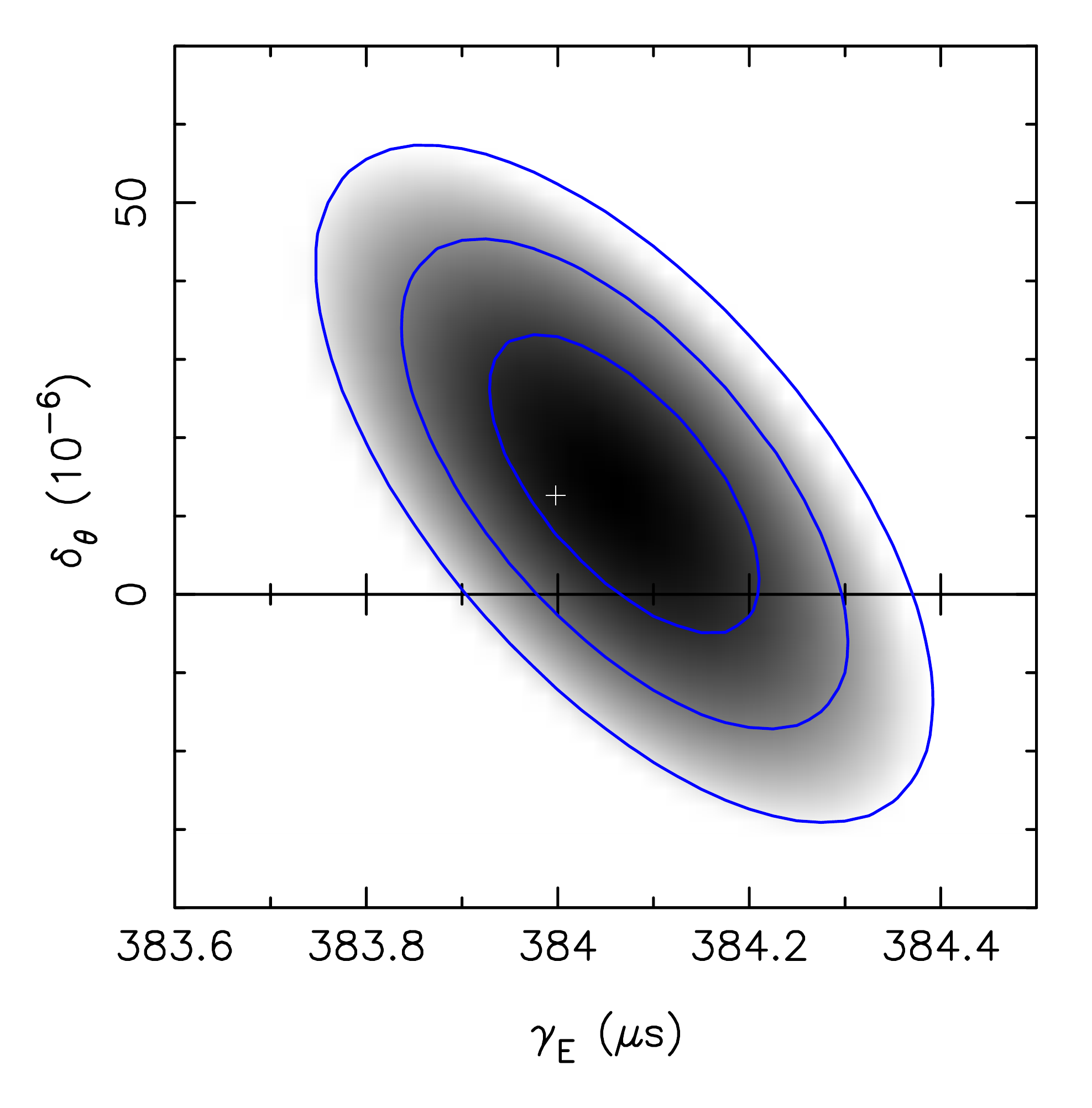}
    \caption{ Map of $\chi^2$-contours demonstrating a correlation between the PK parameters $\gamma_{\rm E}$ and $\delta_\theta$. In order to produce this map, the two parameters were held fixed at their grid position, while fitting for all remaining parameters in the timing model as described earlier. The contours indicate $68.3\%$, $95.4\%$, and $99.7\%$ confidence levels, respectively. The expected GR value is indicated by the cross with uncertainties too small to be visible on this scale: $\gamma_\mathrm{E} = 383.997(5)\,\mu\mathrm{s}$, $\delta_\theta = 12.60889(4) \times 10^{-6}$ (masses from $k$ and $s$, cf.\ Sec.~\ref{subsubsec:masses})}. 
    \label{fig:dtheta}
\end{figure}

\begin{figure}[htp]
    \centering
    \includegraphics[width=7.5cm]{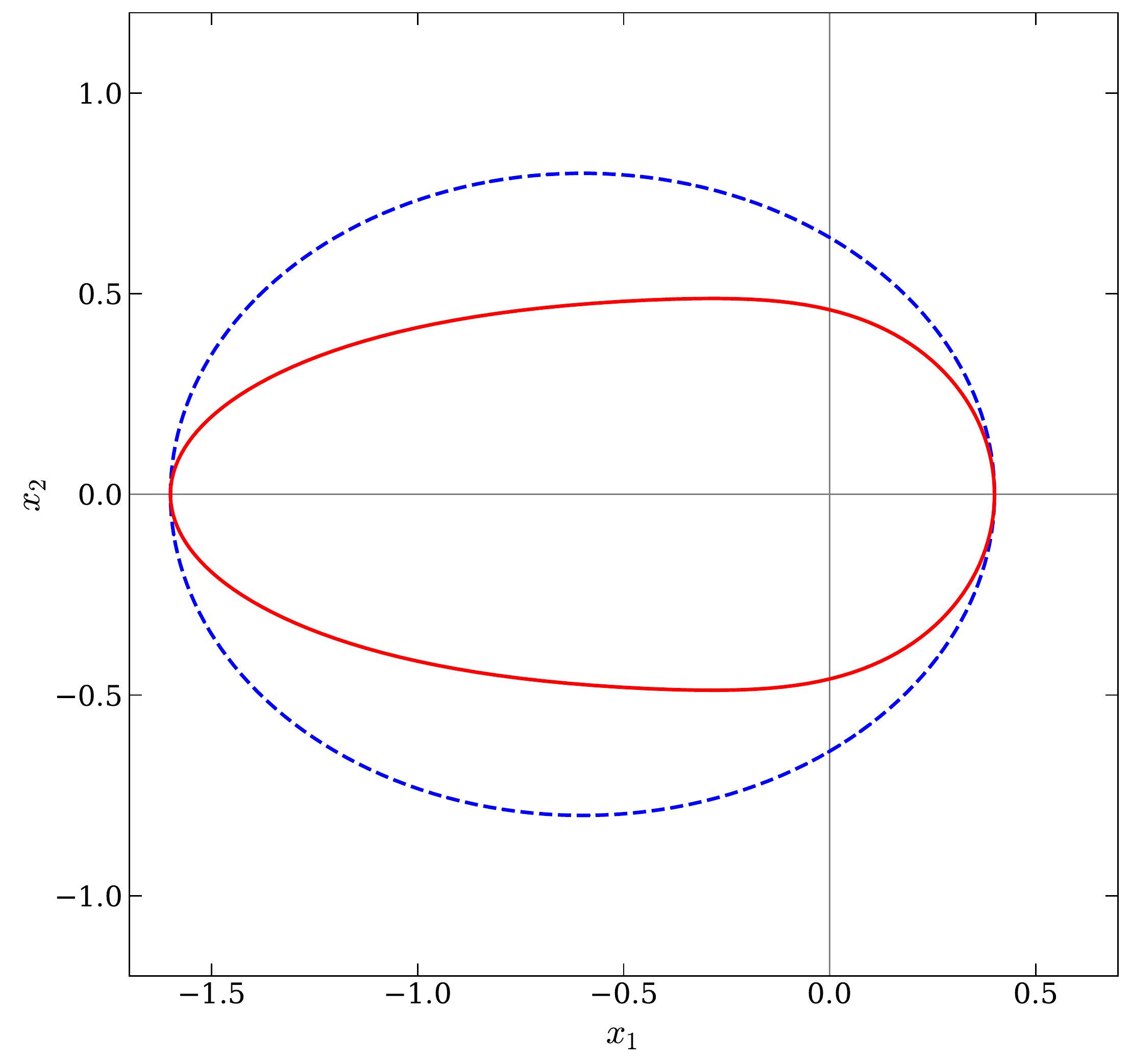}
    \caption{Hugely exaggerated illustration of the effect of the relativistic deformation $\delta_\theta$ on the shape of the orbit,
    which modifies the usually expected elliptical 
    shape. For the red (deformed) orbit we have chosen $a=1$, $e_T = e_r = 0.6$, and $\delta_\theta = 0.5$, while the dashed blue curve has $\delta_\theta = 0$.}
    \label{fig:dtheta-orbit}
\end{figure}

The relativistic deformation of the orbit discussed in context of the R{\o}mer-delay (Section~\ref{subsec:Roemer}) has been detected in our measurements. Similar to the report of a 1.5$\sigma$-detection of $\delta_\theta$ for PSR~B1913+16 \cite{wh16}, our value is also formally detected only just above the 1$\sigma$ level. However, as we demonstrate in Figure~\ref{fig:dtheta}, the parameter is in fact well constrained and in full agreement with the expectation from GR. The figure also demonstrates a correlation between  $\delta_\theta$ and $\gamma_{\rm E}$. Indeed, due to this correlation, the value derived for $\gamma_{\rm E}$ in a timing model ignoring the relativistic deformation, i.e.~assuming $\delta_\theta = 0$, would lead to a value for $\gamma_{\rm E}$ that would be inconsistent with GR at the 2$\sigma$-level while both $\gamma_{\rm E}$ and $\delta_\theta$ are in fact in perfect agreement when including $\delta_\theta$ in our timing model (see white cross in Figure~\ref{fig:dtheta}). In order to illustrate the effect of a non-zero $\delta_\theta$ value, we compare a deformed orbit ($\delta_\theta > 0$) with an elliptical orbit  ($\delta_\theta = 0$) in Figure~\ref{fig:dtheta-orbit} for a hugely exaggerated effect.

\subsubsection{Signal propagation}
\label{subsubsec:test_Shapiro}

\begin{figure}[H]
    \centering
    \includegraphics[width=8cm]{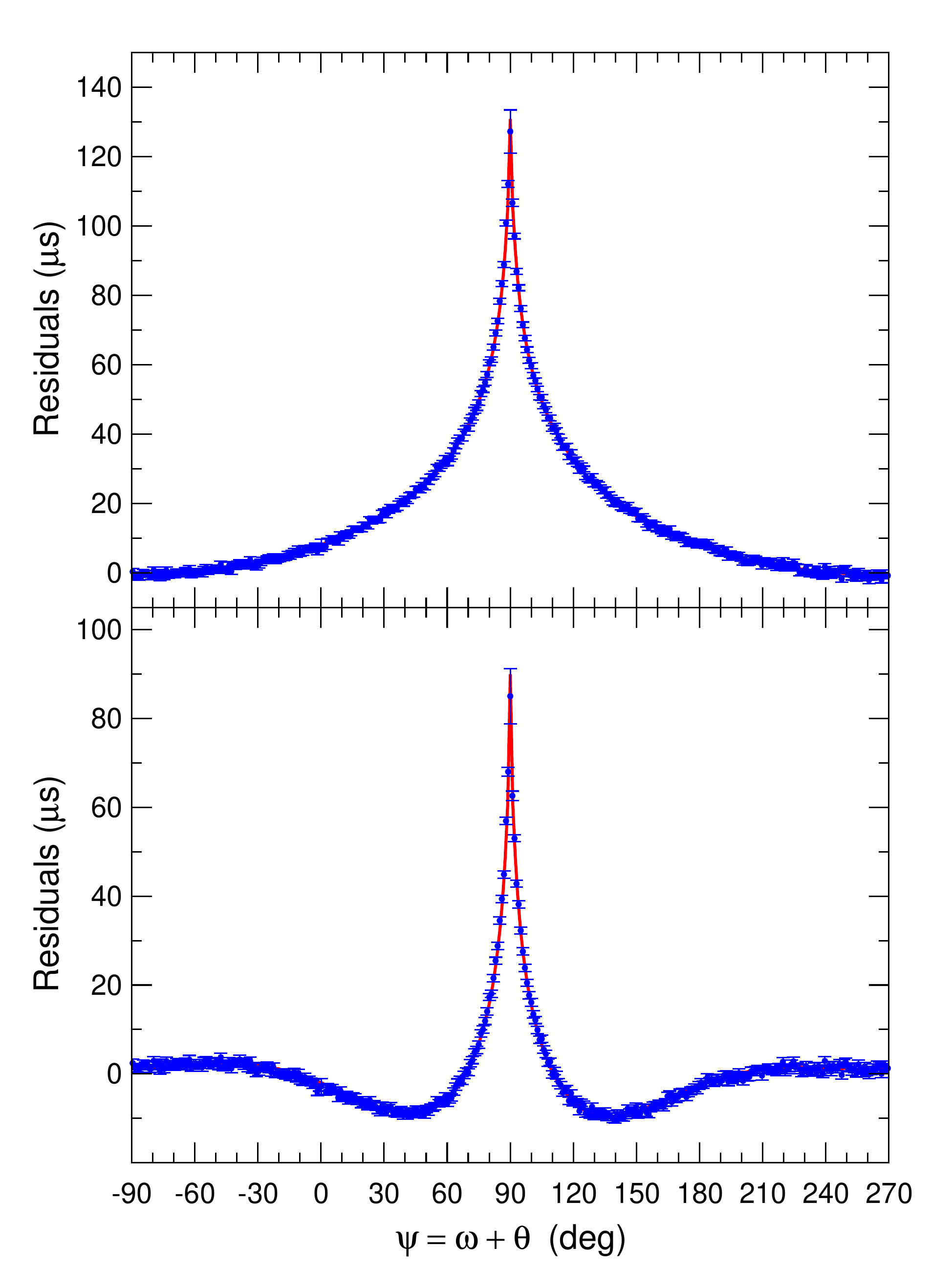}
    \caption{Signature of the Shapiro delay in the timing residuals, plotted against A's angular orbital position $\psi = \omega + \theta$ with respect to the ascending node. The bottom panel shows the aggregated residuals as a function of orbital phase when not including the Shapiro delay parameters, $z_s$ (or equivalently $\sin i$) and $r$. The shape differs from the expected Shapiro delay curve shown in the top panel, as some signal power is absorbed in a fit for the orbital parameters, namely the R{\o}mer delay (see e.g.~\cite{lk04}). The expected shape is recovered by computing the residuals when setting the Shapiro delay parameters to zero in the final timing solution. Note that the shown residuals were corrected for a slow variation in the curve due to the  relativistic orbital precession. The red curves show the GR expectation using values of $\sin i$ and $r$ derived from $\dot \omega$ and $\gamma_{\rm E}$. These are $(\sin i)^{\rm GR} = 0.99988(18)$ and $r^{\rm GR} = 6.1518(11)\,\mu$s, where the uncertainties are determined by those in  $\dot \omega$ and $\gamma_{\rm E}$. Note that the uncertainty in the prediction of $\sin i$ is correspondingly larger than that of the actually measured $\sin i$.}
    \label{fig:shapiro}
\end{figure}

Already in the discovery paper of A \cite{bdp+03} it was suspected that the Double Pulsar system is seen nearly edge-on. This was confirmed by the discovery of a prominent Shapiro delay in the ToAs of A, as well as by the observation of eclipses of A near its superior conjunction, caused by the plasma-filled magnetosphere of B \cite{lbk+04}. The measurement of the Shapiro delay gave access to two additional PK parameters, i.e.\ the ``Shapiro range'' $r$ and the ``Shapiro shape'' $s$ \cite{dd86} (see Eq.~(\ref{eq:ShaLO})). By 2006 these two parameters were measured with high precision, and in particular $s$, as part of an $\dot\omega$-$R$-$s$ test, provided a $10^{-3}$ (95\% C.L.) GR test \cite{ksm+06}. Unfortunately, the measurement of the mass ratio $R$ has not improved since then, mostly because of the fact that B disappeared from view in early 2008 as a result of relativistic spin precession \cite{pmk+10}. In the meantime, the measurement of the Shapiro delay, and therefore $s$, has improved significantly (see Figure~\ref{fig:shapiro}), making it the second most constraining parameter in the mass-mass plane (see Figure~\ref{fig:mm-GR} below), which is also obvious from the mass determinations in Section~\ref{subsubsec:masses}. Since $\gamma_{\rm E}$ by now is more constraining than $R$, one can test $s$ as part of a $\dot\omega$-$\gamma_{\rm E}$-$s$ test. This leads to 
\begin{equation}
     s^{\rm obs}/s^{\rm GR} = 1.00009(18) \,,
     \label{eq:s_test}
\end{equation}
meaning a $4 \times 10^{-4}$ (95\% C.L.) test in agreement with GR, which is somewhat better than the $\dot\omega$-$R$-$s$ test of \cite{ksm+06}, while mainly dominated by the error in $\gamma_{\rm E}$. Moreover, from Eq.~(3.15) in \cite{dt92} it is clear, that for a large range of alternatives to GR, $s$ is only linked to the effective gravitational constant $G_{\rm AB}$ of the orbital dynamics, as $s$ can be identified with the purely geometric quantity $\sin i$. 

A test of the Shapiro range $r$ directly probes the interaction between a strongly self-gravitating NS---in our case pulsar B---and a photon. This can be easily seen within the modified Einstein-Infeld-Hoffmann (mEIH) formalism \cite{will93,dt92}, where $r = (1 + \gamma_{B0})G_{B0}m_{\rm B}/c^3$ (see in particular Eq.~(3.14) in \cite{dt92} with $\varepsilon_{\rm B0} = 2\gamma_{B0} + 1$). $G_{B0}$ is the effective gravitational constant and $\gamma_{B0}$ the strong-field extension of the parametrized post-Newtonian (PPN) formalism parameter $\gamma^{\rm PPN}$ (see e.g.\ \cite{Will:2018}), which enter the leading order spacetime metric $g_{\mu\nu}$ of pulsar B. In GR we have $G_{\rm B0} = G$ and $\gamma_{\rm B0} = 1$. Since $r$ is by far the most uncertain of the PK parameters, apart from $\delta_\theta$ and $\Omega_\mathrm{B}^\mathrm{spin}$, there are several combinations of PK parameters that can be used in the test. In other words, the agreement of $m_{\rm B}^{(r)} \equiv r c^3/G = 1.2510(43)\,{\rm M}_\odot$ (cf.\ Table~\ref{tab:params}) with the companion mass determined in Section~\ref{subsubsec:masses} gives
\begin{equation}
     r^{\rm obs}/r^{\rm GR} = 1.0016(34) \,,
     \label{eq:r_test}
\end{equation}
leading to a $7\times 10^{-3}$ (95\% C.L.) test of $r$ in agreement with GR. 

It is interesting to note that near superior conjunction, the signals of the pulsar propagate through a region with a spacetime curvature (as for instance measured by the Kretschmann scalar) that is many orders of magnitude stronger than in other experiments that test photon propagation in gravitational fields, and therefore the coupling between gravitational and the electromagnetic fields. This is of interest, for instance in the presence of interactions between electromagnetic field and curvature tensor as studied in \cite{rw09}.

\begin{figure}[H]
    \centering
    \includegraphics[width=8.5cm]{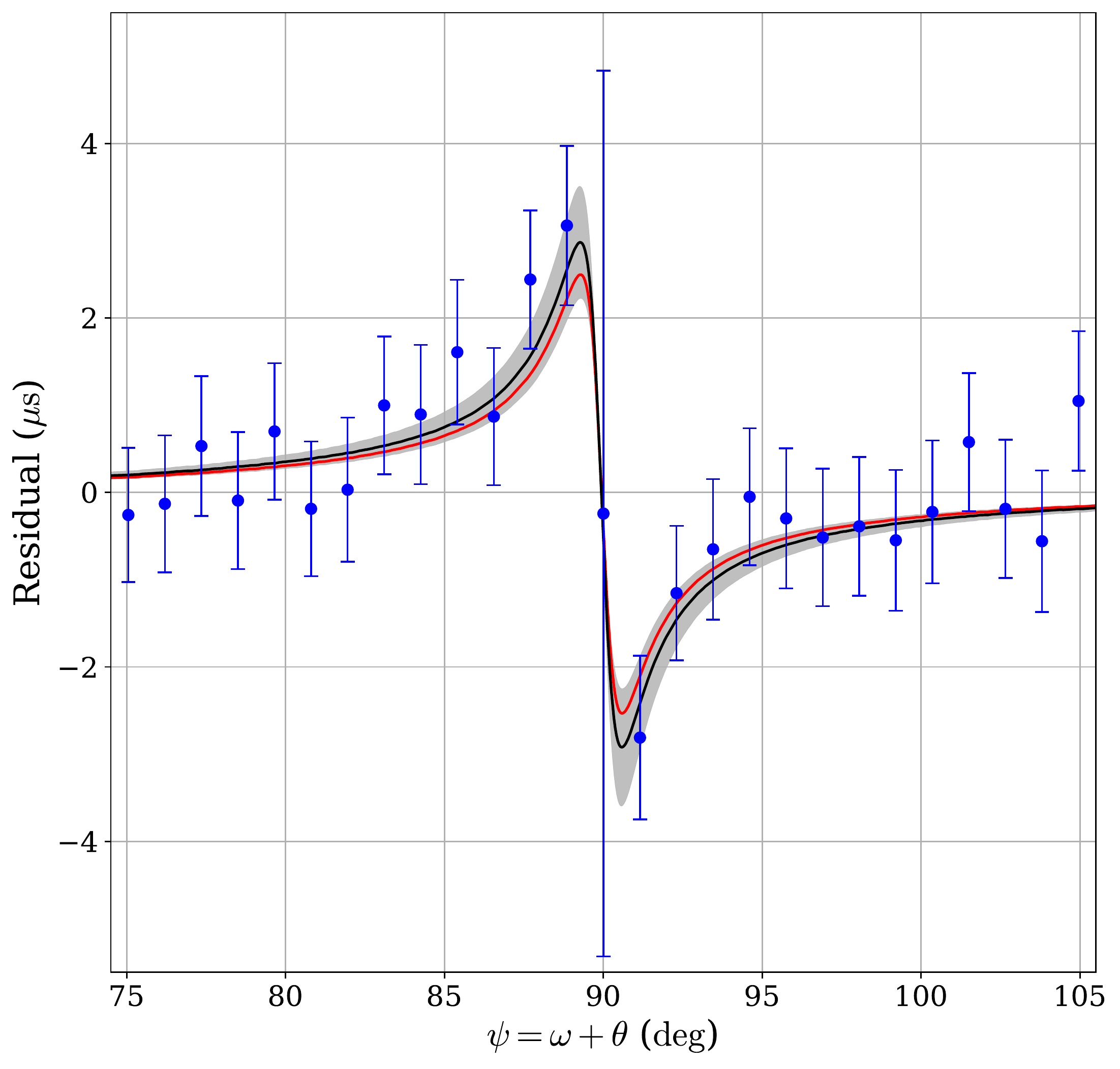}
    \caption{Aggregated residuals (blue) due to NLO contributions in the Shapiro delay and the aberration, plotted against A's angular orbital position $\psi = \omega + \theta$ with respect to the ascending node. The red curve shows the theoretical prediction, and the black curve corresponds to the fitted $q_{\rm NLO}$ (see Table~\ref{tab:params}), with a 2$\sigma$ range indicated by the light grey band. Residuals have been re-scaled by $(1 + e_T\cos\theta)^{-1}$ to account (to leading order) for a secular variation in the amplitude due to the precession of $\omega$. A data point generally results from $\sim 3000$ aggregated residuals. The data point at $\psi=90^\circ$ (i.e.~superior conjunction of A) has a much larger uncertainty, as this bin coincides with the 30-s long eclipse of pulsar A, resulting in both many fewer ToAs available for aggregation and residuals of opposite sign being averaged. }
    \label{fig:shapNLO}
\end{figure}

As it turns out, the leading-order expression (\ref{eq:ShaLO}) does no longer provide a complete description of the signal propagation in the Double Pulsar. Such a model leads to significant residuals near conjunction, as can be seen in Figure~\ref{fig:shapNLO}. These residuals are fully accounted for by the expected NLO contributions discussed in Section~\ref{subsec:Shapiro} and Section~\ref{subsec:Aberration} (red curve in Figure~\ref{fig:shapNLO}). To test the significance of the NLO corrections in $\Delta_{\rm S}$ and $\Delta_{\rm A}$, we have scaled them collectively with a factor $q_{\rm NLO}$ (cf.\ Eqs.~(\ref{eq:Rret}) and (\ref{eq:Rben})), which is unity in GR. As discussed in Section~\ref{subsec:NLOfit}, the similarity of the NLO contributions to $\Delta_{\rm S}$ and $\Delta_{\rm A}$ makes it impossible to test these two contributions separately. For the common factor, which also scales the NLO contribution from lensing (see Eq.~(\ref{eq:Rlen})), we find
\begin{equation}
    q_{\rm NLO} = 1.15(13) \,,
    \label{eq:rhoNLO}
\end{equation}
which can be interpreted as a $\sim$8$\sigma$ measurement of NLO contributions in the signal propagation, in agreement with GR. This clearly shows that the timing of the Double Pulsar has by now reached a precision where NLO contributions to the Shapiro and aberration delay have to be taken into account.

As as consequence of its definition in Section~\ref{subsec:NLOfit}, a fit for $q_{\rm NLO}$ combines two aspects of gravity, 1.5PN corrections to the Shapiro delay related to the motion of the masses (Eq.~(\ref{eq:Rret})) and corrections related to the deflection of the signal beam in the gravitational field of the companion (Eqs.~(\ref{eq:Rlen}) and (\ref{eq:Rben})). Although, as discussed in Section~\ref{subsec:NLOfit}, these contributions cannot be tested separately in a simultaneous fit, in a phenomenological approach one can still test for one at a time, while keeping the other one fixed. If we apply the scaling factor $q_{\rm NLO}$ only to the deflection related contributions, one finds
\begin{equation}
    q_{\rm NLO}[\mbox{deflection}] = 1.26(24) \,.
    \label{eq:RNLOdefl}
\end{equation}
This limit comes solely from the NLO aberration contributions (Eq.~(\ref{eq:Rben})), since a re-scaling of $\delta\Lambda_u^{\rm len}$ is covariant with $s$.  Given the large uncertainty, the above test is admittedly of little interest in all those theories or frameworks where deviations from GR in the signal deflection (caused by a strongly self-gravitating mass) are already well constrained by the measurement of the Shapiro delay PK parameters $r$ and $s$, i.e.\ Eqs.~(\ref{eq:aberrationBend}), (\ref{eq:aberrD}) \footnote{This is for instance the case in the mEIH formalism, where (to leading order) the deflection angle is proportional to $(1 + \gamma_{B0})G_{B0}$ (see e.g.\ Appendix C of \cite{de96a}), which is already tightly constrained by the test of the Shapiro range parameter $r$. More generally, the NLO contribution to $\Delta_{\rm A}$ is parametrized by $r$ and $s$ according to Eqs.~(\ref{eq:aberrationBend}), (\ref{eq:aberrD}).}.

Alternatively, one can keep the deflection contributions fixed, i.e.\ $q_{\rm NLO}=1$ in Eqs.~(\ref{eq:Rlen}), (\ref{eq:Rben}), and only fit for the retardation effect. In this case one obtains
\begin{equation}
    q_{\rm NLO}[\mbox{retardation}] = 1.32(28) \,.
    \label{eq:RNLOret}
\end{equation}
This result clearly demonstrates the significance of the motion of the companion mass while the signal of A propagates across the binary system on its way towards the observer.

Finally, when assuming that the signal deflection is sufficiently well described by GR, then Eq.~(\ref{eq:RNLOdefl}) provides independent evidence (besides \cite{pmk+18,ndk+20}) for the fact that the spin of A is aligned with the orbital angular momentum, and not anti-aligned (${\cal D} \rightarrow -{\cal D}$ in Eq.~(\ref{eq:aberrationBend}))\footnote{It follows from the analysis in \cite{fsk+13} that the rotational axis of A is (practically) perpendicular to the orbital plane. See also Section~\ref{sec:DP_formation}.}.

\subsubsection{Mass-mass diagram}

\begin{figure*}[htp]
    \centering
    \includegraphics[width=17.5cm]{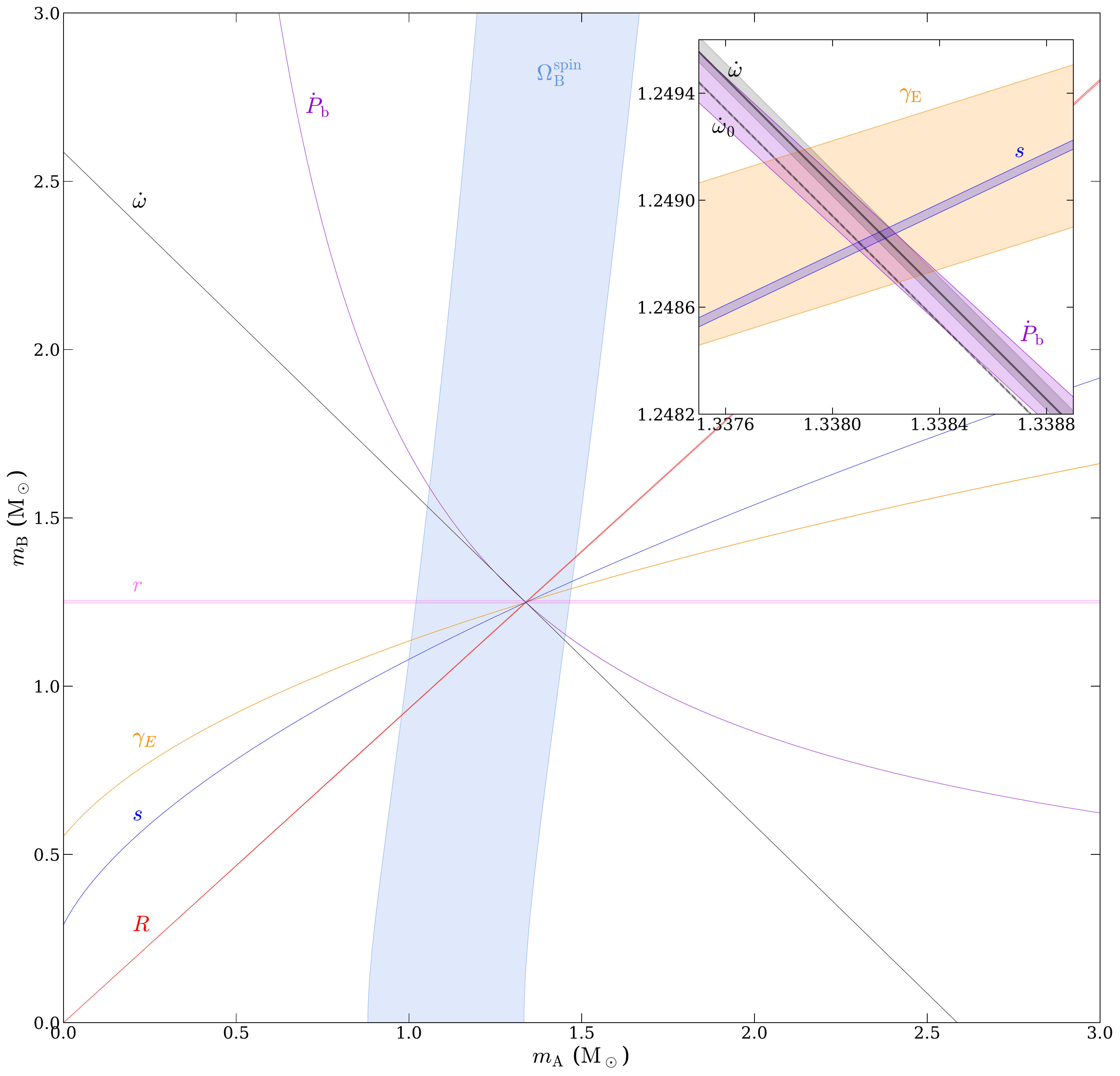}
    \caption{Mass-mass diagram for the Double Pulsar based on GR, for the six PK parameters $\dot\omega\equiv n_\mathrm{b}k$, $\gamma_\mathrm{E}$, $\dot{P}_\mathrm{b}$, $r$, $s$, $\Omega_\mathrm{B}^\mathrm{spin}$, and the mass ratio $R$. The width of each curve indicates the measurement uncertainty of the corresponding parameter. The seventh PK parameter, $\delta_\theta$, is not shown here, since its limits still lie outside the mass ranges shown. For the solid black $\dot\omega$ line $I_{\rm A}^{(45)} = 1.32$ has been used, which corresponds to a NS radius of 12\,km (cf.\ \cite{dcp+20}). The inset is an expanded view of the region of principal interest, where only the four best PK parameters are shown. To illustrate the influence of the LT effect, we have drawn in the inset a dashed black $\dot\omega_0$ line where the LT contribution has been ignored (see Eq.~(\ref{eq:k2PN})). The grey band indicates the range for the $\dot\omega$ line under the variation of $I_{\rm A}$, from the causality limit of \cite{Lattimer:2019AIPC} (left border) to a NS radius of 15\,km (right border). The conversions between radius and MoI are based on the relation in \cite{Lattimer:2019}. There is a small $I_{\rm A}$ dependence of the $\dot{P}_b$ curves, which has been ignored. The intersection of all line pairs is consistent with a small region that corresponds to the masses of A and B.}
    \label{fig:mm-GR}
\end{figure*}

The PK parameters, once external contributions have been removed and a value for $I_\mathrm{A}$ has been chosen, only depend on the a priori unknown masses of pulsars A and B, and the well-measured (and hence determined) Keplerian parameters. Hence, each PK parameter defines a curve in a two-dimensional mass plane displaying their dependence on $m_{\rm A}$ (plotted on the X-axis) and $m_{\rm B}$ (plotted on the Y-axis) for a given theory of gravity. If the theory of gravity used to describe the dependence of the $n$ curves on the masses is consistent with the experimental data, and if no other effects are present, then all curves will meet in a single point. As two curves each define one intersection point, $n-2$ curves have still the potential to miss the intersection point and to falsify the assumed theory. This ``mass-mass'' diagram,  as shown for GR in Figure~\ref{fig:mm-GR}, is a graphical representation of the gravity tests using PK parameters described in Section~\ref{sec:systemintro}, with $\dot P_{\rm b}$ corrected as described in Section~\ref{subsubsec:test_gw}. In the case of the Double Pulsar, the mass ratio adds yet another (largely theory-independent) curve to this diagram in Figure~\ref{fig:mm-GR}. The thickness of the lines represents the uncertainty in the measured PK parameters (68\% c.l.). For all PK-parameters but $\Omega_\mathrm{B}^\mathrm{spin}$, which is determined from eclipse modeling (see Section~\ref{sec:systemintro}), we have reached a point where we need to zoom in to show the intersection point in more detail to recognize the relative uncertainties. While the large panel in Figure~\ref{fig:mm-GR} shows most of the mass-range of interest for NSs, the inset shows such a zoomed-in area, where we choose not to show $R$, $r$ and $\Omega_\mathrm{B}^\mathrm{spin}$ for clarity.  It is obvious that all lines are consistent with a single intersection point given by the masses of Eqs. (\ref{eq:mA}) and (\ref{eq:mB}). The inset, however, highlights the fact that we clearly need to take the Lense-Thirring contribution into account (see discussion in Section~\ref{subsubsec:LT}). With $n=7$ constraints (i.e. the six PK parameters measured here 
and the mass ratio, $R$), we can conduct $n-2=5$ independent tests of theories of gravity, which are all passed with flying colours as shown above.\footnote{An additional test is provided by the PK parameter $\Omega_\mathrm{B}^\mathrm{spin}$, see Table~\ref{tab:grtests}.}


On a final note, in the inset in Figure~\ref{fig:mm-GR} one can see that the measured $\gamma_{\rm E}$ nicely agrees with the intersection of the $s$-curve with either the $\dot\omega$-curve or the $\dot{P}_{\rm b}$-curve. One can convert this into a test of $\gamma_{\rm E}$:
\begin{equation}
    \gamma_{\rm E} / \gamma_{\rm E}^{\rm GR} = 1.00012(25) \,.
\end{equation}
The main interest in such a test lies in the fact that it tests an invariance of the local gravitational constant, which is a specific aspect of local position invariance. We will discuss this in more details in the following section, in the context of two specific alternatives to GR.

\section{Alternative Gravity Theories}
\label{sec:altgrav}

In the previous section we demonstrated that Einstein's GR is in perfect agreement with current observations of the Double Pulsar. This shows that, in contrast to solar-system tests which are in the weak-gravity regime, that GR also is an accurate description of gravitational interactions for two strongly self-gravitating bodies in a mildly relativistic strong-gravity regime. However, since GR was devised in 1915, many other theories of relativistic gravity have been proposed. In this section we consider these alternative theories. Since it is impossible to give an even remotely complete discussion, we focus on two examples. The first, known as ``Damour-Esposito-Far{\`e}se" (DEF) gravity, is a two-parameter class of theories, whereas the second, known as ``Tensor-Vector-Scalar (TeVeS)" gravity, represents the class of ``Modified Newtonian Dynamics" or MONDian gravitational theories. 

These theories are chosen since they are especially suitable for studying the deviations one typically expects when going beyond GR by breaking the SEP, and how they are being constrained by the different PK parameters measured in the Double Pulsar. In this, these theories provide a theory-based approach for putting different GR experiments into context, and allow for a better understanding which structures and symmetries of GR are actually being probed by pulsars, and more specifically the Double Pulsar (see Section~5 in \cite{dam09}).

Before entering the theory-specific discussion, we would like to give a general statement about constraints on deviations from GR in the radiative regime that can come from the precise measurement of the GW emission (see Eq.~(\ref{eq:GR_GW_test})). In alternative theories one often has gravitational dipolar radiation due to a violation of the SEP as a significant contribution to GW damping. In the equations of motion that contribution enters at the 1.5PN level (${\cal O}(\beta_\mathrm{O}^3)$; see e.g.\ \cite{mw13}). If we parametrize a deviation from GR led by dipolar GW damping by $\dot{P}_{\rm b}^{\rm GW} = \dot{P}_{\rm b}^{\rm GW,GR} \left(1 + B_{\rm D} \, \beta_\mathrm{O}^{-2}\right)$, then we find {$B_{\rm D} \lesssim 4 \times 10^{-10}$} (95\% C.L.). This is nearly five orders of magnitude more constraining than the limit from the double NS merger GW170817 \cite{GW170817_GR_test}. Note, however, that the two different limits are coming from different gravity regimes (see the discussion at the end of Section~\ref{subsubsec:test_gw}). How such a limit on $B_{\rm D}$ converts into actual constraints on a given gravity theory depends on the specifics of that theory. In many theories it is expected that due to the similarity of the two masses in the Double Pulsar, the limit is  actually weaker than one would assume directly from the tight limit on $B_{\rm D}$.

A violation of the SEP not only gives rise to dipolar GWs, it generally also results in a temporal modification of the (effective) gravitational constant due to the expansion of the universe (see e.g.\ \cite{nor90}), by this having an addition effect on the orbital period. The orbital period change due to a time-varying gravitational constant is given by $\dot{P}_{\rm b}^{\dot{G}}/P_{\rm b} = -2(\dot{G}/G)\,{\cal F}_{\rm AB}$, where ${\cal F}_{\rm AB}$ accounts for corrections related to the strong gravitational fields of the NSs in the Double Pulsar system \cite{dgt88,nor90,nor93a}\footnote{There are special cases, like Barker's constant-$G$ theory \cite{bar78}, where $\dot{P}_{\rm b}^{\dot{G}}$ cannot be written in this form.}.
For weakly self-gravitating bodies, as in the solar system, ${\cal F}_{\rm AB} \simeq 1$. ${\cal F}_{\rm AB}$ depends on the details of the gravity theory under consideration, as well as on the EoS. While for Jordan-Fierz-Brans-Dicke gravity ${\cal F}_{\rm AB} \approx 0.6$, for some theories ${\cal F}_{\rm AB}$ can be significantly larger than one (cf.\ discussion in \cite{Wex_2014}). If one assumes that $\dot{P}_{\rm b}^{\dot{G}}$ is the only significant non-GR contribution to the orbital period change, then from Eq.~(\ref{eq:GR_GW_test}) one finds that {$|\dot{G}/G| < -0.8(14)\times 10^{-13}\,{\rm yr}^{-1} / {\cal F}_{\rm AB}$}. This limit is about a factor of two better than the currently best pulsar limit \cite{zdw+19}, although the exact factor is theory dependent. In cases where ${\cal F}_{\rm AB}$ is large, this limit can even exceed the best limit obtained in the solar system (i.e.\ $|\dot{G}/G| < 4 \times 10^{-14}\,\mathrm{yr}^{-1}$ \cite{gmg+18}). At this stage, these considerations need to be taken with a grain of salt. As stated above, theories with non-vanishing $\dot{G}$ are expected to also predict dipolar GWs, which in turn leads to an additional non-GR modification of $\dot{P}_{\rm b}$. As a conclusion, in order to perform a $\dot{G}$ test with the Double Pulsar, one has to do a fully consistent analysis within a certain class of gravity theories and combine it with other binary pulsars (similar to the method proposed in \cite{lwj+09}). However, such an analysis goes beyond the scope of this paper and will be presented in a future publication.

\subsection{Damour--Esposito-Far{\`e}se Gravity}
\label{subsec:DEFgrav}

The first alternative to GR which we confront with our Double Pulsar results is the two parameter mono-scalar-tensor gravity $T_1(\alpha_0,\beta_0)$ introduced in \cite{de93} (``{\it DEF gravity}''). This two-parameter class of theories exhibits various effects one typically expects from alternatives to GR, including genuine strong-field effects related to the self-gravity of NSs. Hence, $T_1(\alpha_0,\beta_0)$ can be viewed as an ideal mini-space of well-studied gravity theories in which GR is embedded ($\alpha_0 = \beta_0 = 0$). They are well suited to contrasting the predictions of GR with the predictions of alternatives, and allowing for the comparison of GR tests across different gravity regimes \cite{dam09}.
The constants $\alpha_0$ and $\beta_0$ measure respectively the linear and the quadratic coupling strength between matter and the scalar field. For simplicity we will assume that the potential of the scalar field can be ignored on the typical scales for the Double Pulsar test, while it could still be of importance on Galactic or cosmological scales. We further assume that we can neglect a temporal change in the background scalar field, which amongst others would lead to a long-term variation of the gravitational constant. For sufficiently negative $\beta_0$, DEF gravity shows genuine strong field effects in binary pulsars, related to ``spontaneous scalarization'' \cite{de93,de96}. The presence of the scalar field leads to a modification of the PK parameters in different ways \cite{de92b,de96}.

Already at the Newtonian level, the body-dependent effective scalar couplings, $\alpha_{\rm A}$ and $\alpha_{\rm B}$, enter via the (effective) gravitational constant $G_{\rm AB} = G_\ast(1 + \alpha_{\rm A}\alpha_{\rm B})$, where $G_\ast$ is the bare gravitational constant, linked to the Newton's gravitational constant $G$ (as measured in a Cavendish-type experiment) via $G = G_\ast(1 + \alpha_0^2)$. By this the acceleration of a body depends on the gravitational binding energy, leading to a violation of the SEP \cite{Will:2018}. As a result, one has a strong-field modification of the mass function of the binary system that enters the expression for the Shapiro shape parameter $s$ (cf.\ Eq.~(5.8) in \cite{de96}). 

At the first PN level, one has three body-dependent strong-field counterparts of Eddington's weak-field PPN parameters $\gamma^{\rm PPN}$ and $\beta^{\rm PPN}$ \cite{edd22}, and a periodic modification of the MoI due to a violation of local position invariance. All this leads to genuine strong-field modifications of the corresponding PK parameters (see Appendix~\ref{appdx:pkstg} for details). 

Beyond the first PN level, there are modifications to the conservative dynamics at 2PN \cite{mw13}, which are not (yet) relevant for the test presented here. Hence it is sufficient to use the corresponding GR terms. However, the presence of the dynamical scalar field $\varphi$ considerably modifies the loss of energy and angular momentum due to the emission of scalar GWs. As a result, in addition to the quadrupolar tensor GWs, one has monopolar, dipolar and quadrupolar scalar GWs, entering the equations of motion at the 2.5PN, 1.5PN, and 2.5PN level respectively. These dissipative contributions to the equations of motion result, most importantly, in a modification of $\dot{P}_{\rm b}$ (see Appendix~\ref{appdx:pkstg} for details). The leading term, i.e.\ scalar dipole radiation, is proportional to $(\alpha_{\rm A} - \alpha_{\rm B})^2$, and therefore quite sensitive to the asymmetry in gravitational binding energy in a system\footnote{To leading order one finds in DEF gravity $B_{\rm D} \simeq (5/96)(\alpha_{\rm A} - \alpha_{\rm B})^2[1 + {\cal O}(\alpha_{\rm A},\alpha_{\rm B})][1 + {\cal O}(e^2)]$ (cf.\ \cite{de92b,de98}).}.
Even though the difference in mass between A and B is only 7\%, the small size of the effect is compensated for by the high precision ($1.3\times 10^{-4}$) of the GW emission test (see Section~\ref{subsubsec:test_gw}).
Consequently, the GW damping test provides an important contribution to the constraints on DEF gravity from the Double Pulsar. 

The PK parameters $r$ and $\delta_\theta$ are not used in the tests presented in this section. Due to their large uncertainty they are not relevant for any of those tests. The same is the case for effects related to the scaling parameter $q_{\rm NLO}$.

Finally, we would like to note that the expression for the mass ratio remains unchanged since $R \equiv m_{\rm A}/m_{\rm B} = x_{\rm B}/x + {\cal O}(c^{-4})$ \cite{dam09}.

\begin{figure}[htp]
    \centering
    \includegraphics[width=8.5cm]{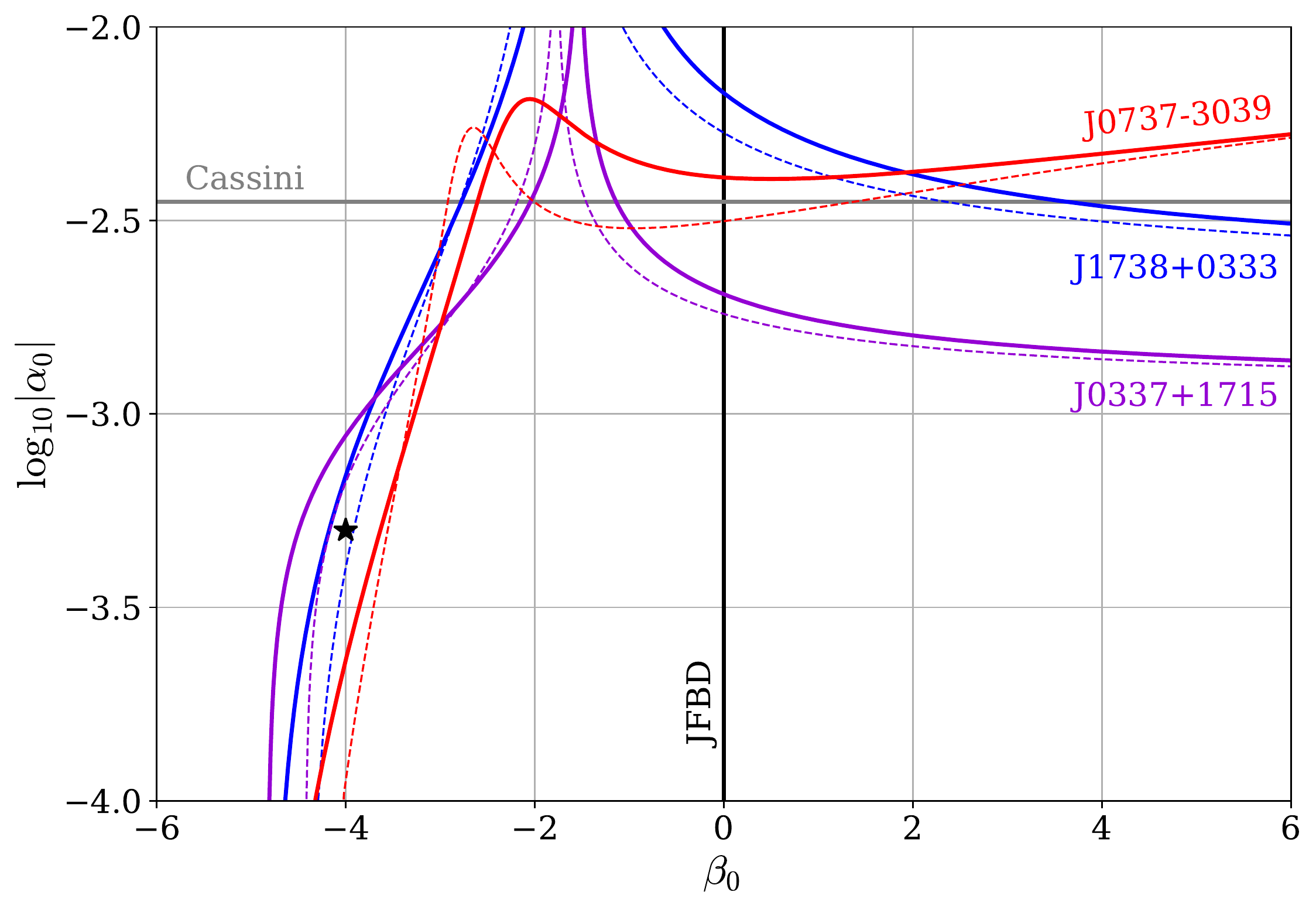}
    \caption{Constraints on the DEF gravity from different experiments: Shapiro delay with the Cassini spacecraft \cite{bit03}, dipolar radiation (J1738+0333) \cite{fwe+12}, gravitational weak equivalence principle (J0337+1715) \cite{vcf+20}, and the Double Pulsar (this paper). Areas above a curve are 
    excluded\footnote{See \cite{de96,de98} for details. We have used $\chi^2 = 4$ as a conservative limit.}.
    Solid lines are for a comparably stiff EoS (MPA1 in \cite{lp01}). The dashed lines are based on a rather soft EoS (WFF1 in \cite{lp01}), to illustrate the EoS dependence of the different limits. GR corresponds to $\alpha_0 = \beta_0 = 0$, and JFBD theory is along the vertical $\beta_0 = 0$ line with Brans-Dicke parameter $\omega_{\rm BD} = (\alpha_0^{-2} - 3)/2$. The black star indicates the parameters used in Figure~\ref{fig:mm-T1}.}
    \label{fig:a0-b0_plane}
\end{figure}

\begin{figure}[htp]
    \centering
    \includegraphics[width=8.5cm]{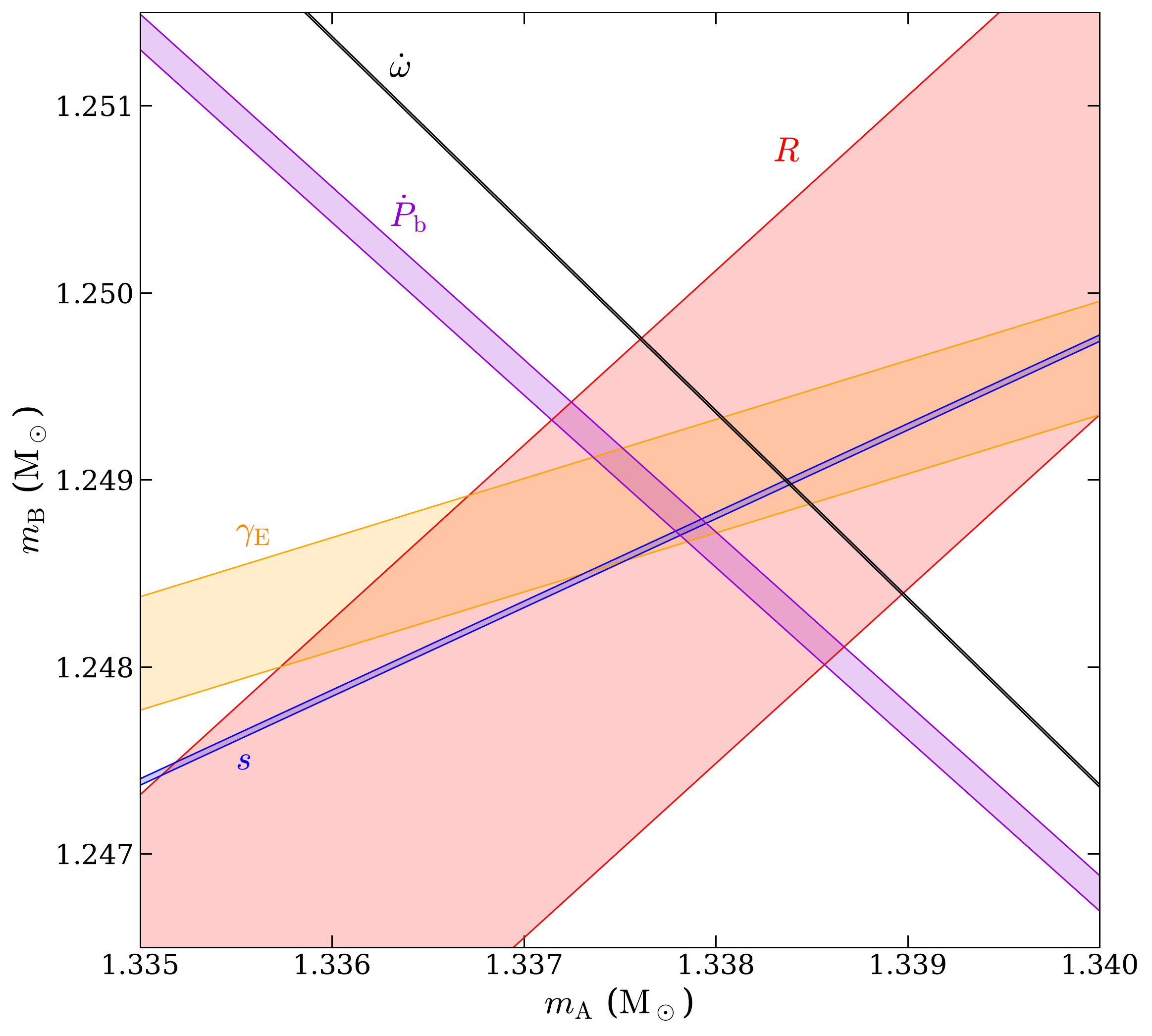}
    \caption{Mass-mass diagram for DEF gravity with $\alpha_0 = 5\times 10^{-4}$, $\beta_0 = -4$. EoS MPA1 has been used to calculate the curves. The curves fail to agree on a common region in the mass-mass plane (see in particular $\dot\omega$ and $\dot{P}_{\rm b}$ curves), meaning that this specific scalar-tensor theory is excluded. The chosen point in the $\alpha_0$-$\beta_0$ plane is not excluded by other experiments (see Figure~\ref{fig:a0-b0_plane}). This figure is merely to illustrate how a specific theory fails the Double Pulsar test. In this case it is due to the additional energy loss from scalar GWs, predominantly the dipolar contribution. The $\dot{P}_{\rm b}$ curve is based on Eq.~(\ref{eq:PbdotGW}). The 2PN and Lense-Thirring contributions to $\dot\omega$ have been calculated assuming GR, since any corrections from the scalar field at this PN level are insignificant.}
    \label{fig:mm-T1}
\end{figure}

Figure~\ref{fig:a0-b0_plane} shows the constraints on DEF gravity from the Double Pulsar, presented as a curve (boundary of allowed region) in the $\alpha_0-\beta_0$ parameter plane. When calculating the limits, we followed the procedure outlined in \cite{de96,de98}. As one can see from Figure~\ref{fig:a0-b0_plane}, the Double Pulsar provides the best limit on DEF gravity for $\beta_0 \lesssim -3$. In Fig.~\ref{fig:mm-T1}, for illustration purposes, we have plotted a mass-mass diagram for a point in the $\alpha_0-\beta_0$ plane that was previous to this paper not excluded: $\alpha_0 = 5\times 10^{-4}$, $\beta_0 = -4$ (marked by `$\ast$' in Figure~\ref{fig:a0-b0_plane}). As one can see for this specific example, $T_1(5\times 10^{-4},-4)$ is falsified by the Double Pulsar, in particular by the GW test.

\subsection{Bekenstein's TeVeS}
\label{subsec:TeVeS}

As a second alternative to GR, we discuss Bekenstein's TeVeS \cite{bek04}. In \cite{fwe+12} TeVeS has been extended to a TeVeS-like class of gravity theories with two coupling parameters $\alpha_0$ and $\beta_0$, similar to the class of standard scalar–tensor theories discussed in Section~\ref{subsec:DEFgrav}. Here we will only discuss Bekenstein's original TeVeS, which corresponds to $\beta_0 = 0$, since this is the case where the Double Pulsar provides the most important contribution. Although Bekenstein's TeVeS has recently been falsified by the multi-messenger observation of the LIGO/Virgo merger event GW170817 \cite{bdkw18}, we still use it here as an example of a MONDian gravity theory that passes solar-system tests but can be excluded by a binary pulsar experiment. At the same time this is a significant update on the result presented in \cite{fwe+12}.

The crucial difference between DEF gravity in Section~\ref{subsec:DEFgrav} and TeVeS is the specific form of the physical metric, which ensures that $\gamma^{\rm PPN} = \beta^{\rm PPN} = 1$, making TeVeS indistinguishable from GR in the weak-field slow-motion regime. Therefore, by design TeVeS passes solar-system tests, like the one from the Cassini spacecraft \cite{bit03}. Interestingly, the special form of the physical metric also leads to the fact that $\gamma_{\rm AB} = 1$ in TeVeS-like gravity theories \cite{fwe+12}, different from Eq.~(\ref{eq:gammaAB})

\begin{figure}[H]
    \centering
    \includegraphics[width=8.5cm]{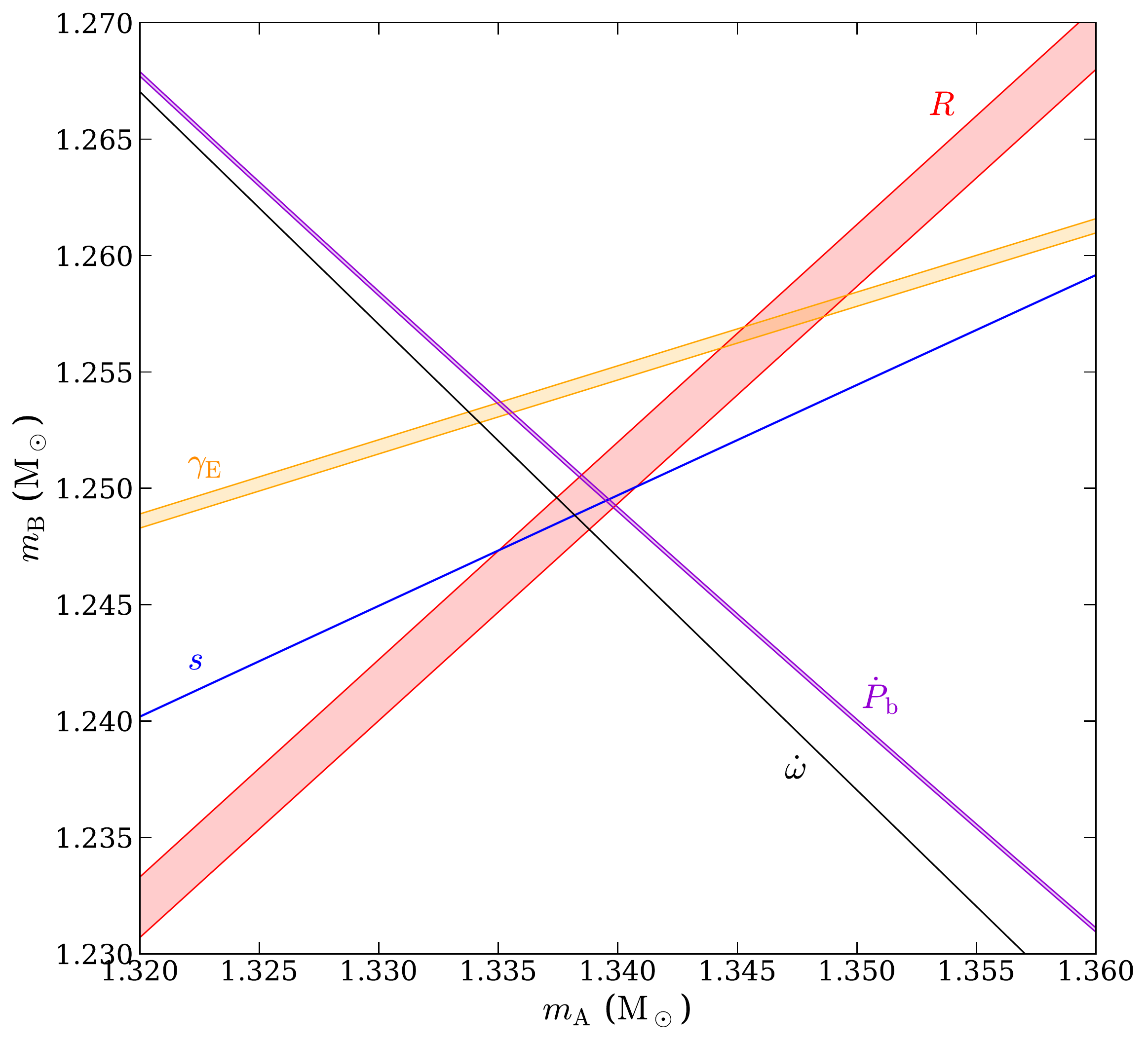}
    \caption{Mass-mass diagram for Bekenstein's TeVeS with  a ``natural'' transition from the Newtonian to the MONDian regime ($\alpha_0 = 0.04$) \cite{be07}. Such a theory is obviously in contradiction with the Double Pulsar observations. The $\dot{P}_{\rm b}$ curve is based on Eq.~(\ref{eq:PbdotGW}). The 2PN and Lense-Thirring contributions to $\dot\omega$ have been calculated assuming GR, since any TeVeS related corrections for them are negligible in this test.}
    \label{fig:mm-TeVeS}
\end{figure}

As discussed in \cite{fwe+12}, the predictions for the Einstein delay amplitude, $\gamma_{\rm E}$, and the GW damping, $\dot{P}_{\rm b}$, keep the same form as in standard scalar–tensor theories. However, $\alpha_{\rm A} = \alpha_{\rm B} = \alpha_0$, and therefore there is no dipolar radiation. As a consequence dipolar radiation tests like in \cite{fwe+12} do not provide any constraints on Bekenstein's TeVeS. The same is the case for the universality of free fall test with the pulsar in the stellar triple \cite{agh+18,vcf+20}. The Double Pulsar, on the other hand, is sensitive to the modification of the MoI due to the periodic variation in the local gravitational constant (modification of $\gamma_{\rm E}$) related to the scalar field of TeVeS, and has a sufficiently precise GW test that is sensitive to modifications of the quadrupolar GW damping due to the presence of the dynamical scalar field. This is obvious from Figure~\ref{fig:mm-TeVeS}, which gives the mass-mass diagram for Bekenstein's TeVeS with a coupling parameter $\alpha_0$ that guarantees a natural transition from the Newtonian to the MONDian regime (see discussion in \cite{be07}). The deviations in $\gamma_{\rm E}$ and $\dot{P}_{\rm b}$ clearly falsify that theory. The current Double Pulsar observations require $|\alpha_0| < 9\times 10^{-3}$ (95\% C.L.) which requires a highly unnatural behaviour of the scalar-field contribution to the gravitational acceleration  (cf.\ Figure~8 in \cite{fwe+12} and Figure~3 in \cite{be07}). 

TeVeS-like theories that modify the conformal factor between the Einstein and the Jordan frame, as introduced in \cite{fwe+12}, are similarly excluded. For $\beta_0$ values significantly different from zero, the constraints do primarily come from dipolar radiation and universality of free fall tests, e.g.\ from limits in \cite{fwe+12,vcf+20}.

As a final comment, we would like to emphasize that like in \cite{fwe+12} we have neglected the contributions of the vector field to the dynamics and the gravitational radiation. We consider this as a conservative approach, since a dynamical vector field is expected to increase the GW damping by extracting additional energy from the orbital dynamics. Furthermore, modifications of TeVeS that only modify the vector part and leave the scalar sector of the theory unchanged, should therefore generally be excluded by this test as well (see discussion in \cite{be07} on the role of the scalar field to reproduce the MONDian dynamics).

\section{Astronomical Implications}
\label{sec:implications}

In this section, we complete our analysis by revisiting several important properties of the Double Pulsar system. We first discuss results relating to the system distance, comparing our best estimate with the DM-based estimates based on the two principal Galactic electron-density models. The observed discrepancies inform how the model parameters of ISM structures in the path to the system should be modified. We next use the observed scintillation properties and DM variations over a 13-yr data span (Sec.~\ref{subsec:astrom_timing}) to investigate the structure of the ISM over six orders of magnitude in fluctuation scale. Finally, we consider the implications of the observed system proper motion and implied transverse velocity on formation models for the system and summarize the evidence for a prograde rotation sense for the A pulsar (Sec.~\ref{subsubsec:test_Shapiro}).

\subsection{The distance to the Double Pulsar}
\label{subsec:impl_astrometry}

The astrometric results derived from timing (Section~\ref{subsec:astrom_timing})
and from interferometric data (Section~\ref{subsec:vlbi})  can be compared against each other and with earlier VLBI observations using the Australian Long Baseline Array  \cite{dbt09}.  A detailed comparison between the two VLBI measurements is presented in Appendix~\ref{appdx:vlbi}, in which we demonstrate that our new VLBA results are more reliable than the previous Long Baseline Array results, in part due to a previously unappreciated source of systematic error (refractive image wander). Accordingly, we focus on the VLBA results presented in this work.

The estimated parallax from the VLBA measurements is $\pi_{\rm v} = 1.30^{+0.13}_{-0.11}$\,mas (Section~\ref{subsec:vlbi}). This contrasts with the value estimated from the pulsar timing of $\pi_{\rm t} = 2.15\pm 0.48$\,mas (Section~\ref{subsec:astrom_timing}). The implied pulsar distances are, respectively, $770\pm 70$\,pc and $465^{+134}_{-85}$\,pc. As described in Appendix~\ref{appdx:vlbi}, we have adopted a weighted mean of these two distances, $735\pm60$\,pc, as our best estimate of the pulsar distance. Note that none of these distance estimates change beyond their uncertainties if one applies the Lutz-Kelker correction (cf., Refs.~\cite{vlm10,vwc+12}). Refractive image wander also potentially has an effect on the timing-based astrometry. We show in Appendix~\ref{appdx:image_wander} that this effect is negligible, changing the values of the astrometric parameters by $\lesssim$ $0.2\,\sigma$. 

We now consider the impact of the pulsar distance on other observables to see if they can assist in determining the most likely parallax value. 

The estimated distance for the pulsar based on the NE2001 Galactic electron-density model \cite{cl02new} is 516\,pc and, based on the YMW16 model \cite{ymw17}, 1105\,pc. We note, however, that the derivation of the YMW16 electron-density model included the earlier VLBI-based distance \cite{dbt09}  (corrected for the Lutz-Kelker bias) of 1100\,pc in its set of independently measured distances used to calibrate the model. Removing the Double Pulsar from the list of independent distances and re-determining the parameters of the YMW16 model had a surprisingly large effect on the estimated distance, changing it from 1105\,pc to 463\,pc (Jumei Yao, private communication). Further investigation showed that the reason for this large change was that, in the modified analysis, the radius of the model component representing the Gum Nebula changed from 125\,pc to 128\,pc. This had the effect of placing the LoS to the Double Pulsar just within the Gum Nebula, which is centered at about 450\,pc from the Sun  \cite{pgs+15}, rather than just outside it. The resulting additional electron column density was sufficient to place the pulsar within the Nebula at 463\,pc. The original YMW16 distance for the Double Pulsar, 1105\,pc, is very close to the adopted independent distance of 1100\,pc. This, and the large change when this independent distance was removed from the model, highlight the fact that the number of independent distances used to build the original YMW16 model (viz., 189) was relatively small. In the intervening years, many more independent estimates of pulsar distances have been published, most notably the VLBI pulsar parallax study PSR$\pi$ \cite{dgb+19}. Consequently, future generations of the YMW16 model will be less susceptible to this problem.

Of course, the real Gum Nebula does not have the relatively sharp and well-defined edge assumed by the YMW16 model. It is a complex region of ionized gas that is believed to be a greatly expanded remnant of a supernova explosion, a fossil H II region, a wind-blown bubble, or some combination of these (see Ref.~\cite{pgs+15} for an H$\alpha$ image of the nebula and a summary of its properties). The Vela supernova remnant is superimposed on the Nebula and probably lies within it, but is not the source of the bulk of the ionization. The Gum Nebula is roughly circular in shape, with an angular radius of about $23^\circ$, corresponding to a physical radius of about 190\,pc. It has large extension to the west (i.e., lower right ascension or Galactic longitude) which covers the projected direction of the Double Pulsar. The identification of this extended nebulosity as the location of the electron density fluctuations responsible for the DM variations and scintillation of the Double Pulsar are discussed in Section~\ref{subsec:impl_ism} below. 

\medskip

\subsection{ISM properties}
\label{subsec:impl_ism}

In addition to the {\em average} DM value used above, we can also make use of the observed {\em variations} of the DM shown in Figure~\ref{fig:dmcurve}, in order to infer properties of the intervening ISM and to compare these with expectations. These variations are consistent with our LoS moving across electron-density enhancements and deficits in the ISM (see, e.g., Ref. \cite{lam2015}). These density fluctuations are part of the same plasma turbulence that causes angular scattering and intensity scintillation. The diffractive intensity scintillations of pulsar A have been thoroughly analyzed \cite{rcn+14} and our observations do not have sufficient frequency resolution to improve on this work. In the following analysis, diffractive parameters are taken from Ref.~\cite{rcn+14}. They show that the scattering screen is located at about 30\% of the distance from the Earth to the pulsar, or about 220\,pc from the Earth for the adopted pulsar distance of 735~pc. If the extended nebulosity on the western side is somewhat closer to the Earth than the Gum Nebula itself, we could identify the scattering screen with this nebulosity.

We also note that the rotation measure (RM) of the Double Pulsar is $+120.8\pm0.2$\,rad\,m$^{-2}$ \cite{ksv+21}, which is consistent both with the RM of other pulsars located near the Gum Nebula (in projection) and with the RM of extra-galactic sources located behind the western nebulosity near the Double Pulsar \cite{pgs+15}. Both of these are consistent with the Double Pulsar being located behind the western extension and much of the Faraday rotation occurring in this region.

Structure functions (SFs) are useful here because the diffractive scattering can be calculated directly from the SF, and stochastic processes with a power-law spectrum have a power-law SF. The SF for DM variations is defined by  
\begin{equation}
    D_{\rm DM}(\tau) = \langle[\mbox{DM}(t+\tau) - \mbox{DM}(t)]^2\rangle,
    \label{eq:strucdm}
\end{equation} 
where $\tau$ is the lag between DM samples (see, e.g.,~\cite{yhc+07}).

To obtain the most reliable estimates of the DM variations and the corresponding SF, we restrict our analysis to the interval MJD 53500 to 58300 which has better sampling in both time and frequency compared to earlier and later times (see Figure~\ref{fig:obsToAs}). We used the results of the Monte Carlo analysis described in Section~\ref{subsec:astrom_timing} to estimate the DM variations at 24-d intervals. The resulting DM-variation curve is fully consistent with that shown in Figure~\ref{fig:dmcurve} for the corresponding interval.

\begin{figure}[H]
    \centering
    \includegraphics[width=8.5cm]{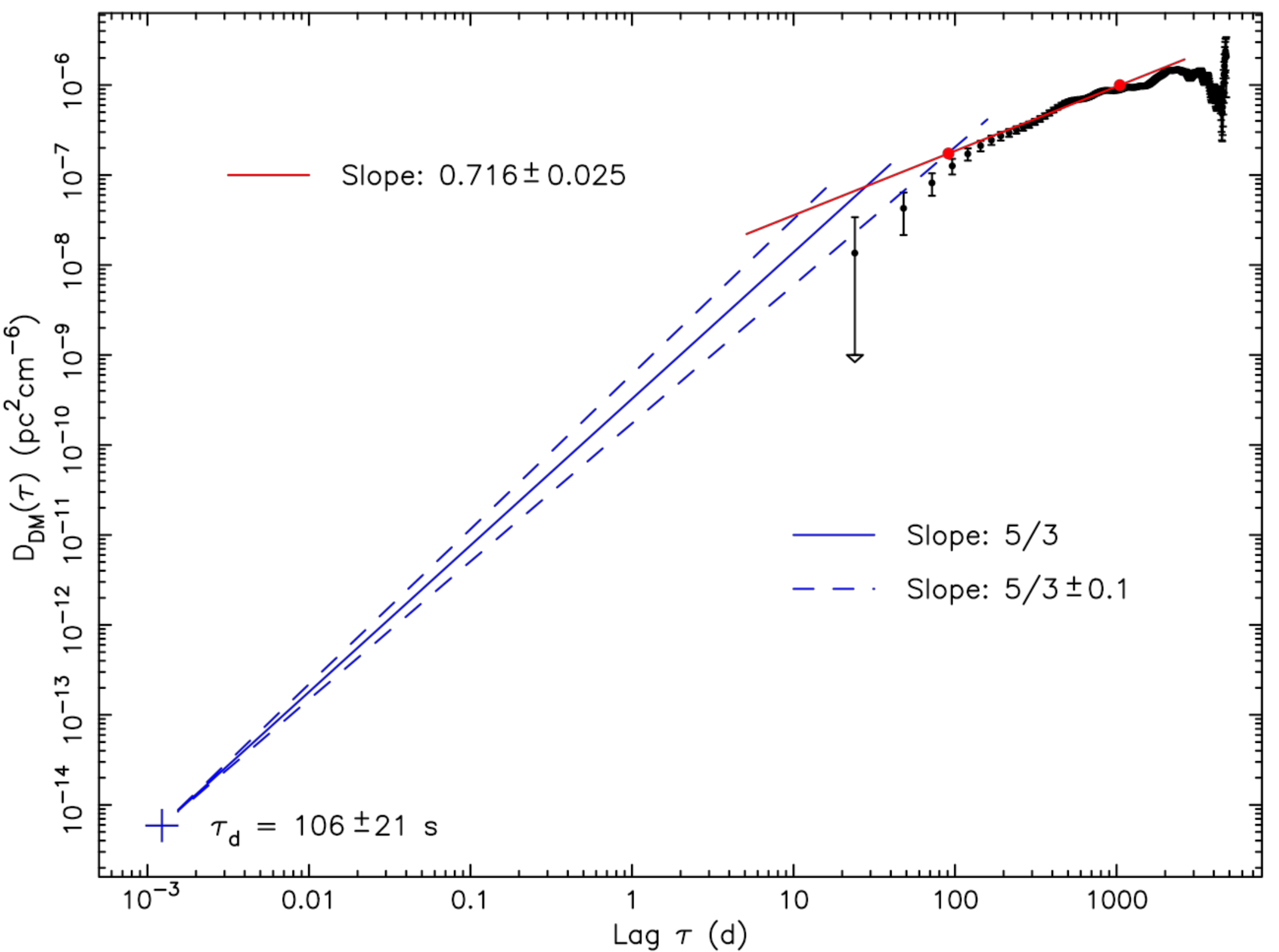}
    \caption{Structure function of DM from MJD 53500 to 58300 with 24-d sampling. The point marked with a $+$ is derived from diffractive time scale  measurements at 820 MHz \cite{rcn+14}. The blue solid line shows the extrapolation from the diffractive scale up to the measured lags assuming a Kolmogorov electron-density fluctuation spectrum. The dashed lines indicate $\pm 0.1$ in the spectral exponent. The red line shows the result of a weighted fit of a power law to the measured structure function for lags between 96\,d and 1008\,d, as indicated by the red dots.}
    \label{fig:extd_struc_fn}
\end{figure}

The structure function estimated directly from Eq.~(\ref{eq:strucdm}) is shown at the upper right of Figure~\ref{fig:extd_struc_fn}. At lags between 100\,d and 1000\,d, the observed SF is approximately power-law with an exponent much smaller than the Kolmogorov value of 5/3.\footnote{Since the DM is the column integral of the electron density along the line-of-sight ($z$) axis, its 2D spatial spectrum is the same as that of the 3D density with the wave-number in $z$ set to zero. Therefore, the 2D spectral exponent of DM ($\beta$) is the same as that of the 3D electron-density fluctuations, i.e., $-11/3$ for a Kolmogorov process. The corresponding 2D structure function has a spectral exponent $\alpha = -(\beta+2) = +5/3$ for Kolmogorov.}  A linear fit to the region between 96\,d and 1008\,d is shown as a red line on the plot. The SF is less useful for lags greater than about 2500\,d (i.e., half the data span), because there are few independent samples to be averaged. 

The LoS velocity (excluding the binary motion) is about 37\,km/s \cite{rcn+14}, so the time scales that are easily measurable in DM($t$), 100 -- 1000\,d, correspond to spatial scales of approximately 2.0 to 20\,AU.\footnote{One astronomical unit (AU) is approximately $1.496\times 10^8$~km.} The intensity scintillations, unlike DM($t$), have a characteristic spatial scale $s_0$ which is very much smaller at $\sim$3700\,km. 
The spatial ACF of the scintillations is given by 
${\rm ACF}(s) = \exp(-D_\phi(s))$ where $D_\phi (s)$ is the structure function of the phase shift $\phi$ caused by the scattering plasma. The phase delay is the negative of the group delay, so we can obtain the phase directly from DM by computing the group delay and multiplying by $2\pi\nu$. A measurement of $s_0$ at $\nu$ provides $D_{\phi}(s_0) = 1$ which defines $D_{\rm DM}(s_0)$. It should be noted that the time scale of the scintillations is extremely variable because the binary velocity is much higher than the mean LoS velocity. The analysis of Ref.~\cite{rcn+14} fits the binary velocity to extract $s_0$ which is not time-variable.

In this way, the observed diffractive intensity scintillations can be used to define the point in Figure~\ref{fig:extd_struc_fn} marked with a $+$ near lag 110\,s. A blue line with the Kolmogorov slope of 5/3 is drawn from the diffractive point to meet the red line at a lag of $\sim$\,30 days. The dashed  lines in Figure~\ref{fig:extd_struc_fn} illustrate the effect of an uncertainty in the slope of the extrapolation from the diffractive point, indicating a likely range for the intersection lag of between 10 and 100~days, or a spatial scale of 0.2 to 2.0~AU using the mean ISM velocity given by Ref.~\cite{rcn+14}. 

The location of the spectral break is important because it can be used to estimate the thickness of the scattering region which is causing both the DM($t$) fluctuations and the intensity scintillations. Accordingly, we have simulated 100 realizations of a pure power-law process with an SF exponent of +0.5 and applied the convolutions by a 100-d triangle implicit in the {\sc Tempo2} analysis and the 24-d binning of the Monte Carlo results. Details of the simulation methods are described in Appendix~\ref{appdx:simulations}.

\begin{figure}[H]
     \centering
     \includegraphics[width=8.5cm]{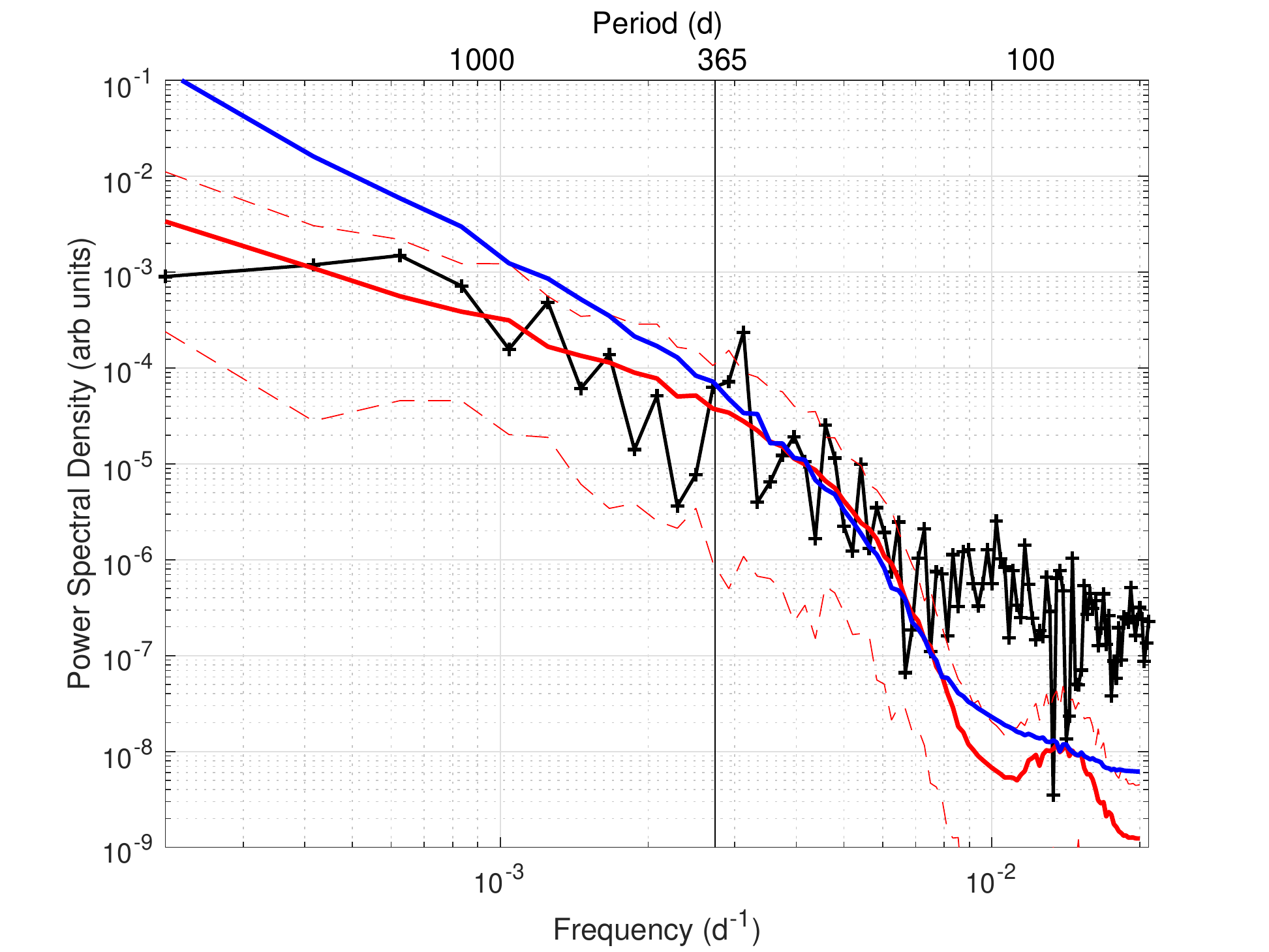}
     \caption{Power spectral density for the DM variations used to form the SF in Figure~\ref{fig:extd_struc_fn} (segmented black line). The blue and red solid lines are means of 100 simulations of the 1D PSD of DM variations given by {\sc Tempo2} for fluctuation spectral exponents of $-8/3$ (Kolmogorov) in blue and $-1.5$ in red. The dashed red lines give the 90\% range of the $-1.5$ simulations. The vertical black line denotes the annual frequency.}\label{fig:dmvar_psd}
 \end{figure}

We first adjusted the amplitude and spectral exponent of the simulated DM($t$) so its power spectral density (PSD) matched that of the observations.\footnote{The fluctuations in DM($t$) are those of a 1D cut across the 2D spatial distribution. Therefore, the 1D PSD of DM($t$) is the 2D spatial spectrum of DM (scaled by the velocity) with the perpendicular wave-number integrated out. This reduction to 1D reduces the Kolmogorov spectral exponent from $-11/3$ to $-8/3$. } The match is much clearer in the PSD than in the SF because the PSD estimation errors are statistically independent and those of the SF are highly correlated.
Figure~\ref{fig:dmvar_psd} shows the PSD of the observed DM variations as a segmented black line. The mean PSD of the simulations is over-plotted with the 90\% confidence limits and both the level and the spectral exponent were adjusted so the observed PSD fits within these confidence limits. A Kolmogorov PSD simulation is also shown to make it clear that the observations cannot be fitted with a Kolmogorov spectrum. The flatter spectrum with exponent $-1.5$ agrees well with the observations up to frequencies of $\sim 7\times 10^{-3}$\;d$^{-1}$. At higher frequencies the observed PSD is dominated by white noise which is not included in the simulations. 

The observed PSD also shows a statistically significant peak at a frequency close to the annual frequency. This peak arises from the annual motion of the Earth moving the LoS to the pulsar across transverse AU-scale gradients in the integrated screen density. For further details on both the effect of the {\sc Tempo2} convolutions and the annual feature in the PSD, see Appendix~\ref{appdx:simulations}.

\begin{figure}[H]
    \centering
    \includegraphics[width=8.5cm]{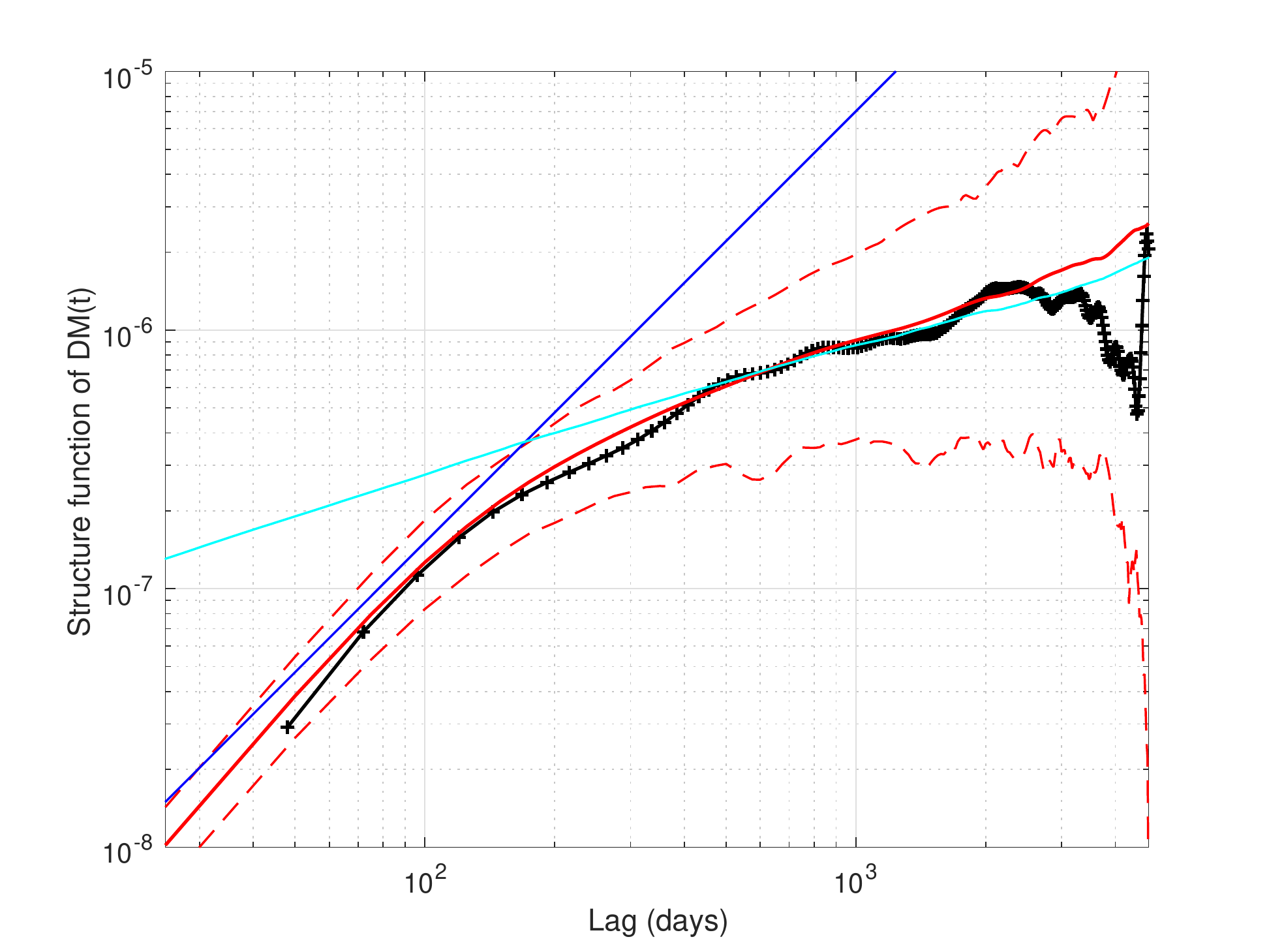}
    \caption{Comparison of the observed structure function for DM variations with white noise removed (in black) with simulations for an input SF exponent of +0.5. The solid red line is the mean of 100 simulations including the effect of the time-domain convolutions and the red dashed lines give the 90\% range. The cyan line is the mean of the simulations before the time-domain convolutions. The blue line is the Kolmogorov  extrapolation (SF exponent 5/3) from the diffractive scale.}
    \label{fig:sf_simulations}
\end{figure}

Figure~\ref{fig:sf_simulations} shows the measured SF from Figure~\ref{fig:extd_struc_fn} in black on an expanded scale. The structure functions of the best-fit simulations with a spectral exponent of $-$1.5 were calculated and the mean and 90\% confidence limits are shown in red. The theoretical SF simulated before convolutions required in the data analysis is indicated by the cyan line and the blue line is the extrapolation from the diffractive time scale, also shown in blue in Figure~\ref{fig:extd_struc_fn}.

The steepening of the SF toward lags shortward of a few hundred days results from the time-domain convolutions. However, there must be a true break in the SF slope from a value close to the Kolmogorov 5/3 at short lags to a flatter slope at lags greater than a few hundred days. The intersection of the blue and cyan lines at a lag of $\sim$170\,d, which corresponds to a linear scale of about 3\,AU, probably gives the best estimate of the frequency of spectral break in the ISM fluctuations.

A break in the PSD from a Kolmogorov spectral exponent of $-$8/3 to an exponent of $-$5/3 occurs naturally in turbulence in a thin layer and the transition frequency is directly related to the thickness of the layer. As discussed by \citet{lay97}, the inverse of the transition frequency for the corresponding PSD corresponds to twice the screen thickness. To relate the SF transition lag and PSD transition frequency we simulated our observed DM($t$) 1000 times with an SF transition at 170\,d and computed the average PSD, which we found had a transition scale of 1/800\,d. Accordingly, the screen depth must correspond to a lag of about 400\,d or a spatial scale of about 8\,AU. The mean electron density integrated through the scattering layer must exceed its rms fluctuation since it is positive definite and the distribution is roughly symmetric. The rms of DM($t$) is about 0.001\,pc\,cm$^{-3}$ (Figure~\ref{fig:dmcurve}) and therefore a lower limit on the mean density in the scattering region is about 25\,cm$^{-3}$. 

As mentioned in Section~\ref{subsec:impl_astrometry} above, the scattering screen can be identified with the nebulosity to the west of the Gum Nebula, which is probably associated with the Nebula but somewhat closer to the Earth. From the observed H$\alpha$ intensity, \citet{pgs+15}  infer emission measures in the fainter parts of the Nebula that correspond to a mean electron density of about 1.5\,cm$^{-3}$. This estimate is appropriate for the nebulosity in front of J0737$-$3039A/B. Recognizing that a filament with a transverse scale of even 10000\,AU would be invisible on the H$\alpha$ image (which has a resolution of 6\,arcmin), an over-density by a factor of 20 or even more in the scattering screen of depth 8\,AU is quite plausible. We therefore conclude that the western nebulosity associated with the Gum Nebula is a likely location for the scattering screen. This is consistent with J0737$-$3039A/B being located at our preferred distance of 735\,pc.

The clear flattening of the spectral exponent at a scale of ~3\,AU is very unusual in ISM observations which usually follow the Kolmogorov exponent. Of the pulsars observed by pulsar timing arrays, which provide DM($t$) with the required precision, only PSR J1713+0747 shows similar behavior. It should be noted that the diffractive scintillation of the Double Pulsar is entirely normal for pulsars with similar DM. Only the flattening of the exponent due to the unusually thin scattering region is abnormal.

\subsection{System velocity \& geometry}
\label{sec:DP_formation}

Using the distance adopted from the weighted mean probability distribution in Figure~\ref{fig:pxpdfs}, i.e.\ $735\pm 60$\,pc, and taking the parameters of Ref.~\cite{McMillan:2017}, in particular, the solar distance to the Galactic mid-plane, $z_{\rm SSB} = +14$\,pc,\footnote{This value agrees well with a pulsar-based estimate \cite{ymw17b} There is some spread in measured values for $z_{\rm SSB}$ (see e.g.\ Table~1 in \cite{ymw17b}), but this has little impact on our conclusions.} we find that the Double Pulsar is about 43\,pc below the Galactic plane.
 
The proper motion of the Double Pulsar is extremely well constrained. Its total value, {$3.304\pm 0.033$\,mas\,yr$^{-1}$}, converts into a transverse velocity with respect to the SSB of just {$11.5\pm1.0\,{\rm km\,s^{-1}}$}, where the uncertainty is dominated by that of the distance measurement. The components in Galactic longitude and latitude are $v_l = -10.7\pm 0.9\,{\rm km\,s^{-1}}$ and $v_b = -4.3\pm 0.4\,{\rm km\,s^{-1}}$, respectively.

\begin{figure}[ht]
    \centering
    \includegraphics[width=8.5cm]{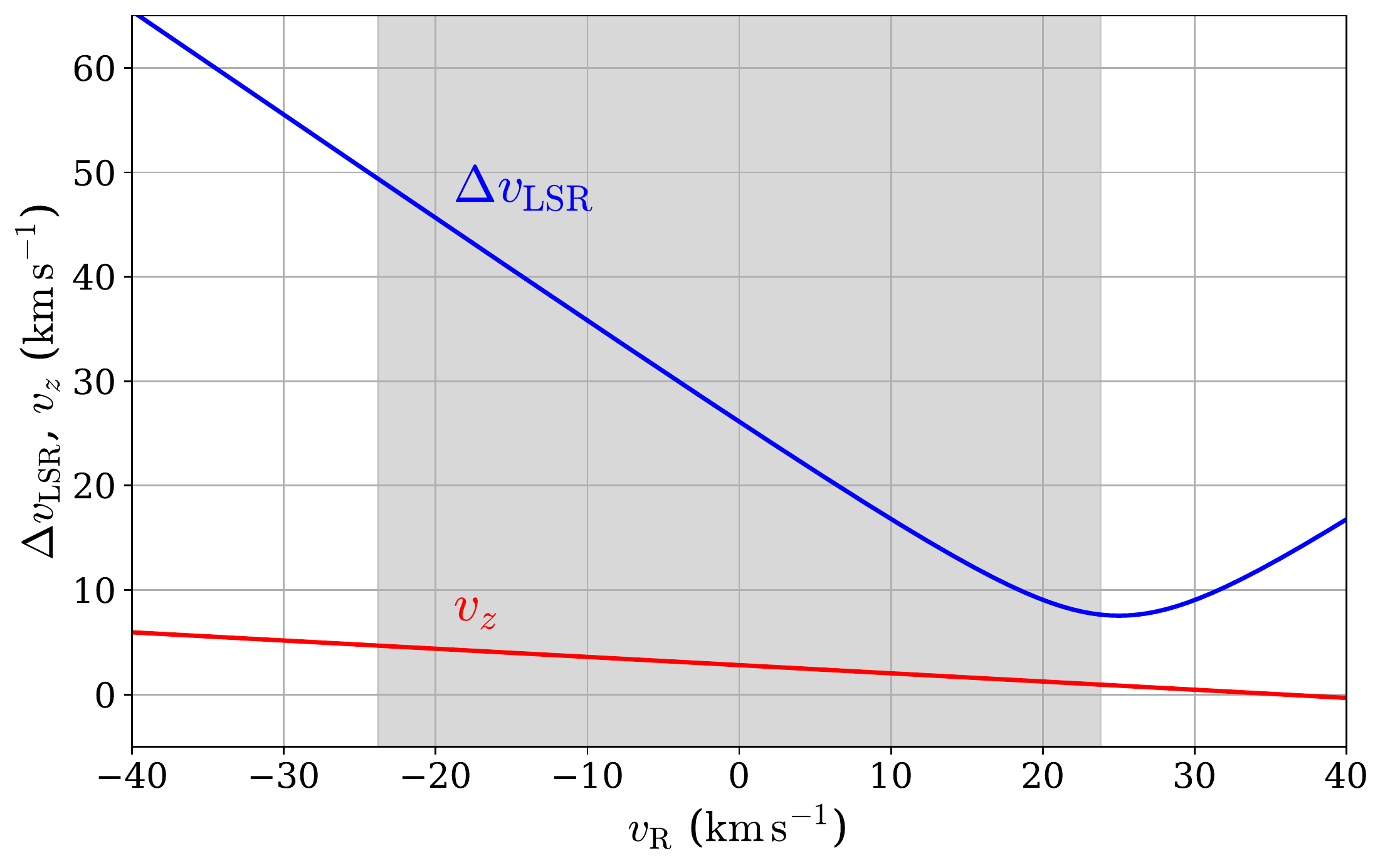}
    \caption{Velocity of the Double Pulsar with respect to its standard of rest ($\Delta v_{\rm LSR}$; blue) and its velocity perpendicular to the Galactic plane  ($v_z$; red), as function of the unknown radial  $v_{\rm R}$. The grey area shows the 90\% probability range for the radial velocity, assuming an (a priori) uniform probability distribution for the direction of the 3D velocity vector with respect to the SSB (cf.\ \cite{std+06}). Calculations are based on the parameters and Galactic potential in \cite{McMillan:2017}, including the components for the peculiar velocity of the Sun $(U_\odot,V_\odot,W_\odot) = (8.6, 13.9, 7.1)\,{\rm km\,s^{-1}}$.}
    \label{fig:Gal_motion}
\end{figure}

The radial velocity $v_{\rm R}$ with respect to the SSB is unknown, but for reasonable values for $v_{\rm R}$, the Double Pulsar has a peculiar velocity of {$\lesssim 50$\,km\,s$^{-1}$} with respect to its standard of rest (see Figure~\ref{fig:Gal_motion}). In particular, the vertical component of the full Galactic velocity $v_z$ is small ({$\lesssim 5$\,km\,s$^{-1}$}), meaning that the Double Pulsar moves practically along the Galactic plane, information that can be used to constrain possible birth places of the Double Pulsar system.

The small peculiar velocity of the Double Pulsar is consistent with the suggested low-kick supernova explosion for the birth of B (e.g.~\cite{std+06,tkf+17}), although this interpretation has been challenged \cite{wkf+06}. A low-kick scenario is, however, also supported by the small misalignment angle, $\delta_{\rm A},$ between A's spin  and the total angular momentum vector, which is based on the lack of profile changes in A \cite{mkp+05,fsk+13}. Visible secular changes in the pulse profile of A are still lacking, even with the doubling of the observation data span, leading to even tighter constraints that will be presented elsewhere. 

The previous argument that the prograde solution, $\delta_{\rm A}<3.2^\circ$, is more likely than the retrograde solution ($176.5^\circ < \delta_{\rm A}< 180^\circ$) was based on the justifiable notion that a large and fortuitous kick would be required for the observed alignment \cite{fsk+13}. Here, we have used our measurement of the relativistic effect of aberrational light-bending to show unequivocally that A is rotating prograde (see Section~\ref{subsubsec:test_Shapiro} for details), supporting the earlier findings of \citet{pmk+18} using a fully independent method based on the emission properties of pulsar B. We note that B is also rotating prograde \cite{ndk+20} but with a significant misalignment angle of $\delta_{\rm B} \simeq 50^\circ$ \cite{bkk+08,ndk+20}. 

We conclude that our results are fully consistent with theoretical considerations for the formation of the Double Pulsar system, as well as with the observed stability of the mean pulse profile.

\section{Prospects}
\label{sec:prospects}

The results presented in the previous sections verify our past expectations on the development of the precision in the timing parameters (cf.~\cite{kw09}). This indicates that not only has the instrumentation improved as expected, but also that possible, as yet undetected, effects have not yet impacted on our timing observations or their interpretation. In particular, this is true for relativistic spin-precession, which could lead to a changing pulse profile, hence making the timing more difficult and potentially less accurate (see e.g.~\cite{mks+10,vbv+19}). In fact, spin precession has rendered B temporarily undetectable \cite{burgay05,pmk+10}, so that the precision of the (direct) mass ratio measurement \cite{lbk+04,ksm+06}, which requires timing of both A and B, cannot easily be improved. We refer to the recent work by Noutsos et al.~\cite{ndk+20} for a detailed study of possible future improvements in this additional constraint. 

The results presented here are not affected by the disappearance of B. The contrasting lack of any profile changes in A further strengthens the notion that the spin direction of A is closely aligned with the total angular momentum vector of the system (see Section~\ref{sec:DP_formation}). In this case, A will continue to be visible and available for precision timing for the foreseeable future, enabling improvements on all parameters and effects studied here. This includes the new effects presented in this work, which will become even more relevant in the future. For instance, in the future the system's acceleration relative to the SSB will become a crucial parameter, while we now also have to take into account the relativistic mass loss due to the pulsar spin-down, and the NS structure. This will allow us eventually to convert our current constraints on the MoI and the NS radius into concrete measurements. Corresponding detailed computations and simulations are presented by Hu et al. \cite{hkw+20} who show that an MoI measurement with 11\% accuracy by 2030 is likely. Combination with results from other sources, such as GW emission from NS-NS mergers, to determine the EoS sufficiently well, would allow for a 7\% test of Lense-Thirring precession, or alternatively provide a $3\sigma$-measurement of the NLO GW damping in GR. 

Obviously, future studies of the Double Pulsar will benefit greatly from more sensitive telescopes. For example, with observations such as those with MeerKAT \cite{ksv+21,hkw+20}, we can study possible profile changes of A due to signal deflection at B near superior conjunction. Moreover, the much improved sensitivity should also result in a better timing performance during the eclipse in general, allowing us to probe this important part of the Shapiro delay curve in more detail for improved NLO tests. Detailed eclipse studies will allow us to track the rotations of B even without seeing its radio signal, also helping to improve the mass-ratio measurement \cite{bkk+08,ksv+21}. 
Finally, continued studies (timing, eclipse and profile studies) of A will improve the precision of all relativistic effects

Future studies will also include a Bayesian-based analysis of the timing data. Currently, the number of ToAs to be studied typically overwhelms the available algorithms and computing power. Optimization work, including studies to deploy smoothly-varying, frequency-dependent templates \cite{ldc+14,pen19} is underway, with initial results on a subset of data being fully consistent with the results shown here.

In addition to timing observations, information obtained via different routes will further aid the exploitation of the system. Firstly, further VLBA astrometric imaging observations can reduce the current uncertainty of the VLBI parallax and hence of the distance.  A factor-of-two reduction in the VLBI parallax uncertainty is within reach of an extended campaign, which would lead to a distance precision of 5\%. Secondly, as mentioned in Section~\ref{sec:timpar}, scintillation measurements can result in an independent measurements of the orbital inclination angle. This is one of the goals of ongoing MeerKAT observations \cite{ksv+21}.

\section{Summary \& Conclusions}
\label{sec:summary}

We have presented results from a timing campaign for the Double Pulsar with a duration of more than 16 years, using data from six different telescopes. 
In order to analyse the data adequately, we have introduced a number of new methods. To facilitate analyses of the system astrometric parameters and DM variations, we formed a data set with 4-min integration time per ToA, which provided the widest possible coverage in terms of time and observing frequency. Analysis of the binary parameters, including the important PK parameters, requires a second data set with ToAs based on 30-s sub-integrations in order to optimally resolve the fast compact orbit of PSR J0737$-$3039A. The analysis of this latter data set required also the development of a new timing model and an improved implementation in {\sc{Tempo}}. This takes into account higher-order effects, including velocity dependent terms in the Shapiro delay. With these efforts, our results not only improve on precision of the previously measured parameters \cite{ksm+06}, but also reveal newly measured effects.

Using a Monte Carlo analysis, we obtained a probability distribution for the weighted mean of the VLBI and pulse timing annual parallaxes to obtain our best estimate of $1.36^{+0.12}_{-0.10}$ mas for the parallax and $735 \pm 60$ pc for the distance of the Double Pulsar. We emphasize again that, because of the fortunate arrangement of the secular (Shklovskii) and the Galactic acceleration being of opposite sign for the Double Pulsar, the impact of distance uncertainties is minimal for our current GW emission test \cite{kwkl18}. This will eventually change, but continued VLBI and timing observations should converge onto a higher precision distance well before any limitations are reached for the GW emission test.

We then discussed in detail the contributions to the timing model that need to be considered given the much improved precision of our measurements. Here, we considered in particular higher-order timing effects, extrinsic modifications to our observed timing parameters and also spin contributions. The latter requires us, for the first time, to pay attention to the EoS of NSs when interpreting pulsar timing data.

\begin{table}[H]
	\centering
	\caption{Summary of the relativistic effects measured in our analysis
	and list of the resulting independent strong-field tests of GR. For each test, the remaining PK parameters and the mass ratio have been used to determine the masses of pulsars A and B as input for GR predictions. In addition, parameters that test the significance of specific higher order contributions in the advance of periastron and the signal propagation are given.}
	\label{tab:grtests}
	\begin{tabular}{lc@{\hspace{1.0em}}l} 
		\hline
		\hline
		\noalign{\smallskip}
		Relativistic effect & Parameter & Obs./GR pred. \\
		\noalign{\smallskip}
		\hline
		\noalign{\smallskip}		
		Shapiro delay shape  &  $s$                &  $1.00009(18)$  \\
		Shapiro delay range  &  $r$                &  $1.0016(34)$   \\
		Time dilation   &  $\gamma_{\rm E}$   &  $1.00012(25)$  \\
		Periastron advance & $\dot\omega \equiv n_\mathrm{b}k$ &  $1.000015(26)$ \\
        GW emission          &  $\dot{P}_{\rm b}$  &  $0.999963(63)$ \\
		Orbital deformation  &  $\delta_\theta$    &  $1.3(13)$      \\
		Spin precession      &  $\Omega_\mathrm{B}^\mathrm{spin}$  &  $0.94(13)^\ast$ \\
		\noalign{\medskip}
        {\it Tests of higher order contributions}\hspace{-20mm} \\
		\noalign{\smallskip}		
		Lense-Thirring contrib.\ to $k$  &  $\lambda_{\rm LT}$  &  $0.7(9)$ \\
		NLO signal propagation  &  $q_{\rm NLO}[\mbox{total}]$     &  $1.15(13)$ \\
		\dots from signal deflection   &  $q_{\rm NLO}[\mbox{deflect.}]$  &  $1.26(24)$ \\
		\dots from signal retardation  &  $q_{\rm NLO}[\mbox{retard.}]$   &  $1.32(24)$  \\
		\noalign{\smallskip}
		\hline
		\noalign{\smallskip}
        \multicolumn{3}{l}{$^\ast$ Determined in Ref.~\cite{bkk+08}.} \\
	\end{tabular}
	
\end{table}

In this work we obtained more independent tests of GR  than are possible 
in any other system. We summarize the six PK parameters measured in this work 
in Table~\ref{tab:grtests}, additionally including 
the test of relativistic spin precession. The measurements allow us to test conservative aspects of the orbital dynamics of two strongly self-gravitating masses up to 2PN order, including a $\sim$1$\sigma$ constraint on the Lense-Thirring contribution, which in turn could be used to constrain the MoI of pulsar A under the assumption of GR. A signal propagation test resulted in a confirmation of GR at the $4\times 10^{-4}$ level for the propagation of photons in the gravitational field of a strongly self-gravitating (material) body (see Section~\ref{subsubsec:test_Shapiro}). Moreover, NLO contributions are clearly present in the timing data, confirmed with a precision of about 10\%. The most precise GR test available with our data probes the radiative aspects of GR, yielding a test at 2.5PN level in the equations of motion with a precision of $1.3\times 10^{-4}$ (95\% C.L., see Section~\ref{subsubsec:test_gw}). In terms of overall fractional precision, this is the most precise test of GR's predictions for GW emission currently available. 

The new effects that we detected include a relativistic deformation of the orbit, and higher-order contributions to the Shapiro and aberration delay. The latter allow us to infer the spin-direction of A as prograde,  confirming earlier results that require a low-kick supernova as the formation process for pulsar B. We can determine the masses of pulsars A and B with a precision of $10^{-5}$ (modulo an unknown Doppler factor, which is expected to deviate by less than $10^{-4}$ from unity), which we can combine with the first constraints on the NS MoI ever obtained from pulsar timing. With the Double Pulsar, we are able to greatly improve the measurement of orbital period decay resulting from GW damping compared to the current best measurement from the Hulse-Taylor binary system. The second largest contribution to the observed orbital period decay is related to the effective mass loss of pulsar A, which now also has to be taken into account. Improving the precision of the Double Pulsar tests even further will, from now on, be constrained by our ability to correct for kinematic effects. 

We pointed out  that a direct comparison of tests of PN inspiral phase coefficients with different compact objects (BHs vs.\ NSs) as well as different gravity regimes (mildly relativistic vs.\ highly relativistic strong field) comes with certain caveats. Stating this, it is nevertheless obvious that our high-precision timing tests superbly complement the LIGO/VIRGO tests, which currently are less precise at low-GW orders but allow us to probe higher-order contributions to the GW emission. 
This can be seen more clearly in approaches based on different theories of gravitation. For instance, in DEF gravity (Section~\ref{subsec:DEFgrav}), the constraints from the Double Pulsar are considerably tighter than those from the double NS merger GW170817 (binary BH mergers do not provide any constraints for DEF gravity). Section~\ref{sec:altgrav} describes two alternatives to GR, namely DEF gravity and Bekensein's TeVeS, in some detail. We show that the new Double Pulsar results constrain effects that one would typically expect from certain modifications of GR including dipolar radiation and a periodic change of the MoI due to a varying local gravitational constant along the orbit of pulsar A. The Double Pulsar observations presented here lead to further constraints in the two-parameter space of DEF gravity, and result in an additional falsification of TeVeS, which is qualitatively different to that from GW170817 \cite{bdkw18}, by improving on the limits in \cite{fwe+12}. We also show that the Double Pulsar can be used to improve constraints on a time-varying gravitational constant, in particular for effects related to strong gravitational fields. However, a more complete analysis is still needed.

\begin{figure}[H]
    \centering
    \includegraphics[width=8.5cm]{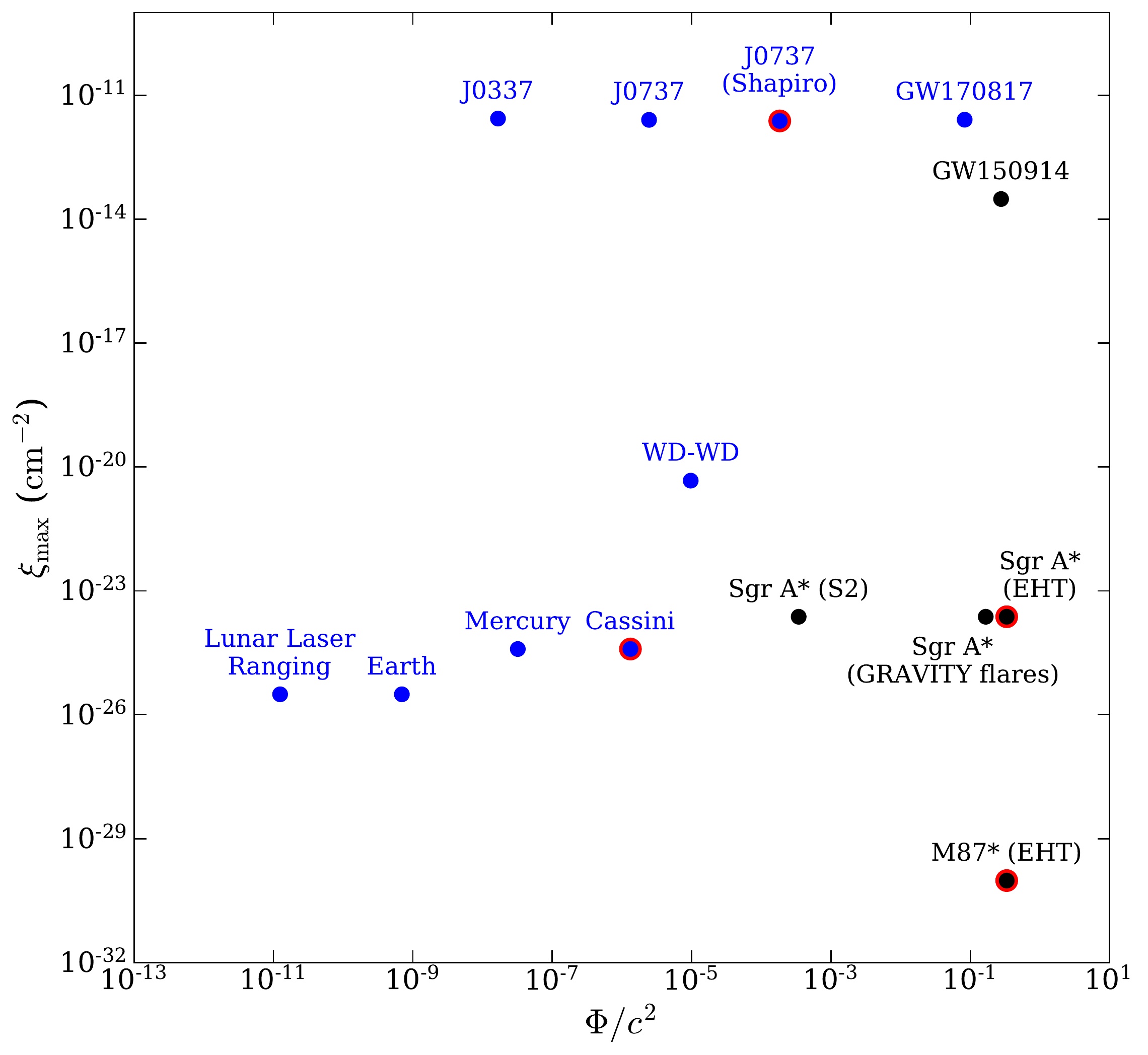}
    \caption{Parameter space for putting the experiments of this paper into context with other gravity experiments. $\Phi$ is the potential of the gravitational interaction, and $\xi_{\rm max}$ is the maximum spacetime curvature in the system. Experiments involving BHs are indicated by black filled circles, whereas other experiments are reported in blue. For pulsar experiments we show the Double Pulsar (`J0737' and `J0737 (Shapiro)') and the universality-of-free-fall test with the triple  system pulsar (`J0337') \cite{agh+18,vcf+20}. Besides these, the following experiments are shown: solar system (cf.\ \cite{Will:2018}), LIGO/Virgo mergers (cf.\ Section~\ref{subsubsec:test_gw}), experiments with the supermassive BHs Sgr~A$^\ast$ and M87$^\ast$ \cite{GRAVITY:2018z, GRAVITY:2018, dhg+19, GRAVITY:2020, EHT:2019, EHT:2020}, and compact double white dwarf systems (`WD-WD'; e.g.\ \cite{hkb+12}). Experiments that probe the propagation of photons in curved spacetime are highlighted by red circles. Details can be found in the text.}
    \label{fig:gravtestsplane}
\end{figure}

The Double Pulsar system is, so far, the only double-NS system where the orientations of both spins relative to the total angular momentum vector can be measured. Combined with a superb knowledge about the systemic velocity and the very high precision measurements of both NS masses, the system is a unique laboratory to study binary evolution, the mechanisms of core-collapse supernovae and the formation and structure of NSs. Obviously, the Double Pulsar is a rich laboratory for a wide variety of physics and astrophysics, but especially for testing  GR and its alternatives. For the latter application, past and future observations of the system provide constraints that are highly complementary to other methods, such as observations with GW detectors, the study of (supermassive) BHs via VLBI imaging \cite{EHT:2019} or orbits of stars \cite{GRAVITY:2020} and flares \cite{GRAVITY:2018}. 
Figure~\ref{fig:gravtestsplane} demonstrates one way of using a two-parameter space to illustrate the complementary of different gravity experiments and put the experiments presented in this paper into context. The first parameter is the potential of the gravitational interaction $\Phi$. This is typically the (external) potential probed by a photon or a mass (e.g.\ a pulsar) in the gravitational field of another mass. $\Phi/c^2$ represents a quantity that typically enters the PN and post-Minkowskian approximation schemes. As a second parameter, we have chosen the maximum spacetime curvature in the system, $\xi_{\rm max}$ defined as the square-root of the Kretschmann scalar. Amongst material bodies, $\xi_{\rm max}$ allows to distinguishing between weakly and strongly self-gravitating masses. For a given EoS, a NS shows a monotonic relation between $\xi_{\rm max}$ and, e.g., its surface potential $\Phi_{\rm s}$. For NSs, one finds $\Phi_{\rm s}/c^2\sim -0.2$, which underlines the strong-field aspect of pulsar tests. For BHs the maximum spacetime curvature (causally connected to its environment) is the one at the horizon, which is a measure for the size and mass of the BH ($\xi_{\rm max} \propto M_\mathrm{BH}^{-2}$, for a non-rotating BH). This is, for instance, of relevance for BHs in alternatives to GR, where in certain cases the  magnitude of the modification decreases with increasing mass, and therefore with decreasing $\xi_{\rm max}$ (see, e.g., \cite{yyp16}). Figure~\ref{fig:gravtestsplane} illustrates well how the Double Pulsar with its strongly self-gravitating components probes the mildly-relativistic strong field regime. The Double Pulsar appears twice in Fig.~\ref{fig:gravtestsplane}, once for the experiments related to orbital dynamics, like GW damping (Sec.~\ref{subsubsec:test_gw}) and periastron precession (Sec.~\ref{subsubsec:test_EoS_LT}), and a second entry (with label ``Shapiro'') for the test related to photon propagation (Sec.~\ref{subsubsec:test_Shapiro}). It is evident that, in terms of coupling between gravitational and electromagnetic fields, the Shapiro delay test of Sec.~\ref{subsubsec:test_Shapiro} is the precision timing experiment which probes the strongest spacetime curvature. When comparing the different gravity experiments in the parameter space of Fig.~\ref{fig:gravtestsplane} , one has to keep in mind the qualitative difference between them, for instance BHs vs.\ material bodies (cf., the discussion at the end of Section~\ref{subsubsec:test_gw}), stationary vs.\ dynamical/radiative situations, etc.. Needless to say, a two-parameter plot cannot capture all quantities relevant for characterizing gravity tests, and therefore always gives an incomplete comparison (for alternative parameter spaces see e.g.\ \cite{bps15,Kramer:2017,GRAVITY:2020,wk20}).
In conclusion, it is clear that studies of the Double Pulsar will continue to be extremely useful, with new applications undoubtedly awaiting us. We believe that this work presents an important milestone in this endeavour. 

\begin{acknowledgments}

We want to acknowledge the contribution of our dear colleague and collaborator Nichi D'Amico, who sadly passed away far too early.

We are indebted to a number of colleagues, who have contributed to our discussions or helped with the observations. These include in alphabetical order: David Champion, Gilles Esposito-Far{\`e}se, Aditya Parthasarathy, Nataliya Porayko, Roman Rafikov,  Gerhard Sch{\"a}fer, and Vivek Venkatraman~Krishnan. We are especially grateful to Jumei Yao for exploring the YMW16 electron density model for our new astrometric constraints, to Michael Keith for his contributions to optimization of {\sc Tempo2} for operation with large data sets, to Victoria Grandy for calibrations of much of the GBT data, and to Lijing Shao for carefully reading the manuscript and many valuable comments that helped to improve it. We would like to thank the staff of all observatories for their continued help over the years.

The Green Bank Observatory is a facility of the NSF operated under cooperative agreement by Associated Universities, Inc. The National Radio Astronomy Observatory is a facility of the NSF operated under cooperative agreement by Associated Universities, Inc.. The Parkes radio telescope is part of the Australia Telescope National Facility (grid.421683.a) which is funded by the Commonwealth of Australia for operation as a National Facility managed by CSIRO. The Nan\c{c}ay Radio Observatory is operated by the Paris Observatory, associated with the French Centre National de la Recherche Scientifique (CNRS). This publication is also based on observations with the 100-m telescope of the Max-Planck-Institut f{\"u}r Radioastronomie at Effelsberg. The Westerbork Synthesis Radio Telescope is operated by the Netherlands Institute for Radio Astronomy (ASTRON) with support from The Netherlands Foundation for Scientific Research NWO. Pulsar research and access to the Lovell telescope is supported by consolidated grants from the Science  and Technology Facilities Council in the UK.

We acknowledge financial support from the Action F\'ed\'eratrice PhyFOG funded by Paris Observatory and from the ``Programme National Gravitation, R\'ef\'erences, Astronomie, M\'etrologie'' (PNGRAM) funded by CNRS/INSU and CNES, France. Pulsar research at UBC is supported by an NSERC Discovery Grant and by the Canadian Institute for Advanced Research. J. W. McKee is a CITA Postdoctoral Fellow: This work was supported by the Natural Sciences and Engineering Research Council of Canada (NSERC), funding reference \#CITA 490888-16. ATD is the recipient of an Australian Research Council Future Fellowship (FT150100415). DRL acknowledges support from the NSF awards AAG-1616042, OIA-1458952. MAM is grateful for support from NSF award number \#1517003 and NSF Physics Frontiers Center award number \#1430284.
P.C.C.F. gratefully acknowledges financial support by the European Research Council, under the European Union’s Seventh Framework Programme (FP/2007-2013) grant agreement 279702 (BEACON) and continuing support from the Max Planck Society. This work is supported by the Max-Planck Society as part of the ``LEGACY'' collaboration on low-frequency gravitational wave astronomy. MK, NW and GD acknowledge financial support by the European Research Council for the ERC Synergy Grant BlackHoleCam (ERC-2013-SyG, Grant Agreement no. 610058). 

Finally, we want to thank three anonymous referees for their kind comments and suggestions, which helped us to
improve the manuscript.
\end{acknowledgments}

\appendix

\section{VLBI observations, data reduction, and verification}
\label{appdx:vlbi}

A total of 18 observations were made using the VLBA between 2016 October and 2018 May (VLBA project code BD193).  Each observation was one hour in duration and recorded 256\,MHz of dual-polarization data with a centre frequency of 1.56\,GHz.  A single scan per observation on the source ICRF~J074533.0+101112 was utilized for calibrating the instrumental bandpass. The source ICRF~J073038.2$-$320820 was used as the primary phase reference calibrator, with NVSS J073709$-$302710 observed in-beam and contemporaneously with the pulsar to refine the calibration solutions and to define a relative position reference point.  Four other background radio sources were also in-beam and observed simultaneously, and were subsequently used as astrometric check sources. NVSS J073709$-$302710 is separated by 15$'$ from PSR~J0737$-$3039A/B on the sky, and by 5$'$ to 15$'$ from the other background sources.

The calibration procedure we employed made use of the pipeline described in detail in \cite{dgb+19}. Because of the southerly location of PSR~J0737$-$3039A/B, we relaxed the elevation-based flagging to times when a source was below an elevation of 15$^{\circ}$, although we tested cutoff elevations in the range 12$^{\circ}$ -- 18$^{\circ}$ and found minimal variation.  After the application of flagging and calibration, the calibrated visibility dataset for each of the in-beam sources was divided by a source model (derived using the concatenated observations from all 18 epochs) and the source position and uncertainty at that epoch was estimated with an image-plane fit to the resultant dataset.  The division by a source model removes any source structure from the image to be fitted, leaving a point-like source whose centroid position is unaffected by changes in resolution due to the absence of different VLBA antennas in some epochs. In addition to the statistical uncertainty recorded from the image-plane fit, an estimate for the systematic position uncertainty was made using Eq.~1 of \cite{dgb+19}, and this uncertainty was added in quadrature to form a total estimated positional uncertainty for each source at every epoch. As discussed in Ref.\,\cite{dgb+19}, this empirical estimator of systematic position uncertainties encapsulates likely contributions such as the differential troposphere and ionosphere between the sightlines to the pulsar and the nearby calibrators.

Again following \cite{dgb+19}, we estimated the astrometric parameters for PSR~J0737$-$3039A/B and the in-beam background sources using a bootstrap method. The orbital reflex motion of the Double Pulsar is negligibly small, and is not included in the fit. We repeated this process twice -- once utilising the proper motion as a free parameter in the fit, and once fixing the proper motion to the expected value (zero for the background sources, and the pulsar timing result reported in Table~\ref{tab:params} for the pulsar). In this latter case, a proper motion value for right ascension and declination was randomly drawn for each bootstrap trial using the corresponding mean and standard deviation from the pulsar timing fit, i.e., the uncertainty in the pulsar timing proper motion was accounted for.  

The best-fit VLBI results for PSR~J0737$-$3039A/B and the background radio sources used as astrometric checks are shown in Table~\ref{tab:fullvlbiresults}.  Given that the background radio sources are expected to be distant radio-emitting active galactic nuclei (AGN) , the parallax and proper motion measured for these sources are expected to be consistent with zero.  Uncertainties for proper motion and parallax are derived from the bootstrap procedure described above.  The uncertainties on the reference position are dominated by the uncertainty in the absolute position of the in-beam reference source NVSS J073709$-$302710.  This quantity is relatively poorly constrained, being derived purely from phase referencing from the primary calibrator ICRF~J073038.2$-$320820, and we estimate a nominal uncertainty of 10\;mas based on the $\sim$2$^\circ$ angular separation between these two sources.  Accordingly, we assign an uncertainty of 10\;mas to the reference position for all sources in Table~\ref{tab:fullvlbiresults}.  While we focus on the proper motion and parallax (which are unaffected by errors in the absolute position that are constant in time) from this point forward, we do note that comparing the reference positions for PSR~J0737$-$3039A/B obtained from timing and VLBI at a common epoch shows agreement within this nominal 10\;mas uncertainty.

\begin{table*}[htp]
	\centering
	\caption{VLBI astrometric results for PSR J0737–3039A/B and in-beam background sources. Listed are reference position in right ascension (R.A.) and declination (Dec.), proper motion in each coordinate, parallax, and reference epoch. The top half of the table shows results when proper motion is fixed to expected values, while the second half shows results when proper motion is freely fitted. See the text for details.}
	\label{tab:fullvlbiresults}
	
	\begin{tabular}{lccccc} 
		\hline
		\hline
		\noalign{\smallskip}
	 & J0737$-$3039A/B & J0736$-$3027 & J0736$-$3037 & J0737$-$3018 & J0738$-$3025   \\
	 \noalign{\smallskip}
		\hline
		\noalign{\smallskip}
RA (J2000)  & 07$^{h}$37$^{m}$51$^{s}$.247(1)
            & 07$^{h}$36$^{m}$45$^{s}$.378(1)
            & 07$^{h}$36$^{m}$51$^{s}$.234(1)
            & 07$^{h}$37$^{m}$09$^{s}$.378(1)
            & 07$^{h}$38$^{m}$19$^{s}$.793(1) \\
Dec (J2000) & $-$30$^{\circ}$39$'$40$''$.68(1)
            & $-$30$^{\circ}$27$'$18$''$.77(1)
            & $-$30$^{\circ}$37$'$46$''$.16(1)
            & $-$30$^{\circ}$18$'$52$''$.68(1)
            & $-$30$^{\circ}$25$'$04$''$.92(1) \\
Parallax, $\pi_{\rm v}$  (mas)  & $\phantom{-}1.30^{+0.13}_{-0.11}$
                        & $\phantom{-}0.02^{+0.06}_{-0.07}$
                        & $-0.24^{+0.18}_{-0.21}$
                        & $-0.03^{+0.11}_{-0.11}$
                        & $-0.07^{+0.11}_{-0.11}$ \\ 
$\mu_\alpha$ (mas yr$^{-1}$, fixed)    & $-2.567\pm0.030$  & 0.0 & 0.0 & 0.0 & 0.0 \\
$\mu_\delta$ (mas yr$^{-1}$, fixed)   & $\phantom{-}2.082\pm0.038$     & 0.0 & 0.0 & 0.0 & 0.0 \\
Position epoch (MJD)    & 58000 & 58000 & 58000 & 58000 & 58000 \\
\noalign{\smallskip}
		\hline
		\noalign{\smallskip}
RA (J2000)  & 07$^{h}$37$^{m}$51$^{s}$.247(1)
            & 07$^{h}$36$^{m}$45$^{s}$.378(1)
            & 07$^{h}$36$^{m}$51$^{s}$.234(1)
            & 07$^{h}$37$^{m}$09$^{s}$.378(1)
            & 07$^{h}$38$^{m}$19$^{s}$.793(1) \\
Dec (J2000) & $-$30$^{\circ}$39$'$40$''$.68(1)
            & -30$^{\circ}$27$'$18$''$.77(1)
            & -30$^{\circ}$37$'$46$''$.16(1)
            & -30$^{\circ}$18$'$52$''$.68(1)
            & -30$^{\circ}$25$'$04$''$.92(1) \\
Parallax, $\pi_{\rm v}$ (mas) & $\phantom{-}1.43^{+0.14}_{-0.13}$
                         & $-0.05^{+0.07}_{-0.08}$
                         & $-0.35^{+0.27}_{-0.34}$
                         & $-0.12^{+0.17}_{-0.17}$
                         & $-0.10^{+0.13}_{-0.14}$ \\
$\mu_\alpha$ (mas yr$^{-1}$, fitted)    & $-2.23^{+0.22}_{-0.21}$
                                & $-0.20^{+0.15}_{-0.14}$
                                & $-0.32^{+0.42}_{-0.48}$
                                & $-0.25^{+0.26}_{-0.25}$
                                & $-0.12^{+0.21}_{-0.22}$ \\
$\mu_\delta$  (mas yr$^{-1}$, fitted)   & $\phantom{-}2.79^{+0.45}_{-0.43}$
                                & $\phantom{-}0.01^{+0.11}_{-0.11}$
                                & $\phantom{-}0.06^{+0.61}_{-0.57}$
                                & $-0.50^{+0.30}_{-0.32}$
                                & $-0.03^{+0.48}_{-0.46}$ \\
Position epoch (MJD)    & 58000 & 58000 & 58000 & 58000 & 58000 \\
\noalign{\smallskip}
		\hline
	\end{tabular}
\end{table*}

\begin{figure*}
\centering
\begin{tabular}{cc}
\includegraphics[width=0.50\textwidth]{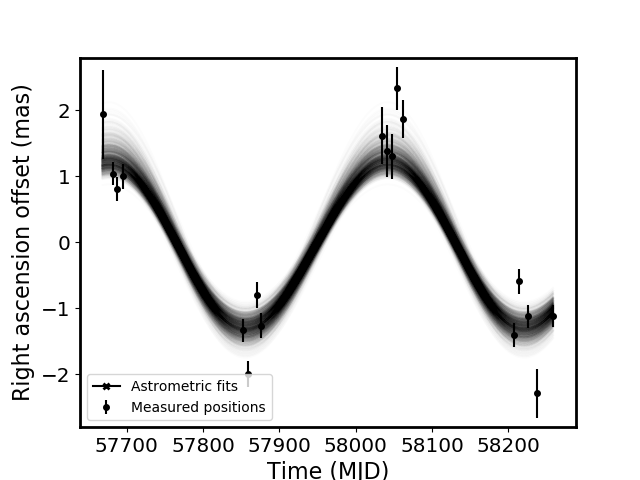} &
\includegraphics[width=0.46\textwidth]{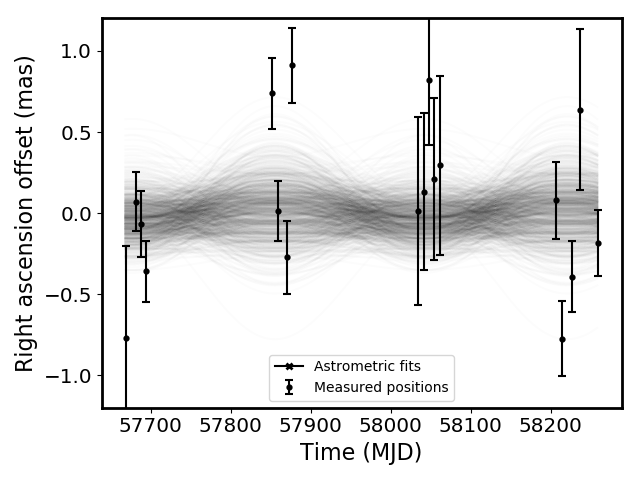}
\end{tabular}

\caption{\label{fig:vlbi} Bootstrap fit plots of right ascension vs time, with the fixed proper motion subtracted.  Each line shows the result of a single bootstrap trial.  The left panel shows PSR~J0737$-$3039A/B, and the right panel shows the in-beam background source NVSS~J073709$-$301853, which shows a parallax consistent with zero as expected.}
\end{figure*}

As our reference case, we utilise the fit in which the proper motion has been constrained based on the timing value (top section of Table~\ref{tab:fullvlbiresults}), but note that the best-fit value is consistent to  within 1$\sigma$ regardless of whether the proper motion was fixed or not.  The results were similarly insensitive to other potential choices in the data reduction such as a scaling factor applied to the estimated systematic uncertainty added to the position measurements, or the elevation flagging limit. As an illustration of the results, we show the offset in the position in right ascension as a function of time for both PSR~J0737$-$3039A/B (where the parallax signature is clearly seen) and the in-beam background source NVSS~J073709$-$301853 (which shows no parallax signature, as expected) after the subtraction of proper motion for our reference case in Figure~\ref{fig:vlbi}.  Right ascension is shown for three reasons: first, because the elongated VLBA synthesized beam leads to greater precision in the right ascension axis than in declination; second, the parallax signature is larger in right ascension than declination; and third, because of the first two reasons, observations were scheduled at times of maximum parallax displacement in right ascension (and hence minimal parallax displacement in declination).

As we are primarily concerned with the parallax (and the corresponding uncertainty) for PSR~J0737$-$3039A/B, the consistency of the astrometric results for the  contemporaneously observed background radio sources with expectations is of considerable interest.  As shown in Table~\ref{tab:fullvlbiresults}, the fitted parallax and proper motion are consistent with expectations at the 1$\sigma$ level in 75\% of cases, and at the 2$\sigma$ level in 100\% of cases.  This offers a high degree of reassurance that the parallax uncertainty for PSR~J0737$-$3039A/B has been well estimated.  It is however necessary to note that the background sources (and PSR J07373$-$3039A/B) span a range of angular separations from 5$'$ to 15$'$ to the in-beam calibrator, and at 15$'$, the angular separation between PSR~J0737$-$3039A/B and the in-beam calibrator is the equal-highest of all source pairs considered.  While systematic errors are likely to scale with angular separation \citep{dgb+19}, the bootstrap technique employed both here and in the study made by \citet{dgb+19} offers a high degree of robustness in incorporating these into the estimated uncertainties.

The parallax probability distributions resulting from the bootstrap fits described above are not perfectly Gaussian, exhibiting a slight skew with a tail towards larger values. 

We now consider the potential effects of refractive image wander on the VLBI results. Refraction caused by large-scale gradients in the ionized ISM can cause the apparent position of a radio source to shift with a characteristic timescale usually on the order of months to years, depending on the observing frequency and relative pulsar--ISM--Earth velocity \cite{ric90}.  For most radio sources, this astrometric offset would be uncorrectable but, as discussed in Section~\ref{subsec:impl_ism}, for radio pulsars the time variability of the pulsar DM can be related back to gradients in the ionized ISM and hence used to estimate (a component of) the necessary correction. Figure~\ref{fig:image_wander} shows both the DM variations and the derived image wander (see Appendix~\ref{appdx:image_wander} for details) over the entire observing span.

We have applied these corrections to the VLBA positions presented above and re-fitted the dataset for parallax and reference position.  When doing so, the best-fit parallax (from our reference-case fit with the proper motion constrained by the timing values) becomes $1.27^{+0.11}_{-0.10}$\;mas -- a negligible change.  This result is unsurprising when considering Figure~\ref{fig:image_wander}, in which the derived offsets can be seen to be small at the times sampled by our VLBA observations (MJD 57670 to 58260.)

Since the refractive wander corrections that we can derive based on DM variations are necessarily incomplete, we also estimated a ``worst--case" impact of refractive wander by sampling the corrections that would have been applied had the VLBA observations began in earlier years.  We sampled corrections from the VLBA observation dates shifted by an integer number of years in the past, generating a total of 10 potential sets of corrections.  Across these 10 sets of potential corrections, the mean change to parallax was 0.03~mas, with a maximum change of 0.09~mas -- smaller in all cases than the currently-estimated uncertainty.

\begin{figure}
\centering
\includegraphics[width=0.47\textwidth]{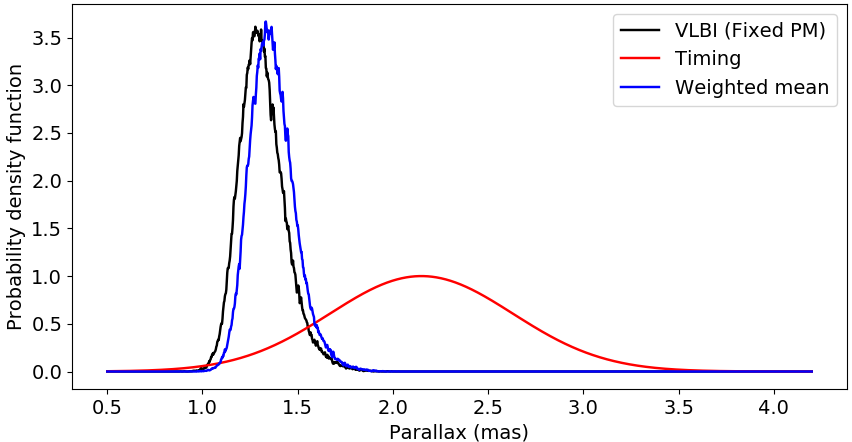} 
\caption{\label{fig:pxpdfs} Probability distribution for parallax resulting from VLBI with proper motion fixed to the timing value (black), from pulsar timing (red), and the weighted mean of the two (blue).}
\end{figure}

These refractive wander corrections can also be applied to historical VLBI datasets, in particular, to the Long Baseline Array work from \cite{dbt09}.  As can be seen from Figure~\ref{fig:image_wander}, the impact during the time of these observations (MJD 53970 -- 54500) was considerably greater than during our VLBA observations.  Moreover, the LBA parallax fit is particularly sensitive to the observation at MJD 54308 (which is the sole point taken near the parallax maximum in declination), and the estimated refractive shift is particularly pronounced at this time.  After correcting the estimated refractive offsets, the best-fit parallax from the LBA dataset changes appreciably, from 0.87~mas to 0.96~mas.  We also note that the original fit to the LBA data \cite{dbt09} did not have the advantage of the now well-constrained proper motion, and used a least-squares fit to estimate the parallax uncertainty, rather than a bootstrap fit as we have done for the newer VLBA data.  Repeating the analysis of the LBA data using the same approach taken here yields a 68\% confidence interval for the parallax of 0.61\;mas to 1.17\;mas, which becomes consistent at the $\sim$1$\sigma$ level with the VLBA results (but still favours a larger distance, in contrast to the timing, which favours a smaller distance).  We note that the increased number of observations (by a factor of almost three), a more compact primary calibrator source, and the presence of in-beam background sources that can be used to check for unmodeled systematic astrometric offsets, all act to add extra confidence in the accuracy of the uncertainty estimation for the VLBA dataset, compared to the LBA dataset.

Finally we combine the VLBA parallax $\pi_v$ with the timing parallax $\pi_t$ into a single estimate $\pi_c$. As $\pi_v$ and $\pi_t$ are independent, a weighted mean, with weights equal to the inverse variance of each measurement, is the best linear unbiased estimator of $\pi_c$. The variances are well estimated in both cases. In a weighted mean one would normally compute a $\chi^2$ value as a goodness of fit measure. A large $\chi^2$ would prompt a search for an error. In this case we have only two estimates to combine, so we simply take the difference $D = \pi_v - \pi_t = 0.85$. The variance of $D$ is the sum of the variances of $\pi_v$ and $\pi_t$, which gives a standard deviation of 0.51. Thus the measured $D$ represents a 1.7-$\sigma$ event - not improbable enough to cause a search for errors in  $\pi_v$ and $\pi_t$. We can put it as $\chi^2(1) = (0.85/0.51)^2$ and use the $\chi^2(1)$ distribution to show that the probability of $D > 0.85 =$ 10\%, confirming that the two independent estimates are adequately consistent.

The probability distribution for $\pi_c$ was derived using a Monte Carlo approach, in which we randomly drew a timing parallax and a VLBI parallax from their distributions, and then took a weighted average to form a combined sample. This process was repeated 100,000 times to form the combined parallax probability distribution. The final result for the parallax was $\pi_{\rm c} = 1.36^{+0.12}_{-0.10}$\,mas (68\% confidence levels), a shift of $\sim$0.5$\sigma$ from the VLBA-only estimate. For each iteration, we also computed the distance. Because of the compensating effects of the tail on the high side of the parallax distribution and the bias resulting from the parallax inversion, the distance probability distribution is close to symmetric, giving our best estimate for the distance of $735\pm 60$\,pc. We show the parallax probability distributions from our VLBA observations, from pulsar timing, and for the weighted mean in Figure~\ref{fig:pxpdfs}.

\section{Analysis of ISM electron-density fluctuations}
\label{appdx:dmvar}

\subsection{PSD/SF Simulations}
\label{appdx:simulations}

The measured fluctuations DM($t$) (Sec.~\ref{subsec:astrom_timing}) are distorted by the fitting technique in two ways. Firstly, the {\sc Tempo2} analysis fits a piece-wise linear model of DM($t$) with a sampling of 100 days. This is equivalent to convolving the data with a triangle of baseline $\pm 100$~d. Secondly, the Monte Carlo scheme adjusts the phase of the DM($t$) sampling and then bins it in 24-d bins. This is equivalent to convolving the data by a 24-d rectangle. To estimate the spectral exponent of the original DM($t$) variations we simulated various pure power-law processes, convolved them by the two measurement effects, and compared their power spectra with that of the observations. 

We make the comparison in the spectral domain (rather than the SF) because the errors in the spectral harmonics are independent and we can adjust the level and the exponent to minimize the number of harmonics in the observed spectrum that exceed the 90\% confidence limits. The samples of the SF, like those of a covariance function, are highly correlated and this makes them appear smoother than the errors would suggest. The mean and 90\% confidence limits of the best match are shown in Figure~\ref{fig:dmvar_psd} in red. We also show the mean of the expected Kolmogorov exponent in blue. Note that the spectra are relatively steep so the PSD must be computed with pre-whitening and post-darkening to avoid a bias due to spectral leakage. We pre-whiten the time series with a first difference, compute the power spectrum, and correct the spectrum with the Fourier transform of the difference operator \cite{chc+11}.

The match is good from $2\times 10^{-4}\,$d$^{-1}$ up to about $7\times 10^{-3}$\,d$^{-1}$, where white noise begins to dominate the spectrum. It is very clear that a Kolmogorov spectrum cannot fit the data. The PSD also shows a peak near the annual frequency (at a period of 335 days). This is probably due to the Earth’s orbital motion through phase gradients in the ISM. As the ISM is also moving, these phase gradients can essentially Doppler-shift the annual motion to a higher frequency. The same simulation was used to compare with the observed structure function in more detail. This is shown in Figure~\ref{fig:sf_simulations}.

\subsection{Refractive Image Wander}
\label{appdx:image_wander}

The basic phenomenon causing intensity scintillations is multipath propagation, also referred to as scattering. This both increases the apparent angular diameter of the pulsar and broadens its pulse width. Interference between the scattered rays causes the intensity scintillation. The scattering contribution to the pulse width can be estimated directly from the bandwidth of the intensity scintillations \cite{rcn+14} as $\tau_p = 1/(2 \pi\, \nu_{0.5}) \sim 3\,\mu$s at 820\,MHz for the Double Pulsar. Both diffractive and refractive intensity scintillations cause fluctuations in the ToAs of the pulses on short time scales, and phase gradients cause fluctuations of the ToAs on longer scales. However all of these fluctuations are comparable with, but less than, the pulse width \cite{crg+10}, so we do not correct for them in this work. Phase gradients do have an important effect because they displace the apparent position of the pulsar by $\theta_r = \nabla \phi /k$ where $k=2\pi/\lambda$ is the propagation constant. This displacement can be comparable with the uncertainty in parallax measurements, so we have examined the effect of refractive displacement in detail.

\begin{figure}[H]
    \centering
    \includegraphics[width=8.5cm]{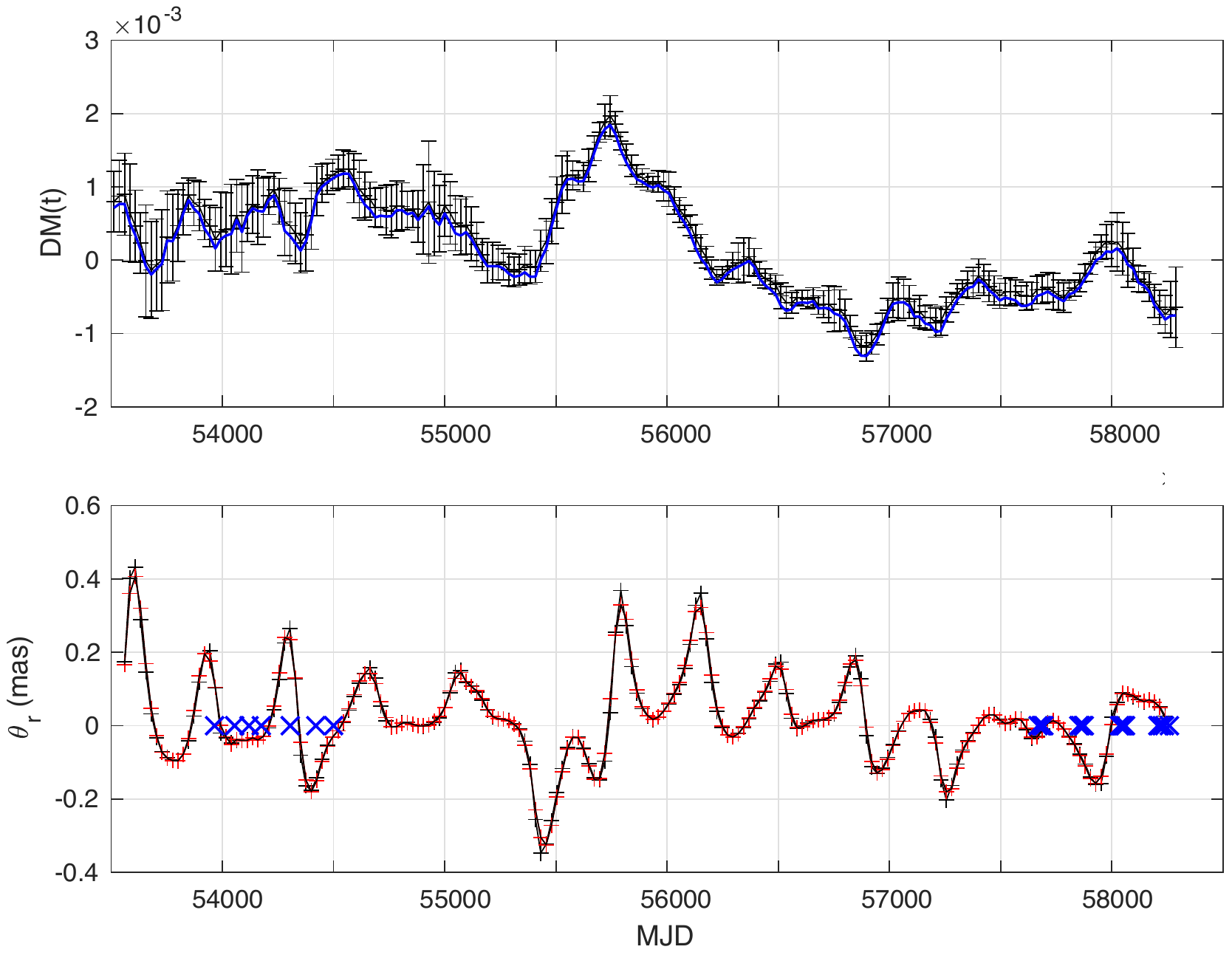}
    \caption{Top: Dispersion measure of the Double Pulsar as a function of time. The black line shows the interpolated DM variations and the blue line shows these interpolated variations with the worst-case solar-wind contribution subtracted. Bottom: The image wander in the direction of the velocity $\theta_r$ (mas) at 1.56 GHz.  The times of VLBI observations are marked with the blue x symbols, with the early observations being made with the LBA and the more recent observations with the VLBA and presented in this paper.}
    \label{fig:image_wander}
\end{figure}

The phase gradients for PSR~J0737$-$3039A/B have been estimated directly from intensity scintillations \cite{rcn+14}. Unfortunately, few of our observations have sufficient frequency resolution to measure the dynamic spectra of intensity scintillations accurately. However, we can use a daily interpolation of the observed DM variations to estimate their temporal gradient which can be scaled to phase by 
\begin{equation}
\phi = 2 \pi\times 10^6\, {\rm DM} / 2.41\times 10^{-4} \nu_{\rm MHz}.
\end{equation}
We then scale the temporal gradient to a spatial gradient in the direction of the velocity by dividing by the velocity ${\bf{v}}(t)$.  We must assume that there is a comparable gradient perpendicular to the velocity since the turbulence is known to be roughly isotropic \cite{rcn+14}; consequently the rms gradient should be direction independent. In Figure~\ref{fig:image_wander} we show the displacement $\theta_r$ at 1.56\,GHz in the direction of $\bf{v}$. The rms($\theta_r$) = 0.125 mas so the total rms image wander should be $\sqrt{2}$ times greater, $\sim$ 0.177 mas.

The dates of the VLBI observations used to estimate the parallax (and thus the pulsar distance) are marked as blue x symbols on Figure~\ref{fig:image_wander}. One can see that the later VLBI observations, which were used in this paper, are at a time of relatively low phase gradients (at least in the direction of the velocity). To estimate the effect on parallax we resolved the vector displacement $\theta_r$ into components parallel to the R.A. and Dec. axes. The actual effects of these displacements were already discussed in Appendix~\ref{appdx:vlbi}.

Refractive image wander also potentially has an effect on the astrometric parameters derived from pulsar timing. The timing signature of a position offset is an annual sinusoid, with the phase determined by the ratio of the perpendicular components of the offset (in R.A. and Dec., for example). For proper motion the signature is an annual sinusoid of uniformly increasing amplitude and for parallax it is a six-month sinusoid. Since the refractive image wander (Figure~\ref{fig:image_wander}) has a strong annual component, attributed to the annual motion of the LoS  over a gradient in the screen electron density (Appendix~\ref{appdx:simulations}), there is a possibility of a significant covariance between image wander and the astrometric parameters.

To assess the significance of this effect, we took the variations in refractive angle shown in Figure~\ref{fig:image_wander}, interpolated to daily values, and transformed these to daily values of ToA offset using
\begin{equation}
  \Delta t_{\rm refr} = 0.5\;\theta_r^2\;\frac{d}{c}\; \frac{s}{1-s}
\end{equation}
where $\theta_r$ is the refractive angle in radians, $d$ is the pulsar distance (735\,pc), $c$ is the velocity of light and $s$ is the distance of the scattering screen from the Earth as a fraction of the pulsar distance. These daily ToA offsets were linearly interpolated to the MJD of each ToA in the 4-min data set and then scaled by the ratio $(f_{\rm ToA}/f_{\rm Ref})^{-2}$, where $f_{\rm ToA}$ is the radio frequency for the ToA and $f_{\rm Ref}$ is the reference frequency for the refractive angles shown in Figure~\ref{fig:image_wander}, 1.56\,GHz. These time offsets were then added to the corresponding ToA. Comparison of the results of a standard {\sc Tempo2} analyses solving for the astrometric parameters with and without the refractive delay corrections showed that the uncertainties in the astrometric parameters were unaffected by the refractive image wander and that the changes in the values were all less than or about $0.2\,\sigma$.

In this discussion we have assumed that DM($t$) is entirely due to the ISM. However the observed DM($t$) will include a solar-wind component. The worst-case solar-wind component would be a spherical solar wind with $n_e \propto R_{\rm Sun}^{-2}$ and $n_e  = 10$\, cm$^{-3}$ at the Earth. We have subtracted this from the observations and plotted the result on as a blue line in Figure~\ref{fig:image_wander}. Evidently the worst-case solar-wind contribution is negligible and our assumption is justified.

\section{PK parameters in scalar-tensor gravity}
\label{appdx:pkstg}

In this section we provide expressions for the PK parameters valid for a wide class of scalar-tensor gravity (STG), including DEF gravity \cite{de93,de96} used in Section~\ref{subsec:DEFgrav} of this paper. More details can be found in reviews like \cite{dt92,de92b,dam09,Will:2018,wk20}, and the corresponding references given therein. Here we closely follow the representation used in \cite{de96}. We will restrict the discussion to PK parameters that we actually use for the tests in Section~\ref{sec:altgrav}, i.e.\ $k$, $\gamma_{\rm E}$, $s$, and $\dot{P}_{\rm b}$.

\subsection{Advance of periastron: \texorpdfstring{$k$}{k}}

For the advance of periastron in the presence of two strongly self-gravitating bodies, one finds in fully conservative, boost-invariant gravity theories
\begin{equation}
    k = \left(2 + 2\gamma_{\rm AB} - 
            X_{\rm B} \beta_{\rm BB}^{\rm A} - 
            X_{\rm A} \beta_{\rm AA}^{\rm B} \right) 
        \frac{\hat\beta_\mathrm{O}^2}{1 - e^2} 
        + {\cal O}(\hat\beta_\mathrm{O}^4)\,,
    \label{eq:omdot_STG}
\end{equation}
where $\hat\beta_\mathrm{O} \equiv (G_{\rm AB}Mn_{\rm b})^{1/3}/c$. $G_{\rm AB}$ is the effective gravitational constant, entering the orbital dynamics of the binary system already at the Newtonian level. In GR, $\hat \beta_\mathrm{O}$ is equal to $\beta_\mathrm{O}$ of Eq.~(\ref{eq:bO}). In STG $G_{\rm AB} = G_\ast(1 + \alpha_{\rm A}\alpha_{\rm B})$, where $G_\ast$ is the bare gravitational constant, and $\alpha_a$ ($a = {\rm A,B}$) denotes the effective scalar coupling of body $a$. $\alpha_a$ is a measure for the change of the inertial mass of body $a$ with respect to the (asymptotic) scalar field $\varphi_0$, i.e.\ $\alpha_a = \partial \ln m_a / \partial\varphi_0$, where the number of baryons of the NS is kept fixed when taking the partial derivative. The parameters $\gamma_{\rm AB}$, $\beta^{\rm A}_{\rm BB}$, and $\beta^{\rm B}_{\rm AA}$ represent three body-dependent strong-field generalizations of Eddington's two weak-field PPN parameters $\gamma^{\rm PPN}$ and $\beta^{\rm PPN}$ \cite{edd22}. They enter the modified Einstein-Infeld-Hoffmann equations of motion for a binary system consisting of strongly self-gravitating bodies at the 1PN level \cite{will93,dam09}. In GR, as a result of the effacement principle \cite{dam87}, $G_{\rm AB} = G$ and $\gamma_{\rm AB} = \beta^{\rm A}_{\rm BB} = \beta^{\rm B}_{\rm AA} = 1$. In STG they are given by \cite{dam09}
\begin{eqnarray}
    \gamma_{\rm AB} &=& 1 - \frac{2 \alpha_{\rm A}\alpha_{\rm B}}
        {1 + \alpha_{\rm A}\alpha_{\rm B}} \,,
        \label{eq:gammaAB} \\
    \beta^{\rm A}_{\rm BB} &=& 1 + \frac{\beta_{\rm A}\alpha_{\rm B}^2}
        {2(1 + \alpha_{\rm A}\alpha_{\rm B})^2} \,,\\
    \beta^{\rm B}_{\rm AA} &=& 1 + \frac{\beta_{\rm B}\alpha_{\rm A}^2}
        {2(1 + \alpha_{\rm A}\alpha_{\rm B})^2} \,.
\end{eqnarray}
The quantity $\beta_a$ ($a = {\rm A,B}$) is calculated according to $\beta_a = \partial\alpha_a / \partial\varphi_0$, where the number of baryons of the NS is kept fixed when taking the partial derivative with respect to the asymptotic scalar field $\varphi_0$. Depending on the coupling parameters $\alpha_0$ and $\beta_0$ of DEF gravity, $\beta_a$ can assume rather large values. More generally, in a system consisting of two NSs, these generalized PPN parameters can be very different from the corresponding weak-field PPN parameters. Even if there are only very small (or no) deviations from GR in the weak field, one can have order unity deviations in double NS system (see e.g.\ \cite{de93}).

As discussed in Section~\ref{subsec:DEFgrav}, although the ${\cal O}(\hat\beta_\mathrm{O}^4)$ corrections in Eq.~(\ref{eq:omdot_STG}) do become relevant in the analysis of the Double Pulsar, presently it is still sufficient to use the corresponding GR expressions in the gravity tests presented here.

\subsection{Amplitude of the Einstein delay: \texorpdfstring{$\gamma_{\rm E}$}{gammaE}}

In STG, the amplitude $\gamma_{\rm E}$ of the time dilation (Einstein delay) gets modified as well. Apart from changes in the expressions for the time dilation, there is a modulation of the spin period of the pulsar caused by a periodic variation of the local gravitational constant, at the location of the pulsar as it moves around its companion. This effect has the same orbital dependence as the time dilation, and hence can be absorbed in $\gamma_{\rm E}$. The resulting (total) expression for A is \cite{de96}
\begin{equation}
    \gamma_{\rm E} = \frac{e}{n_{\rm b}} \, X_{\rm B} \left(X_{\rm B} +
        \frac{1 + k_{\rm A}\alpha_{\rm B}}{1 + \alpha_{\rm A}\alpha_{\rm B}}
        \right) \hat\beta_\mathrm{O}^2 \,,
    \label{eq:gammaE_STG}    
\end{equation}
where $k_{\rm A} \equiv -\partial\ln I_{\rm A} / \partial\varphi_0$. Here again, the number of baryons of A is kept fixed when taking the partial derivative. Note, $\alpha_{\rm A} = \alpha_{\rm B} = k_{\rm A} = 0$ in GR.

\subsection{Shapiro shape parameter: \texorpdfstring{$s$}{s}}

The Shapiro shape parameter, $s$, can simply be identified with $\sin i$. When expressed as a function of the masses of the binary system, one uses the mass function, which is linked to the theory dependent third Kepler law. To leading order one finds in STG \cite{de96}
\begin{equation}
    s = \frac{x n_{\rm b}}{X_{\rm B} \hat\beta_\mathrm{O}} \,.
    \label{eq:s_STG}
\end{equation}
NLO corrections to the mass function depend on $\gamma_{\rm AB}$ (cf.\ Eq.~(3.9) in \cite{dt92}), where $\varepsilon = 2\gamma_{\rm AB} + 1$. Since such NLO corrections are at the one-sigma level of the measured $s$, it is sufficient to use the corresponding GR expression in Eq.~(\ref{eq:mf1PN}) in the PK-parameter test.

\subsection{GW damping: \texorpdfstring{$\dot{P}_{\rm b}$}{Pbdot}}

Concerning GW damping, the (dynamical) scalar field leads to various modifications of the expression for $\dot{P}_{\rm b}$. In terms of PN order, the leading contribution comes from scalar dipole radiation, and enters already at the 1.5PN level, i.e.\ $\hat\beta_\mathrm{O}^3$. In STG, the leading order term for dipolar GW damping is given by
\begin{equation}
    \dot{P}_{\rm b}^{\rm dipole} = -2\pi \, X_{\rm A} X_{\rm B} \,
    \hat\beta_\mathrm{O}^3 \, \frac{1 + \frac{1}{2}e^2}{(1 - e^2)^{5/2}} \,
    \frac{(\alpha_{\rm A}-\alpha_{\rm B})^2}{1+\alpha_{\rm A}\alpha_{\rm B}} \,.
    \label{eq:PbdotD_STG}
\end{equation}  
In our test we have also used those NLO corrections which are of order $(\alpha_{\rm A} - \alpha_{\rm B}) \times \hat\beta_\mathrm{O}^5$, even though $(\alpha_{\rm A} - \alpha_{\rm B})$ has to be small, given the tight agreement with GR as seen e.g.\ in Eq.~(\ref{eq:GR_GW_test}). The rather lengthy expression for these NLO corrections can be found in Eq.~(6.52b) in \cite{de92b}. There are also terms of order $(\alpha_{\rm A} - \alpha_{\rm B})^2 \times \hat\beta_\mathrm{O}^5$, which can be safely ignored.

At the 2.5PN order one finds a modified quadrupole formula, that combines quadrupolar contributions from the tensor and the scalar field. For STG, it is given by \cite{de92b}
\begin{eqnarray}
   \dot{P}_{\rm b}^{\rm quadrupole} &=&  
   -\frac{192\pi}{5} \, X_{\rm A}X_{\rm B} \, \hat\beta_\mathrm{O}^5 \,
   \frac{1+\frac{73}{24}e^2+\frac{37}{96}e^4}{(1-e^2)^{7/2}} 
   \nonumber\\&&
   \times\,\frac{1 + \frac{1}{6}\left(\alpha_{\rm A} X_{\rm B} + \alpha_{\rm B} X_{\rm A}\right)^2}{1+\alpha_{\rm A}\alpha_{\rm B}} \,.
\end{eqnarray}
In addition, the leading contribution from the scalar monopole radiation also enters at the 2.5PN order, and in STG is given by \cite{de92b}
\begin{eqnarray}
   &&\dot{P}_{\rm b}^{\rm monopole} = -3\pi\, X_{\rm A}X_{\rm B} \,
   \hat\beta_\mathrm{O}^5 \,\frac{e^2(1 + \frac{1}{4}e^2)}{(1 - e^2)^{7/2}} \nonumber\\ && \qquad
   \times\,\frac{\left[
   \alpha_{\rm A}(1 + \frac{2}{3}X_{\rm B}) + 
   \alpha_{\rm B}(1 + \frac{2}{3}X_{\rm A}) +
   \frac{\beta_{\rm A}\alpha_{\rm B} + \beta_{\rm B}\alpha_{\rm A}}
   {1+\alpha_{\rm A}\alpha_{\rm B}}\right]^2}{1+\alpha_{\rm A}\alpha_{\rm B}} \,.
   \nonumber\\&&
\end{eqnarray}
Note that in the Double Pulsar this contribution is greatly suppressed, compared to the other ${\cal O}(\hat\beta_\mathrm{O}^5)$ contributions, as it is proportional to $e^2 \approx 0.008$.

\section{Glossary of frequently used symbols and abbreviations}
\label{appdx:glossary}

For the convenience of the reader, we provide a list of frequently used symbols and their meaning. We also list abbreviations frequently used in the paper.


\vspace{12pt}
\noindent
{\bf Symbols}

\begin{description}[leftmargin=!,labelwidth=3em ]
\item[$\alpha$ ] Right ascension (RA) in equatorial coordinates; Spectral index
\item[$\delta$ ] Declination (DEC) in equatorial coordinates
\item[$l$ ] Galactic longitude 
\item[$b$ ] Galactic latitude 
\item[$\mu_\alpha$ ] Proper motion in RA 
\item[$\mu_\delta$ ] Proper motion in DEC
\item[$\mu_l$ ] Proper motion in $l$ 
\item[$\mu_b$ ] Proper motion in $b$ 
\item[$v_{\rm T}$ ] Transverse velocity 
\item[$\pi$ ] Annual parallax 
\item[$\nu$ ] Pulsar spin frequency 
\item[$\dot{\nu}$ ] First spin frequency derivative  
\item[$\ddot{\nu}$ ] Second spin frequency derivative 
\item[$\dddot{\nu}$ ]  Third spin frequency derivative 
\item[$\ddddot{\nu}$ ]  Fourth spin frequency derivative 
\item[$\phi$ ] Rotational phase 
\item[$N_0$ ] Pulse number at a reference epoch $t_0$
\item[$c_i$ ] Profile evolution ``FD parameters'' 
\item[{$D$} ] Doppler factor between SSB and pulsar system
\item[$P_{\rm b}$ ] Orbital period 
\item[$n_{\rm b}$ ] Orbital angular frequency, $2\pi/P_{\rm b}$
\item[$x$ ] Projected semimajor axis 
\item[$e_T$ ] Eccentricity (Kepler equation) 
\item[$e_\theta$, $e_r$ ] Additional eccentricities, see Eqs.~(\ref{eq:Roemer}), (\ref{eq:ecc})  
\item[$T_0$ ] Epoch of periastron 
\item[$\omega$, $\omega_0$  ] Longitude of periastron, at $T_0$
\item[$\Delta_{\rm R}$ ] R{\o}mer delay
\item[$\Delta_{\rm E}$ ] Einstein delay
\item[$\Delta_{\rm S}$ ] Signal propagation delay
\item[$\Delta_{\rm A}$ ] Aberration delay
\item[$u$ ] Relativistic eccentric anomaly
\item[$\theta$ ] True anomaly
\item[$\beta_{\rm O}$ ] Dimensionless orbital velocity $(G M n_{\rm b})^{1/3}/c$ 
\item[$k$ ] Periastron advance parameter
\item[$\dot\omega$ ] Secular periastron advance ($\dot\omega \equiv n_\mathrm{b}k$) 
\item[$\dot{P}_{\rm b}$ ] Change of orbital period 
\item[$\gamma_{\rm E}$ ] Einstein delay amplitude 
\item[$r$ ] Range parameter of the Shapiro delay
\item[$s$ ] Shape parameter of the Shapiro delay; Scintillation scale; Fractional scattering screen distance 
\item[$t$ ] Time
\item[$z_s$ ] Logarithmic Shapiro shape, $z_s \equiv -\ln(1-s)$ 
\item[$\Lambda_u$ ] Argument of the Shapiro delay, see Eq.~(\ref{eq:ShaLO})
\item[$\lambda_\mathrm{LT}$ ] Scaling parameter of Lense-Thirring precession
\item[$q_{\rm NLO}$ ] NLO factor for signal propagation
\item[$\delta_\theta$ ] Relativistic deformation of orbit
\item[${\cal A}$, ${\cal B}$ ] aberration coefficients, see Eq.~(\ref{eq:aberration})
\item[$\dot{x}$ ] Change of projected semimajor axis
\item[$\dot{e}_T$ ] Change of eccentricity
\item[$\Omega_\mathrm{B}^\mathrm{spin}$ ] Rate of relativistic spin precession of B
\item[$i$ ] Orbital inclination
\item[$M$ ] Total system mass 
\item[$m_{\rm A}$ ] Mass of pulsar A 
\item[$m_{\rm B}$ ] Mass of pulsar B 
\item[$X_{\rm A,B}$ ] $m_{\rm A, B}/M$
\item[$R$ ] Mass ratio $m_{\rm A}/m_{\rm B}$
\item[$I_{\rm A}$ ] MoI of pulsar A 
\item[$\alpha_0$, $\beta_0$ ] Coupling strength parameters in STG
\item[$T_1(\alpha_0$, $\beta_0)$] Damour--Esposito-Far{\`e}se STG
\item[$\alpha_a$ ] Effective scalar coupling of body $a$ ($a = \mathrm{A,B}$)
\item[$\beta_a$  ] Derivative of $\alpha_a$ w.r.t.\ scalar field
\item[$k_\mathrm{A}$ ] Derivative of $-\ln I_\mathrm{A}$ w.r.t.\ scalar field
\item[$\gamma^\mathrm{PPN}$] Spatial-curvature parameter of PPN formalism
\item[$\beta^\mathrm{PPN}$] Nonlinearity parameter of PPN formalism
\item[$\gamma_{ab}$] Body-dependent strong-field generalization of $\gamma^\mathrm{PPN}$ ($a,b \in \{\mathrm{A,B,0}\}$)
\item[$\beta_{ab}^c$] Body-dependent strong-field generalization of $\beta^\mathrm{PPN}$ ($a,b,c \in \{\mathrm{A,B}\}$)
\end{description}

\vspace{12pt}
\noindent
{\bf Abbreviations}

\begin{description}[leftmargin=!,labelwidth=3em ]
\item[{\rm BH} ]  Black hole
\item[{\rm DM} ]  Dispersion Measure
\item[{\rm EoS} ] Equation of state
\item[{\rm GLSQ} ] Generalized least-squares timing method
\item[{\rm GR} ]  General relativity 
\item[{\rm GW} ]  Gravitational wave 
\item[{\rm ISM} ] Interstellar medium
\item[{\rm LoS} ] Line of sight
\item[{\rm LT} ]  Lense-Thirring
\item[{\rm MJD} ] Modified Julian Date 
\item[{\rm MoI} ] Moment of inertia 
\item[{\rm NLO} ] Next-to-leading-order
\item[{\rm NS} ]  Neutron star
\item[{\rm PK} ]  Post-Keplerian
\item[{\rm PN} ]  Post-Newtonian
\item[{\rm PPN} ] Parametrized post-Newtonian
\item[{\rm PSD} ] Power spectral density
\item[{\rm SEP} ] Strong equivalence principle
\item[{\rm SSB} ] Solar system barycentre
\item[{\rm SF} ]   Structure function
\item[{\rm S/N} ]  Signal-to-noise ratio
\item[{\rm STG} ]  Scalar-tensor gravity
\item[{\rm ToA} ]  Time of arrival
\item[{\rm VLBA} ] Very Long Baseline Array
\item[{\rm VLBI} ] Very Long Baseline Interferometry
\end{description}


%

\end{document}